\documentclass{aa}
\usepackage{natbib}
\usepackage{graphics}
\usepackage{amssymb}
\begin{document}
\title{Molecular inventories and chemical evolution of low-mass
protostellar envelopes} \author{J.K. J{\o}rgensen\inst{1}
\and F.L. Sch\"{o}ier\inst{1,2} \and E.F. van Dishoeck\inst{1}}
\titlerunning{Molecular inventories of low-mass
protostellar envelopes}
\institute{Leiden Observatory, P.O. Box 9513, NL-2300 RA Leiden, The
Netherlands \and Stockholm Observatory, AlbaNova, SE-106 91 Stockholm,
Sweden}
\offprints{Jes K.\,J{\o}rgensen}
\mail{joergensen@strw.leidenuniv.nl} \date{Received <date> / Accepted
<date>}

\abstract{This paper presents the first substantial study of the
chemistry of the envelopes around a sample of 18 low-mass pre- and
protostellar objects for which physical properties have previously
been derived from radiative transfer modeling of their dust continuum
emission. Single-dish line observations of 24 transitions of 9
molecular species (not counting isotopes) including HCO$^+$,
N$_2$H$^+$, CS, SO, SO$_2$, HCN, HNC, HC$_3$N and CN are reported. The
line intensities are used to constrain the molecular abundances by
comparison to Monte Carlo radiative transfer modeling of the line
strengths. In general the nitrogen-bearing species together with
HCO$^+$ and CO cannot be fitted by a constant fractional abundance
when the lowest excitation transitions are included, but require
radial dependences of their chemistry since the intensity of the
lowest excitation lines are systematically underestimated in such
models. A scenario is suggested in which these species are depleted in
a specific region of the envelope where the density is high enough
that the freeze-out timescale is shorter than the dynamical timescale
and the temperature low enough that the molecule is not evaporated
from the icy grain mantles. This can be simulated by a ``drop''
abundance profile with standard (undepleted) abundances in the inner-
and outermost regions and a drop in abundance in between where the
molecule freezes out. An empirical chemical network is constructed on
the basis of correlations between the abundances of various
species. For example, it is seen that the HCO$^+$ and CO abundances
are linearly correlated, both increasing with decreasing envelope
mass. This is shown to be the case if the main formation route of
HCO$^+$ is through reactions between CO and H$_3^+$, and if the CO
abundance still is low enough that reactions between H$_3^+$ and N$_2$
are the main mechanism responsible for the removal of H$_3^+$. Species
such as CS, SO and HCN show no trend with envelope mass. In particular
no trend is seen between ``evolutionary stage'' of the objects and the
abundances of the main sulfur- or nitrogen-containing species. Among
the nitrogen-bearing species abundances of CN, HNC and HC$_3$N are
found to be closely correlated, which can be understood from
considerations of the chemical network. The CS/SO abundance ratio is
found to correlate with the abundances of CN and HC$_3$N, which may
reflect a dependence on the atomic carbon abundance. An
anti-correlation is found between the deuteration of HCO$^+$ and HCN,
reflecting different temperature dependences for gas-phase deuteration
mechanisms. The abundances are compared to other protostellar
environments. In particular it is found that the abundances in the
cold outer envelope of the previously studied class 0 protostar
IRAS~16293-2422 are in good agreement with the average abundances for
the presented sample of class 0 objects.

} \maketitle

\section{Introduction}
Understanding the chemistry of protostellar environments is important
in order to build up a complete and consistent picture for the
star-formation process. Detailed knowledge about the chemistry is
required in order to fully understand the physical processes since it
regulates the ionization balance and the gas temperature through
cooling of the molecular gas, for example. At the same time the
chemistry may potentially serve as a valuable tool, both as a time
indicator for the protostellar evolution and as a diagnostic of the
properties of different components of young stellar objects. In this
paper a molecular inventory for the envelopes around a sample of
low-mass protostars is presented. Compared to previous studies this is
the first substantial study of the chemical composition of a
significant sample of low-mass protostars. These objects have formed a
central protostar, but are still deeply embedded in their envelope and
represent the first stage after the collapse of the dark cloud
cores. Such objects differ from the pre-stellar cores for which
chemical studies have previously been performed by, e.g.,
\cite{bergin01}, \cite{tafalla02} and \cite{lee03}, in that the
central source heats the envelope and dominates the energy balance
rather than, e.g., the external interstellar radiation field.

Other studies of the chemistry of low-mass protostellar objects
include those in the Serpens region by \cite{hogerheijde99} and of
specific molecules such as the sulfur-bearing species and deuterated
molecules \citep[e.g.,][]{buckle03,roberts02}. Single objects, such as
the low-mass protostar \object{IRAS~16293-2422}, have been the target
of numerous studies
\cite[e.g.,][]{blake94,vandishoeck95,ceccarelli98,schoeier02,cazaux03}. This
object is particularly interesting because of its rich spectrum and
evidence for evaporation of ices in the inner hot regions
\citep{ceccarelli00a,ceccarelli00b,schoeier02}. One of the questions
that can be addressed with this study is how representative the
chemistry of \object{IRAS~16293-2422} is compared to that in other
low-mass protostellar objects.

One of the major steps forward in this line of research in recent
years has been the observations and analysis of the (sub)millimeter
continuum emission from the dust around pre- and protostellar objects
using bolometer cameras such as SCUBA on the JCMT
\citep[e.g][]{chandler00, hogerheijde00sandell, evans01, motte01,
jorgensen02, shirley02, schoeier02} and infrared extinction studies
\citep[e.g.,][]{alves01,harvey01}. By fitting the radial distributions
of the continuum emission and SEDs of the objects, the dust component
and physical structure of the envelopes can be constrained, and, with
assumptions about the gas-to-dust ratio and gas-dust temperature
coupling, the physical properties of the gas in the envelope can be
derived. Such physical models can then be used as the basis for
determining the molecular excitation and for deriving abundances
relative to H$_2$ by comparing to molecular line observations
\citep[e.g.,][]{bergin01,jorgensen02,schoeier02,tafalla02,lee03}.

Based on these methods it has become increasingly clear that large
variations of molecular abundances can occur in protostellar
environments. Examples are depletion of molecules at low temperatures
due to freeze-out on dust grains
\citep[e.g.,][]{caselli99,jorgensen02,tafalla02} and enhancements of
molecular species in warm regions where ices evaporate
\citep{ceccarelli00a,ceccarelli00b,schoeier02} or in shocked gas
associated with protostellar outflows or jets
\citep{bachiller95,bachiller97,i2art}.

\cite{jorgensen02}\defcitealias{jorgensen02}{Paper~I} (Paper I in the
following) established the physical properties of the envelopes around
a sample of low-mass protostars from 1D radiative transfer modeling of
SCUBA dust continuum maps. The derived density and temperature
structure and size was used as input for modeling CO (sub)millimeter
line emission. It was found that the CO lines could be reproduced with
the physical models assuming constant fractional abundances with
radius. The derived values for the envelopes with the most massive
envelopes - typically interpreted as the ``youngest'' class 0
protostars - were found to be lower than abundances quoted for nearby
dark clouds by an order of magnitude. In contrast the potentially more
evolved class I objects were found to have envelopes with CO
abundances closer to the dark cloud value. It was suggested that this
was related to CO freezing out on dust grains at low temperatures and
in dense environments.

This paper is a continuation of \citetalias{jorgensen02} and the
analysis of the class 0 YSO, \object{IRAS~16293-2422} presented by
\cite{schoeier02}. Based mainly on JCMT and Onsala 20~m observations,
abundances for a large number of molecules are derived using detailed
Monte Carlo radiative transfer for the full set of pre- and
protostellar objects presented in \citetalias{jorgensen02} with the
envelope parameters derived in that paper as input. The combination of
low $J$ 3~mm observations (from the Onsala telescope) and higher $J$
lines from the JCMT allows the radial variation of the chemistry to be
discussed with the low $J$ lines mainly sensitive to the outer cold
part of the envelope and the high $J$ lines to the inner dense
regions. Similar analyzes for H$_2$CO, CH$_3$OH and more complex
organic species, which are particularly sensitive to the innermost hot
core region, are presented in separate papers (\citealt{maret03};
J{\o}rgensen et al. in prep.)

The paper is laid out as follows: in Sect.~\ref{observations} the
details of the observations and reduction are presented. The modeling
approach is described in Sect.~\ref{modeling} and caveats and
implications for the radial structure described. Relations between the
different molecular species are discussed in Sect.~\ref{discussion}.

\section{Observations}\label{observations}
\subsection{Observational details}
The principal data set forming the basis of this work was obtained
with the James Clerk Maxwell Telescope (JCMT) on Mauna Kea,
Hawaii\footnote{The JCMT is operated by the Joint Astronomy Centre in
Hilo, Hawaii on behalf of the parent organizations: PPARC in the
United Kingdom, the National Research Council of Canada and The
Netherlands Organization for Scientific Research} where 15 sources
were observed between February 2001 and February 2003. In addition
archival data for 3 class I sources - \object{L1551-I5}, \object{TMC1}
and \object{TMC1A} - observed previously in a number of these settings
were used.

The A3 and B3 receivers at 210-270~GHz and 315-370~GHz, respectively,
were used with the digital autocorrelation spectrometer (DAS) in
setups with bandwidths ranging from 125~MHz to 500~MHz with resulting
resolutions of 0.1 to 0.6~km~s$^{-1}$. Each setting was observed with on
source integration times ranging from 10 to 60 minutes per mixer
reaching a typical RMS (on $T_A^\ast$ scale) of 0.03 to 0.05~K in 30
minutes. The pointing accuracy for the JCMT was found to be a few
arcseconds. The calibration was checked by comparison to spectral line
standards and was estimated to be accurate to approximately 20\%, when
comparing data taken in separate runs. For most sources beam switching
with a chop of 180\arcsec\ was used. The only exception was N1333-I4A
and -I4B for which position switching to an emission-free position at
(-120\arcsec, 250\arcsec) was used.

Further observations at 3~millimeter (85 to 115~GHz) were obtained
with the Onsala Space Observatory 20~m telescope\footnote{The Onsala
20~m telescope is operated by the Swedish National Facility for Radio
Astronomy, Onsala Space Observatory at Chalmers University of
Technology.} in observing runs in March 2002 and May 2003. The entire
sample was observed at Onsala in the same species, except the two
$\rho$ Ophiuchus sources \object{L1689B} and \object{VLA1623} which
are located too far south. These two sources were observed in early
April 2003 at 3~mm using the Swedish-ESO Submillimeter Telescope
(SEST)\footnote{The SEST is operated by the Swedish National Facility
for Radio Astronomy on behalf of the Swedish Natural Science Research
Council and the European Southern Observatory.} at La Silla in
Chile. Finally CS and C$^{34}$S 2--1 and 3--2 spectra were taken for a
few sources in November 2001 with the IRAM 30~m telescope\footnote{The
IRAM 30~m telescope is operated by the Institut de Radio Astronomie
Millim\'etrique, which is supported by the Centre National de
Recherche Scientifique (France), the Max Planck Gesellschaft (Germany)
and the Instituto Geogr\'afico Nacional (Spain).} at Pico Veleta,
Spain in the range 90 to 250~GHz.

In addition to the observed settings the public JCMT archive was
searched for useable data and included to constrain the models
together with previously published observations. All spectra were
calibrated at the telescopes onto the natural antenna temperature
scale, $T_A^\ast$, using the chopper-wheel method
\citep{kutner81}. The spectra were corrected for the telescope beam
and forward scattering efficiencies and brought onto the main beam
brightness scale, $T_{\rm mb}$, by division with the appropriate main
beam efficiencies, $\eta_{\rm mb}$ (or $F_{\rm eff}/B_{\rm eff}$ in
the terminology adopted at the IRAM 30~m telescope). Finally a low
order polynomial baseline was subtracted for each spectrum. An
overview of the observed lines is given in Table~\ref{datasum}.

\begin{table}
\caption{Summary of the observed lines.}\label{datasum}
\begin{tabular}{llll}\hline\hline
Molecule & Line          & Frequency & Telescope       \\ \hline
CS       & 2--1          & \phantom{1}97.9810 & OSO, SEST \\
         & 3--2          & 146.9690           & IRAM            \\
         & 5--4          & 244.9356           & JCMT, IRAM      \\
         & 7--6          & 342.8830           & JCMT            \\
C$^{34}$S    & 2--1          & \phantom{1}96.4129 & OSO, IRAM, SEST \\
         & 5--4          & 241.0161           & JCMT            \\
H$^{13}$CO$^+$  & 1--0          & \phantom{1}86.7543 & OSO, SEST       \\
         & 3--2          & 260.2555           & JCMT            \\
         & 4--3          & 346.9985           & JCMT            \\
DCO$^+$   & 3--2          & 216.1126           & JCMT            \\
N$_2$H$^+$   & 1--0$^a$      & \phantom{1}93.1737 & OSO, SEST       \\
HCN      & 4--3          & 354.5055           & JCMT            \\
H$^{13}$CN   & 1--0$^a$      & \phantom{1}86.3402 & OSO, SEST       \\
         & 3--2          & 259.0118           & JCMT            \\
DCN      & 3--2          & 217.2386           & JCMT            \\
HNC      & 1--0          & \phantom{1}90.6636 & OSO, SEST       \\
         & 4--3          & 362.6303           & JCMT            \\
CN       & 1--0$^a$      & 113.4910           & OSO, SEST       \\
         & 3--2$^a$      & 340.2478           & JCMT            \\
HC$_3$N  & 10--9         & \phantom{1}90.9790 & OSO, SEST       \\
SO       & $2_3-1_2$     & \phantom{1}99.2999 & OSO, SEST       \\ 
         & $8_7-7_6$     & 340.7142           & JCMT            \\
SO$_2$   & $3_{1,3}-2_{0,2}$ & 104.0294       & OSO, SEST       \\
         & $9_{3,7}-9_{2,8}$ & 258.9422       & JCMT            \\  \hline
\end{tabular}

Notes: $^a$Hyperfine splitting observable.
\end{table}

\subsection{Resulting spectra}
\begin{figure*}
\resizebox{\hsize}{!}{\includegraphics{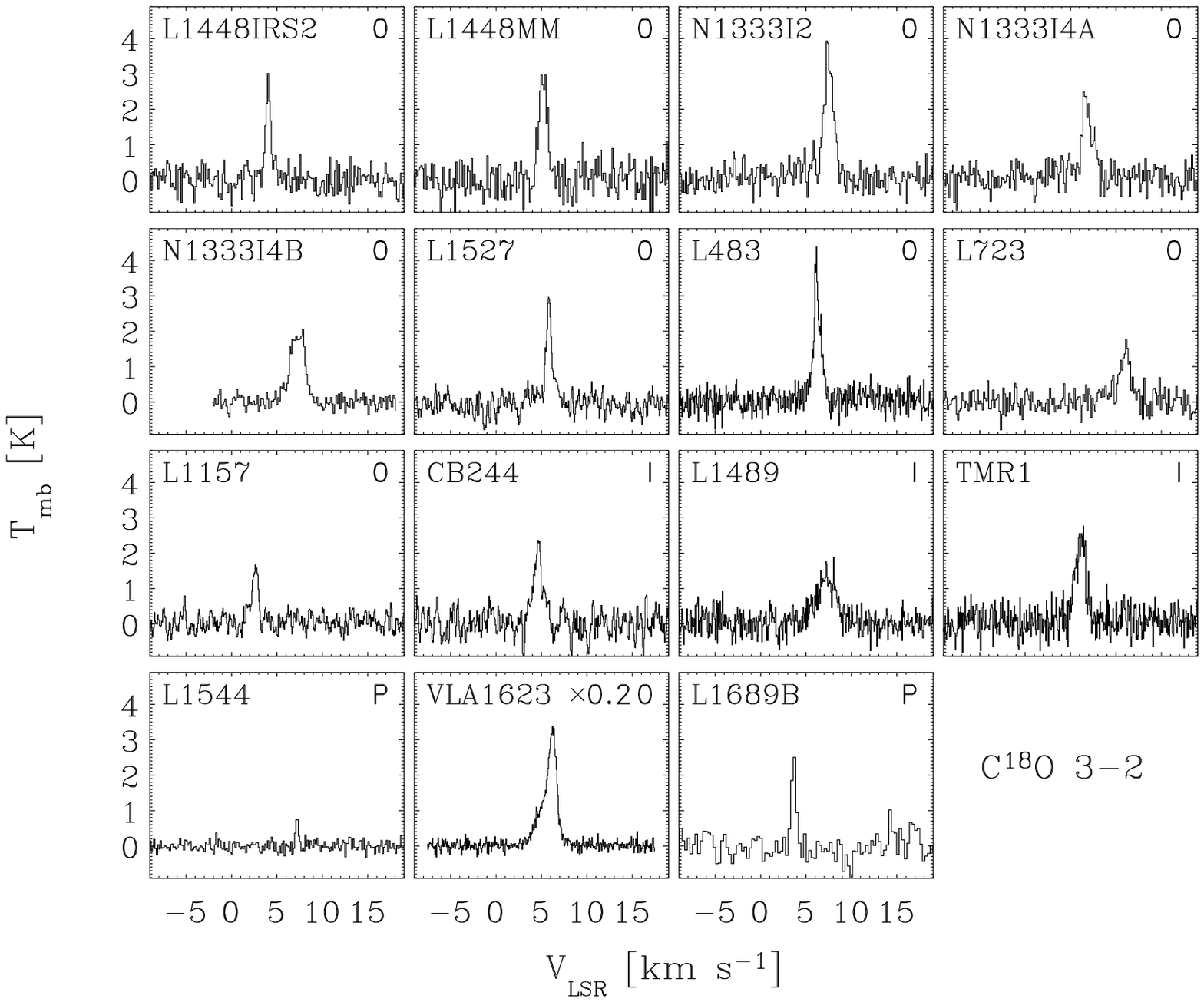}\includegraphics{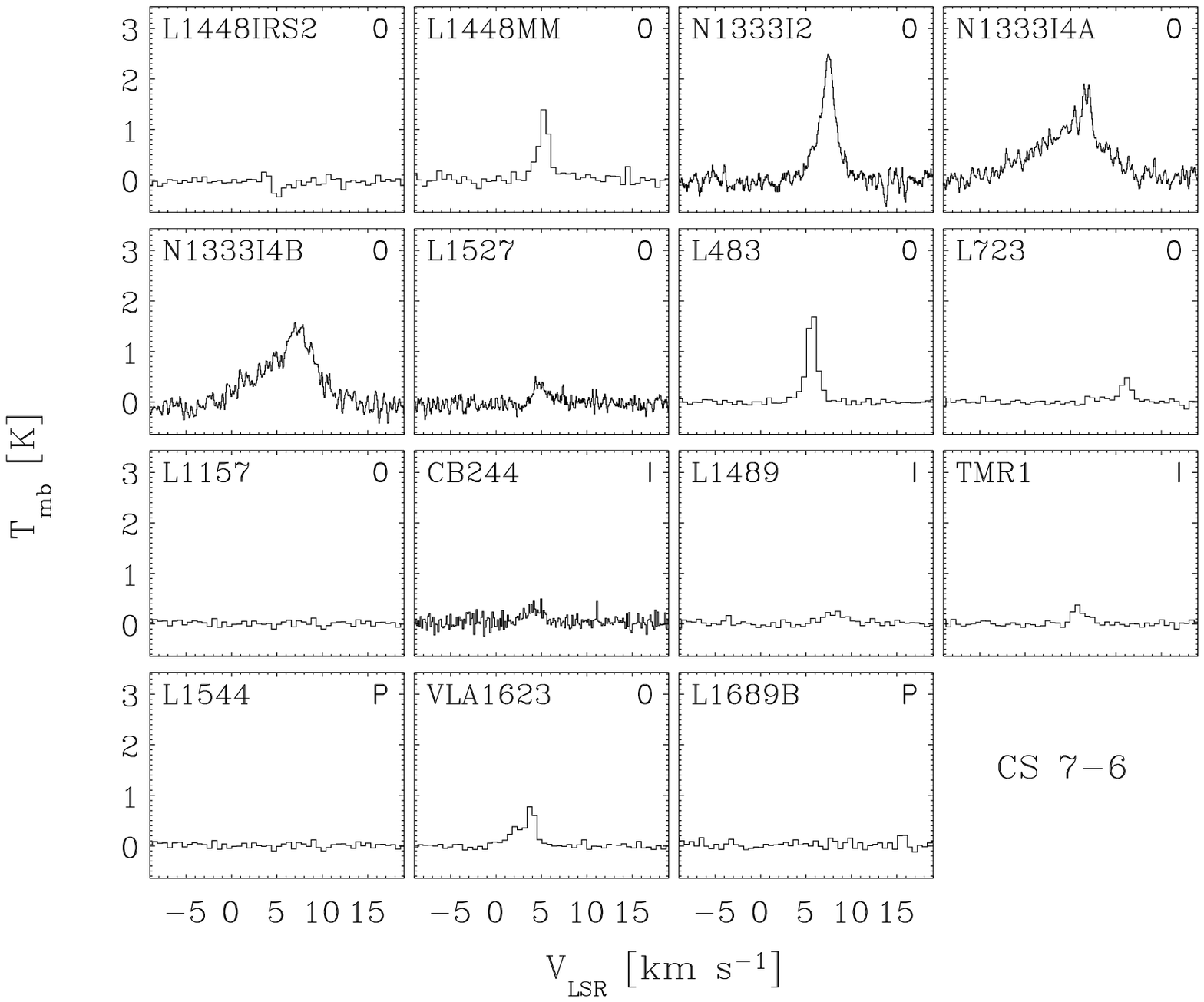}}
\caption{Spectra of C$^{18}$O $J=3-2$ (left) and CS $J=7-6$ (right)
from JCMT observations. In this figure, and
Fig.~2-\ref{overview_last}, the classes of the individual objects are
indicated in the upper right corner of each plot by ``0'' for the
class 0 objects (envelope mass $> 0.5 M_\odot$), ``I'' for the class I
objects (envelope mass $< 0.5 M_\odot$) and ``P'' for the pre-stellar
cores.}\label{overview_first}
\end{figure*}

\begin{figure*}
\resizebox{\hsize}{!}{\includegraphics{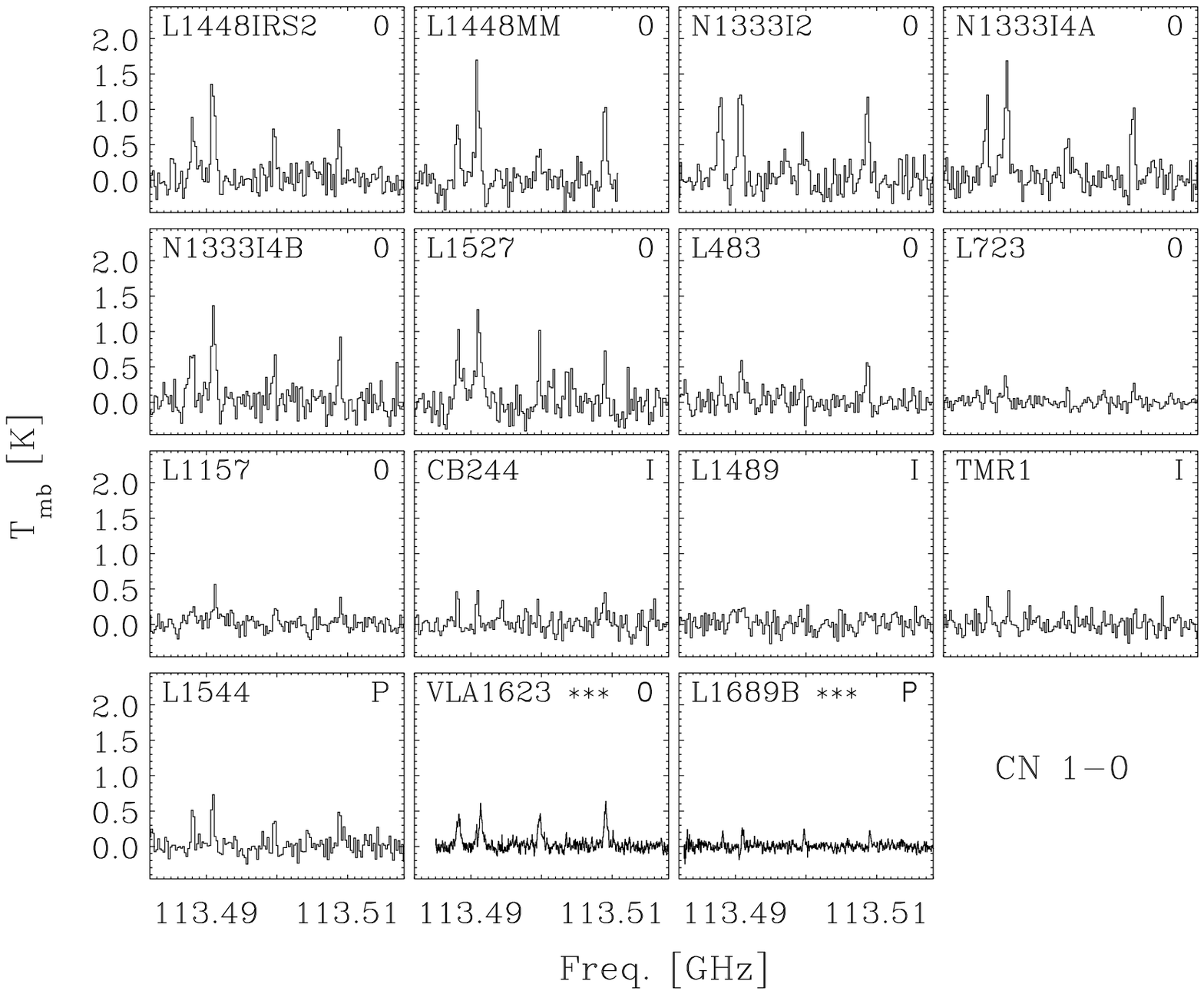}\includegraphics{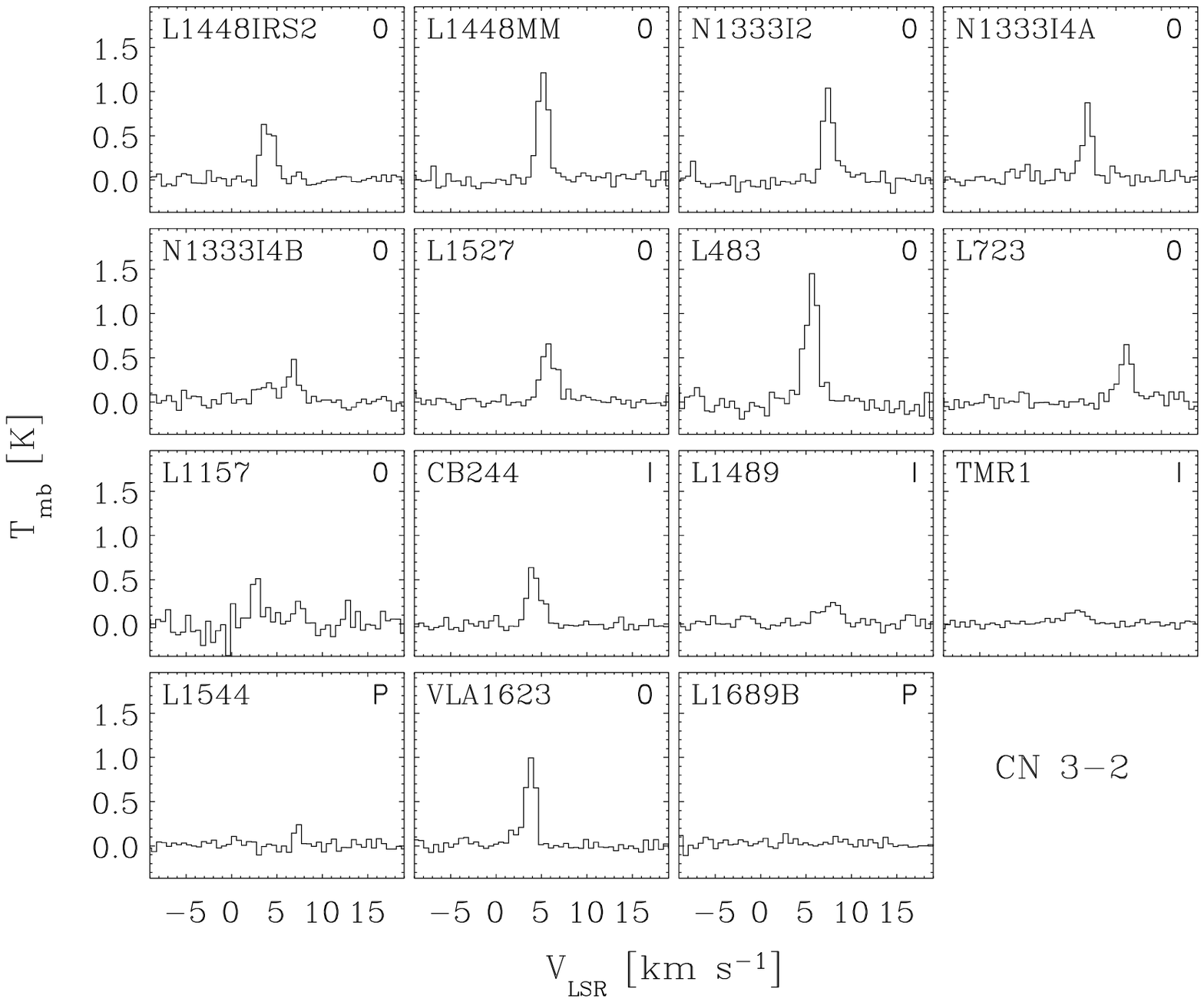}}
\caption{Spectra of CN $J=1-0$ (left) and $J=3-2$ (right). The $J=1-0$
observations are from the Onsala 20~m telescope and the SEST (marked
with ***), the $J=3-2$ observations are from the JCMT.}
\label{overview_last}\label{cn_spectra_fig}
\end{figure*}

Spectra of selected molecular transitions are presented in
Fig.~\ref{overview_first}-\ref{overview_last}. In order to derive the
line intensities, Gaussians were fitted to each line. For the few
asymmetric lines the emission was integrated over $\pm$~2~km~s$^{-1}$ from
the systemic velocity of the given source. The integrated line
intensities are listed in
Tables~\ref{intens_first}-\ref{intens_last}. In case of non-detection,
the 2$\sigma$ upper limit is given where $\sigma=1.2\sqrt{\Delta
v\,\delta v}\,\sigma_{\rm rms}$ with $\Delta v$ the expected line
width ($\approx$~1~km~s$^{-1}$ for the observed sources/molecules), $\delta
v$ the channel width in the given spectral line-setup and $\sigma_{\rm
rms}$ the rms noise in the observed spectra for the specific channel
width. The factor 1.2 represents the typical 20\% calibration
uncertainty found by comparing to spectral line standards and
observations from different nights.

For most sources the line profiles are quite symmetric and can be
well-represented by the Gaussians: the main exceptions are the strong,
optically thick HCN 4-3 and CS lines toward especially N1333-I4A and
-I4B. The HCN and CS lines toward these objects seem to be dominated
by outflow emission. SO$_2$ is only detected in the low excitation
$3_{1,3}-2_{0,2}$ line toward N1333-I4A and -I4B, and the two objects
in $\rho$ Oph, \object{VLA1623} and \object{L1689B}. The higher
excitation $9_{3,7}-9_{2,8}$ line was also observed in a setting
together with H$^{13}$CN 3--2 but was not detected toward any
source. Furthermore the high $J$ lines of SO were also only detected
toward the objects in NGC~1333 and toward \object{VLA1623}, suggesting
a chemical effect.

Some systematic trends can be seen from the Tables and Figures. In
general the lines are significantly weaker than those found in
\object{IRAS~16293-2422} \citep{blake94,vandishoeck95}. Especially for
the Class I objects in our sample (i.p., \object{L1489} and
\object{TMR1}) a number of usually quite strong lines (e.g., HCN 4--3)
were not detected. The effects of the chemistry are also hinted at by
comparing, e.g., the source to source variations of the HNC 4--3 and
CN 3--2 spectra. An interesting effect can be seen for the
deuterium-bearing species: note that the DCO$^+$ 3--2 lines are
detected toward the pre-stellar cores but not the H$^{13}$CO$^+$ 3--2 lines
and vice versa for the class I objects, clearly indicating a higher
degree of deuteration in the colder pre-stellar cores. How much of
this can be attributed to excitation and simple mass or distance
effects is, however, not clear. \object{IRAS~16293-2422} for example
has the most massive envelope compared to the other class 0 objects
and is also located closer than, e.g., the other massive sources in
the NGC~1333 region. In contrast the class I objects by their very
definition have the least massive envelopes, so the absence of lines
toward some of these objects may simply reflect lower column densities
for the observed species for these sources. In order to address this
in more detail it is necessary to model the full line radiative
transfer as was done in \cite{jorgensen02} and \cite{schoeier02}.

\begin{table*}
\caption{Line intensities ($\int T_{\rm MB}\,{\rm d}v$) for CS,
C$^{34}$S and SO transitions from the JCMT and Onsala 20~m
telescope.}\label{intens_first}
\begin{tabular}{lllllll} \hline\hline
            & \multicolumn{2}{c}{CS} & \multicolumn{2}{c}{C$^{34}$S}          & \multicolumn{2}{c}{SO}\\
            & 5--4     & 7--6     & 2--1     & 5--4     & $2_3-1_2$ & $8_7-7_6$ \\ \hline
  L1448-I2  &   0.31   & $<0.18$  &    0.39  & $\ldots$ &    1.7   & $<0.053$   \\
   L1448-C  &   2.2    &  2.0     &    0.47  &   0.26   &    1.7   & $<0.096$   \\
 N1333-I2   &   4.0    &  4.9     &  IRAM$^a$&   0.66   &    2.9   &    1.7     \\
 N1333-I4A  &   7.9    &  4.8     &  IRAM$^a$&   1.25   &    7.4   &    5.8     \\
 N1333-I4B  &   6.3    &  4.7     &  IRAM$^a$&   0.79   &    6.3   &    1.8     \\
     L1527  &   1.8    &  0.45    &  IRAM$^a$& $<0.061$ &    0.42  & $<0.067$   \\
   VLA1623  &   4.3    &  1.6     &  SEST$^a$ &   0.28  &  SEST$^a$&    0.98    \\
      L483  &   4.1    &  2.5     &    0.38  &   0.47   &    1.4   & $<0.13$    \\
      L723  &   4.3    &  0.67    &    0.18  &   0.29   &    1.4   & $<0.12$    \\
     L1157  &   0.93   & $<0.094$ &    0.20  & $<0.075$ &    2.1   & $<0.21$    \\
     CB244  &   1.9    &  0.87    &    0.16  & $<0.12$  &    0.87  & $<0.074$   \\
     L1489  &   0.62   &  0.71    & $<0.040$ & $<0.098$ &    0.26  & $<0.083$   \\
      TMR1  &   0.77   &  0.60    & $\ldots$ & $<0.086$ &    0.28  & $<0.054$   \\
     L1551  &   3.2    &  3.4     & $\ldots$ &   0.35   &    0.59  & $\ldots$   \\
      TMC1  &   0.46   & $\ldots$ & $\ldots$ & $\ldots$ &    0.39  & $\ldots$   \\
     TMC1A  &   0.58   & $\ldots$ & $\ldots$ & $\ldots$ &    0.12  & $\ldots$   \\
     L1544  & $<0.14$  & $<0.094$ &   0.22   & $\ldots$ &    0.82  & $<0.13$    \\
    L1689B  & $<0.16$  & $<0.12$  & SEST$^a$ & $<0.028$ &  SEST$^a$& $<0.099$   \\ \hline\hline
\end{tabular}

$^a$3~mm observations from the IRAM 30~m and SEST telscopes given in
Table~\ref{iramobs} and \ref{sestobs}, respectively.
\end{table*}

\begin{table*}
\caption{Line intensities ($\int T_{\rm MB}\,{\rm d}v$) for the
H$^{13}$CO$^+$, N$_2$H$^+$, DCO$^+$ and DCN transitions from the JCMT
and Onsala 20~m telescope.}
\begin{tabular}{lllllll} \hline\hline
            & \multicolumn{3}{c}{H$^{13}$CO$^+$} & N$_2$H$^+$ & DCO$^+$ & DCN  \\
            &   1--0   &   3--2      &  4--3       & 1--0             &   3--2   & 3--2 \\ \hline
  L1448-I2  &   1.6    &   0.89      & $\ldots$    & 10.5             &   0.91   &  0.13    \\ 
   L1448-C  &   2.0    &   1.9       & $\ldots$    & 11.7             &   1.9    &  0.40    \\
  N1333-I2  &   1.8    &   2.1       &    2.7      & 14.2             &   0.95   &  0.46    \\
 N1333-I4A  &   2.3    &   1.4       & $\ldots$    & 15.9             &   3.2    &  0.40    \\ 
 N1333-I4B  &   2.1    &   0.57      & $\ldots$    & 13.5             &   2.0    &  0.31    \\
     L1527  &   2.2    &   1.1       &    0.47     & 4.4              &   0.70   &  0.10    \\
   VLA1623  &  SEST$^a$&   4.7       &    3.0      &SEST$^a$          &   3.6    &  0.43    \\
      L483  &   1.5    &   1.7       &    1.2      & 13.7             &   0.85   &  0.25    \\
      L723  &   0.61   &   0.94      &    0.70     & 4.8              &   0.36   & $<0.064$ \\
     L1157  &   0.89   &   0.61      &    0.52     & 7.3              &   0.64   & $<0.078$ \\
     CB244  &   0.87   &   0.73      &    0.43     & 6.3              &   0.34   & $<0.083$ \\
     L1489  &   0.96   &   0.82      & 0.61$^b$    & 0.26             & $<0.092$ & $<0.089$ \\
      TMR1  &   1.1    &   0.51      & 0.20$^b$    & 0.37             & $<0.095$ & $<0.089$ \\
     L1551  & $\ldots$ & 2.4$^b$     & 2.2$^b$     & 15.7             &   1.3   & $\ldots$ \\
      TMC1  & $\ldots$ & $<$0.27$^b$ & $<$0.12$^b$ & 5.8 & $<0.16$    & $\ldots$ \\
     TMC1A  & $\ldots$ & $<$0.27$^b$ & $<$0.13$^b$ & 7.1 & $<0.12$    & $\ldots$ \\
     L1544  &   0.92   & $<$0.10     & $\ldots$    & 6.0 	      &   0.56   &  0.12    \\
    L1689B  &  SEST$^a$&   0.22      & $\ldots$    &SEST$^a$          &   0.55   & $<0.107$ \\ \hline\hline
\end{tabular}

$^a$SEST 3~mm observations of the southern sources given in
Table~\ref{sestobs}. $^b$from \cite{hogerheijde97}.
\end{table*}
 
\begin{table*}
\caption{Line intensities ($\int T_{\rm MB}\,{\rm d}v$) for the HCN,
H$^{13}$CN, HNC and CN transitions from the JCMT and Onsala 20~m
telescope.}
\begin{tabular}{lllllllllll} \hline\hline
            & HCN      & \multicolumn{2}{c}{H$^{13}$CN} & \multicolumn{2}{c}{HNC} & \multicolumn{5}{c}{CN}\\
            &  4--3    &  1--0    &   3--2   &  1--0  &  4--3  & $1_{022}-0_{011}$ & $1_{023}-0_{012}$   & $1_{021}-0_{011}$ & $1_{022}-0_{012}$ & 3--2$^{b}$\\  \hline
  L1448-I2  &    1.4   &   0.44   &  $<0.13$ &  8.2   &   0.60 &    1.3   &    2.7   &   0.88   & 0.92    &   1.2    \\
   L1448-C  &    5.3   &   0.82   &    0.50  &  8.6   &   3.4  &    1.4   &    2.5   &   0.87   & 1.6     &   1.9    \\
  N1333-I2  &    5.3   &   0.59   &    0.62  &  8.9   &   3.3  &    2.5   &    2.8   &   1.1    & 1.7     &   1.5    \\
 N1333-I4A  &    6.6   &   1.5    &    0.65  & 11.6   &   2.8  &    1.9   &    3.0   &   1.2    & 1.8     &   1.1    \\
 N1333-I4B  &    9.6   &   0.96   &    0.70  &  4.4   &   2.0  &    1.8   &    2.5   &   0.88   & 1.2     &   0.73   \\
     L1527  &    1.4   &   0.43   & $<0.067$ &  3.8   &   0.40 &    1.4   &    2.2   &   1.0    & 0.81    &   1.3    \\
   VLA1623  &    2.2   &  SEST$^a$& $<0.153$ &SEST$^a$&   2.5  & \multicolumn{4}{l}{\rule[0.5ex]{2.5cm}{0.1mm}~SEST$^a$~\rule[0.5ex]{2.5cm}{0.1mm}} &   1.4    \\
      L483  &    3.5   &   0.69   &    0.52  &  3.7   &   2.2  &    0.63  &    1.4   &   0.46   & 0.99    &   2.4    \\
      L723  &    1.7   &   0.16   & $<0.080$ &  2.8   &   1.3  &    0.28  &    0.51  &   0.21   & 0.31    &   1.1    \\
     L1157  &    1.1   &   0.23   & $<0.087$ &  3.7   &   1.0  &    0.57  &    0.87  &   0.37   & 0.35    &   0.40   \\
     CB244  &    3.2   &   0.14   & $<0.093$ &  3.7   &   0.74 &    0.54  &    0.62  &   0.31   & 0.43    &   1.1    \\
     L1489  &    0.85  &   0.29   & $<0.087$ &  2.1   &   1.0  &  $<0.14$ &    0.63  &  $<0.14$ & $<0.14$ &   0.67   \\
      TMR1  &    0.63  &   0.15   & $<0.15$  &  2.3   &   0.33 &  $<0.14$ &    0.37  &  $<0.14$ & $<0.14$ &   0.47   \\
     L1551  & $\ldots$ &  $<0.12$ & $\ldots$ &  9.0   &   2.97 &    1.4   &    1.8   &   1.2    & 1.1     & $\ldots$ \\
      TMC1  & $\ldots$ & $<0.094$ & $\ldots$ &  3.2   &$<$0.27 &    0.63  &    0.72  &   0.82   & 0.30    & $\ldots$ \\
     TMC1A  & $\ldots$ & $<0.092$ & $\ldots$ &  3.8   &   0.29 &    0.55  &    0.79  &   0.49   & 0.49    & $\ldots$ \\ 
     L1544  & $<0.12$  &   0.70   &  $<0.11$ &  4.0   &   0.35 &    0.56  &    0.90  &   0.44   & 0.65    &   0.21   \\
    L1689B  & $<0.11$  &  SEST$^a$&  $<0.10$ &SEST$^a$&$<$0.24 & \multicolumn{4}{l}{\rule[0.5ex]{2.5cm}{0.1mm}~SEST$^a$~\rule[0.5ex]{2.5cm}{0.1mm}} & $<0.092$ \\ \hline\hline
\end{tabular}

Notes: $^{a}$SEST 3~mm observations of the southern sources given in
Table~\ref{sestobs}. $^{b}$Hyperfine components blended.
\end{table*}

\begin{table*}[!h]
\caption{Line intensities ($\int T_{\rm MB}\,{\rm d}v$) for CS and
C$^{34}$S transitions from the IRAM~30~m telescope.}\label{iramobs}
\begin{tabular}{llll} \hline\hline
          & \multicolumn{2}{c}{CS} & C$^{34}$S \\  
          & 3--2       & 5--4  & 2--1 \\ \hline
L1448-C   & 2.7        & 2.9   & 0.45 \\
N1333-I2  & 6.2        & 9.4   & 1.3  \\
N1333-I4A & 11.7       & 15.1  & 1.6  \\
N1333-I4B & 6.7        & 10.3  & 0.73 \\
L1527     & $\ldots^a$ & 1.8   & 0.11 \\ \hline
\end{tabular}

Notes: $^a$2$\sigma$ upper limit of C$^{34}$S 3--2 intensity of
0.064~K~km~s$^{-1}$.
\end{table*}

\begin{table*}[!h]
\caption{Line intensities ($\int T_{\rm MB}\,{\rm d}v$) for the 3~mm
observations of the southern sources from the
SEST.}\label{sestobs}\label{intens_last}
\begin{tabular}{llll} \hline\hline
           &                   & VLA1623 & L1689B  \\ \hline
CN         & $1_{022}-0_{011}$ & 0.73    & 0.13    \\
           & $1_{023}-0_{012}$ & 1.2     & 0.20    \\
           & $1_{021}-0_{011}$ & 0.70    & 0.16    \\
           & $1_{022}-0_{012}$ & 0.81    & 0.15    \\
C$^{34}$S  & 2--1              & 0.41    & 0.33    \\
H$^{13}$CO$^+$ & 1--0           & 3.1     & 1.2     \\
H$^{13}$CN & 1--0              & 0.98    & 0.10    \\
HC$_3$N    & 10--9             & 0.38    & 0.052   \\
HNC        & 1--0              & 1.5     & 1.1     \\
N$_2$H$^+$ & 1--0              & 8.0     & 6.0     \\
SO         & $2_3-1_2$         & 3.1     & 2.9     \\ \hline
\end{tabular}
\end{table*}

\begin{figure*}
\resizebox{\hsize}{!}{\includegraphics{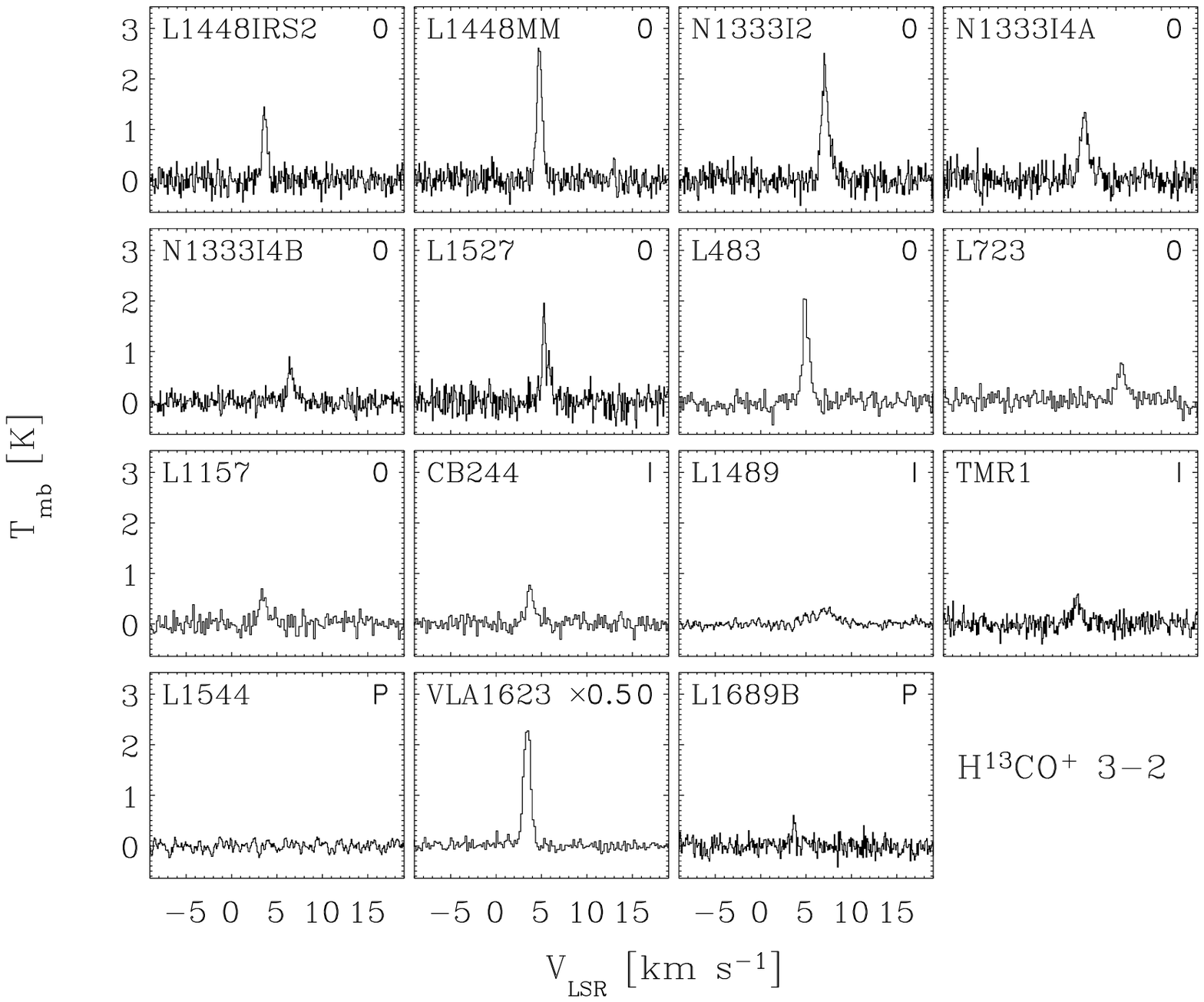}\includegraphics{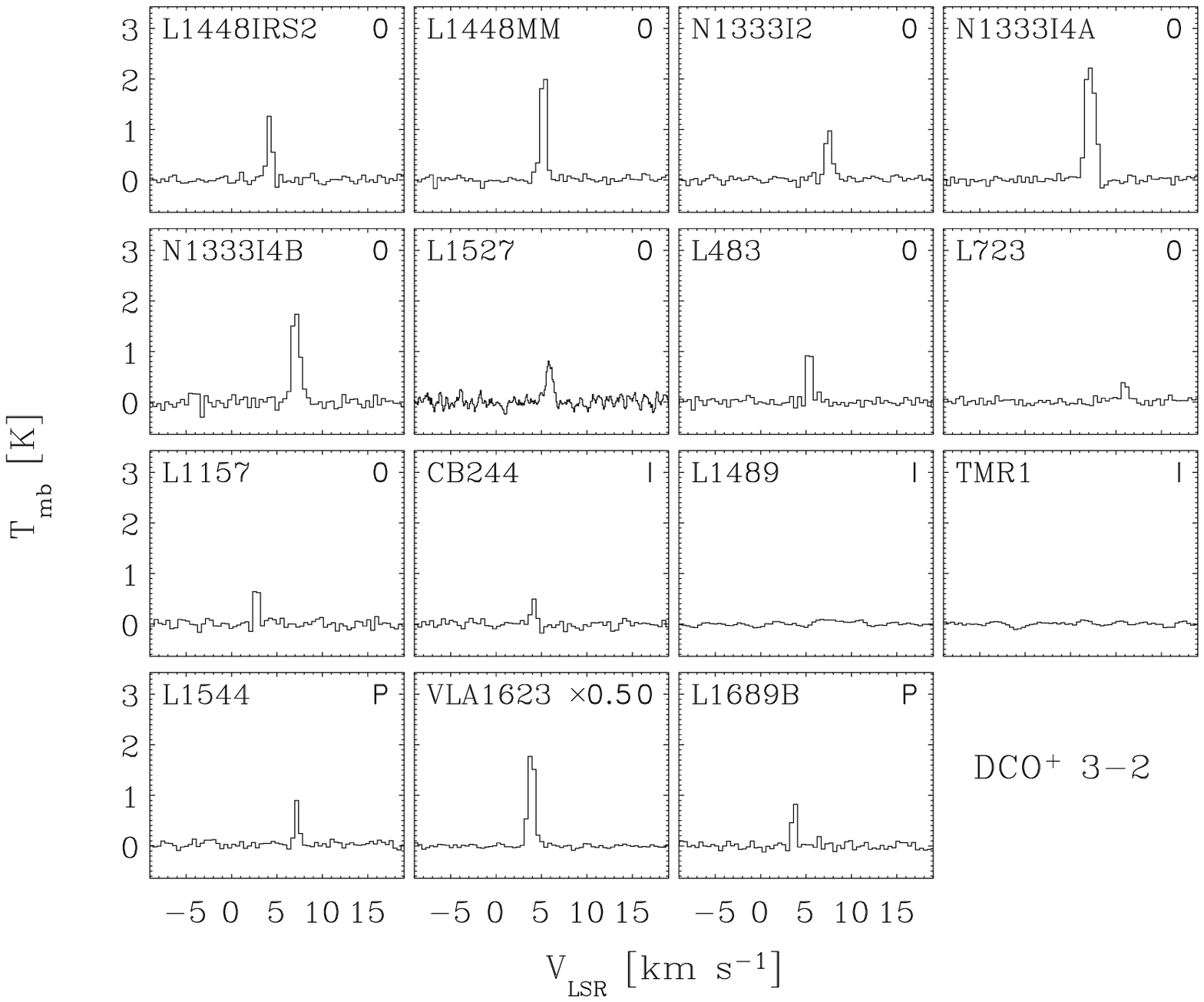}}
\caption{Spectra of H$^{13}$CO$^+$ (left) and DCO$^+$ $J=3-2$ (right) from JCMT
observations.}
\end{figure*}
\begin{figure*}
\resizebox{\hsize}{!}{\includegraphics{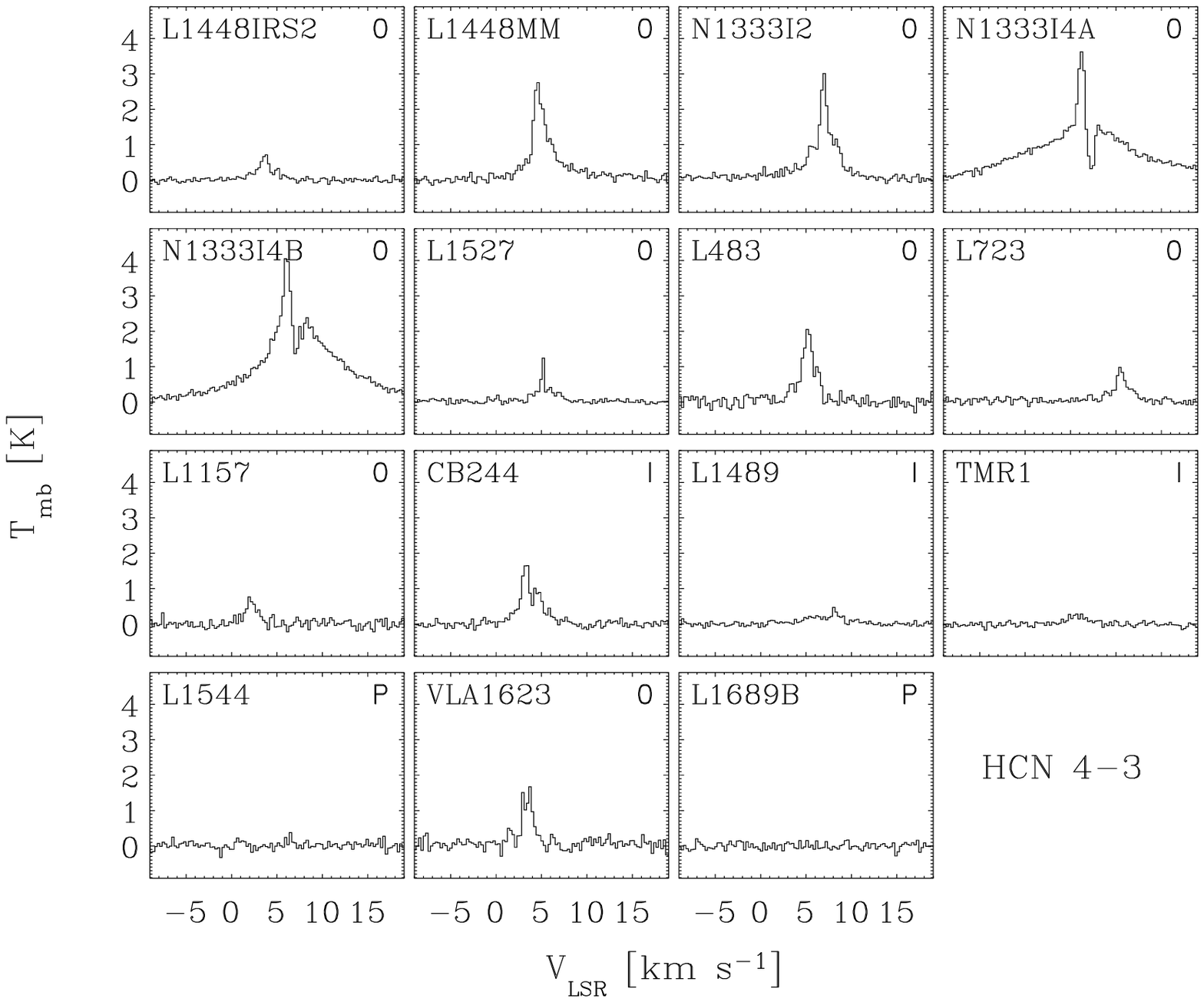}\includegraphics{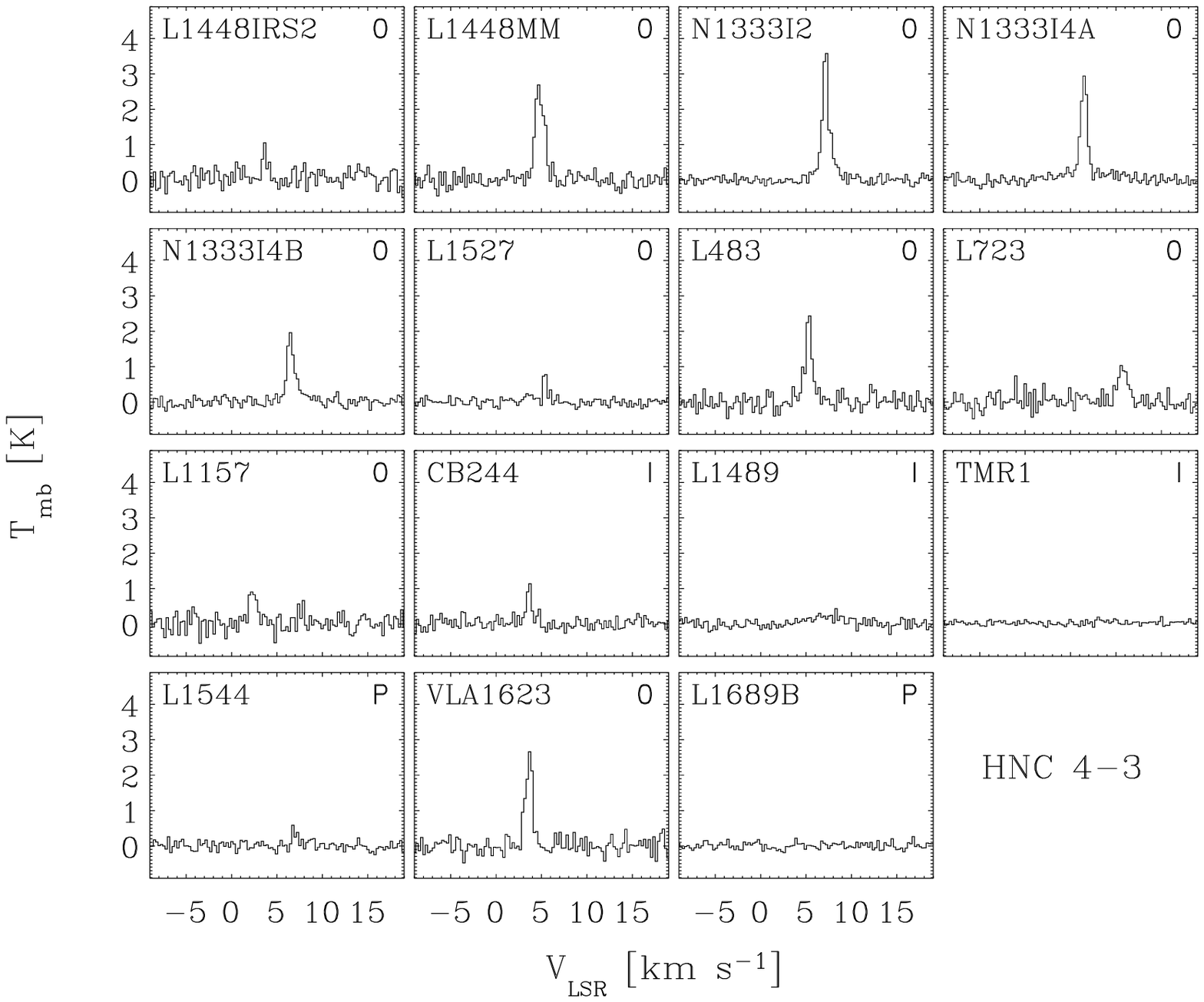}}
\caption{Spectra of HCN (left) and HNC (right) $J=4-3$ from JCMT
observations.}
\end{figure*}

\section{Modeling}\label{modeling}
\subsection{Constant abundances in static models}
To model the chemical abundances the same approach as in
\citetalias{jorgensen02} and \cite{schoeier02} was adopted: the
physical structure for each individual source presented in
\citetalias{jorgensen02} was used as input for calculating the
molecular excitation and the detailed radiative transfer using the
Monte Carlo code of \cite{hogerheijde00vandertak}. This code agrees
within the Monte Carlo noise with the code of \cite{schoeier02} that
was used to derive the abundances for \object{IRAS~16293-2422}. Both
codes were furthermore benchmarked by \cite{vanzadelhoff02} together
with other line radiative transfer codes. The same molecular data as
in \cite{schoeier02} were used. These are summarized in the database
of Sch\"{o}ier et al. (in prep.).

A few species, e.g., CN and N$_2$H$^+$, show clear hyperfine splitting
of the lines. For the CN 1--0 line the individual hyperfine components
can easily be disentangled (see Fig.~\ref{cn_spectra_fig}) and each of
these can be modeled as separate lines with individual excitation
rates. In general the model fits the individual hyperfine
components well, although in the poorest fits the strongest hyperfine
component is overestimated in the modeling. For the CN 3--2 lines at
340.248~GHz three hyperfine components are
overlapping. These transitions are optically thin, however, and can
therefore be modeled as one line. For N$_2$H$^+$ molecular data only
exist for the main rotational transitions. Again this does not pose a
problem if the emission is optically thin. For the 1--0 transition
this is true for most sources as also indicated by the observed ratios
of their hyperfine components.

For a given line the resulting sky brightness distribution was
convolved with the appropriate beam and the resulting spectrum
compared to the observed one. The envelope was assumed to be static
and the integrated line intensity and line width fitted by varying the
abundance profile and turbulent line broadening. In the first
iteration, a constant fractional abundance of each molecule relative
to H$_2$ was assumed. It is found that most lines are fitted well with
such a description, except for some of the low $J$ 3~mm
lines. Abundance jumps, e.g., due to evaporation of ice mantles as
found for \object{IRAS~16293-2422}, are not excluded by the present
observations. However, the region where the ice mantles would
evaporate (T $\gtrsim$ 90~K) is typically less than 100~AU
($\approx$~0.5-1\arcsec) for our sources, and therefore heavily
diluted in the beam. Furthermore, the lines presented in this study
are predominantly sensitive to the material at low to intermediate
temperatures in the envelope.

Tables~\ref{abund_first}-\ref{abund_last} list abundances together with
the number of observed lines and the reduced $\chi^2$ for each
individual source for species for which more than one line was
observed. A summary of the abundances for all molecules assuming
standard isotope ratios (Table~\ref{isotoperatio}) is given in
Table~\ref{abundoverview}. In each of these tables the abundances were
taken to be constant over the entire extent of the envelope.
\begin{table*}
\caption{Inferred abundances for CS and C$^{34}$S and reduced $\chi^2$ where
applicable.}\label{abundcs}\label{abund_first}
\begin{tabular}{lllll}\hline\hline
Source    & Abundance   & $\chi^2_{\rm red}$ & $n_{\rm lines}$$^a$ & Lines$^b$ \\ \hline
\multicolumn{5}{c}{CS} \\ \hline
L1448-I2  & 4.3$\times 10^{-11}$          & $\ldots$   & 1     & 5--4 (JCMT) \\
L1448-C   & 2.5$\times 10^{-9}$           & 2.6        & 2     & 5--4, 7--6 (JCMT) \\
N1333-I2  & 1.3$\times 10^{-9}$           & 0.91       & 2     & 5--4$^c$, 7--6$^c$ (JCMT) \\
N1333-I4A & 4.7$\times 10^{-10}$          & 0.019      & 2     & 5--4$^c$, 7--6$^c$ (JCMT) \\
N1333-I4B & 2.5$\times 10^{-9}$           & 1.1        & 2     & 5--4$^c$, 7--6$^c$ (JCMT) \\
L1527     & 3.0$\times 10^{-9}$           & 2.9        & 4     & 5--4, 7--6 (JCMT, CSO$^d$) \\
VLA1623   & 2.6$\times 10^{-9}$           & 2.6        & 2     & 5--4, 7--6 (JCMT) \\
L483      & 1.5$\times 10^{-9}$           & 1.7        & 2     & 5--4, 7--6 (JCMT) \\
L723      & 1.3$\times 10^{-9}$           & 13$^{e}$   & 2     & 5--4 (JCMT), 7--6 (JCMT) \\
L1157     & 1.9$\times 10^{-10}$          & $\ldots$   & 1+(1) & 5--4 (JCMT), 7--6 (JCMT; nd) \\
CB244     & 4.0$\times 10^{-9}$           & 0.22       & 2     & 5--4, 7--6 (JCMT) \\
L1551     & 4.1$\times 10^{-10}$          & 0.64       & 4     & 5--4, 7--6 (JCMT, CSO$^d$)  \\
L1489     & 2.8$\times 10^{-9}$           & 2.7        & 4     & 5--4, 7--6 (JCMT, CSO$^d$)  \\
TMR1      & 1.0$\times 10^{-8}$           & 3.2        & 4     & 5--4, 7--6 (JCMT, CSO$^d$)  \\
TMC1      & 6.5$\times 10^{-9}$           & 7.6        & 2     & 5--4 (JCMT), 7--6 (CSO$^d$) \\
TMC1A     & 4.9$\times 10^{-10}$          & 1.9        & 2     & 5--4 (JCMT), 7--6 (CSO$^d$) \\
L1544     & $<$2.5$\times 10^{-10}$       & $\ldots$   & 1     & 5--4 (JCMT) \\
L1689B    & $<$3.0$\times 10^{-10}$       & $\ldots$   & 1     & 5--4 (JCMT) \\ \hline
\multicolumn{5}{c}{C$^{34}$S} \\ \hline
L1448-I2  & 4.1$\times 10^{-11}$    & $\ldots$       & 1     & 2--1 (OSO) \\
L1448-C   & 1.1$\times 10^{-10}$    & 1.6            & 3     & 2--1 (OSO, IRAM), 5--4 (JCMT) \\
N1333-I2  & 1.4$\times 10^{-10}$    & 0.50           & 2     & 2--1 (IRAM), 5--4 (JCMT) \\
N1333-I4A & 4.6$\times 10^{-11}$    & 0.10           & 2     & 2--1 (IRAM), 5--4 (JCMT) \\
N1333-I4B & 5.5$\times 10^{-11}$    & 4.1            & 2     & 2--1 (IRAM), 5--4 (JCMT) \\
L1527     & 1.5$\times 10^{-11}$    & $\ldots$       & 1+(1) & 2--1 (IRAM), 5--4 (JCMT ; nd) \\
VLA1623   & 1.8$\times 10^{-10}$    & 3.0            & 2     & 2--1 (SEST), 5--4 (JCMT) \\
L483      & 3.1$\times 10^{-11}$    & 9.2            & 2     & 2--1 (OSO), 5--4 (JCMT) \\
L723      & 1.0$\times 10^{-10}$    & 2.7            & 2     & 2--1 (OSO), 5--4 (JCMT) \\
L1157     & 3.7$\times 10^{-11}$    & $\ldots$       & 1+(1) & 2--1 (OSO), 5--4 (JCMT ; nd) \\
CB244     & 7.2$\times 10^{-11}$    & $\ldots$       & 1+(1) & 2--1 (OSO), 5--4 (JCMT ; nd) \\
L1551     & 3.7$\times 10^{-11}$    & $\ldots$       & 1     & 5--4 (JCMT) \\
L1489     & $<$1.1$\times 10^{-10}$ & $\ldots$       & (2)   & 2--1 (OSO; nd), 5--4 (JCMT; nd) \\
TMR1      & $<$7.9$\times 10^{-10}$ & $\ldots$       & (1)   & 5--4 (JCMT; nd) \\
L1544     & 3.9$\times 10^{-11}$    & $\ldots$       & 1     & 2--1 (OSO) \\ 
L1689B    & 1.2$\times 10^{-10}$    & $\ldots$       & 1     & 2--1 (SEST) \\ \hline
\end{tabular}

$^a$Number of observed lines; number in parentheses indicate number of
lines not detected. $^b$Observed lines; lines not detected indicated
by ``nd''. $^c$Complex line profile - intensity defined as line
integrated over $\pm 2$~km~s$^{-1}$ relative to systemic velocity. $^{d}$CSO
line intensities from \cite{moriartyschieven95}. $^{e}$The 7--6 line
alone corresponds to an abundance of 5$\times 10^{-10}$, the 5--4 line taken
alone to 4$\times 10^{-9}$. The latter is in agreement with the results from
modeling of the C$^{34}$S lines.
\end{table*}
\begin{table*}
\caption{Inferred abundances for SO and reduced $\chi^2$ where
applicable.}\label{abundso}
\begin{tabular}{llllll}\hline\hline
Source    & Abundance     & $\chi^2_{\rm red}$    & $n_{\rm lines}$ & Lines \\ \hline
L1448-I2  &  7.0$\times 10^{-10}$ & $\ldots$ &  1+(1) & 2$_3$--1$_2$ (OSO), 8$_7$--7$_6$ (JCMT ; nd) \\
L1448-C   &  1.4$\times 10^{-9}$  & $\ldots$ &  1+(1) & 2$_3$--1$_2$ (OSO), 8$_7$--7$_6$ (JCMT ; nd) \\
N1333-I2  &  3.4$\times 10^{-9}$  & 1.3      &      2 & 2$_3$--1$_2$ (OSO), 8$_7$--7$_6$ (JCMT)$^b$ \\
N1333-I4A &  4.6$\times 10^{-9}$  & 0.66     &      2 & 2$_3$--1$_2$ (OSO), 8$_7$--7$_6$ (JCMT)$^b$ \\
N1333-I4B &  3.0$\times 10^{-9}$  & 0.82     &      2 & 2$_3$--1$_2$ (OSO), 8$_7$--7$_6$ (JCMT)$^b$ \\
L1527     &  1.4$\times 10^{-10}$ & 1.4      &  2+(1) & 2$_3$--1$_2$ (OSO), 4$_3$--3$_2$ (NRAO$^a$), 8$_7$--7$_6$ (JCMT; nd) \\
VLA1623   &  1.2$\times 10^{-8}$  & 1.6      &      3 & 2$_3$--1$_2$, 6$_5$-5$_4$ (SEST), 8$_7$--7$_6$ (JCMT) \\
L483      &  2.9$\times 10^{-10}$ & $\ldots$ &  1+(1) & 2$_3$--1$_2$ (OSO), 8$_7$--7$_6$ (JCMT; nd) \\
L723      &  2.4$\times 10^{-9}$  & $\ldots$ &  1+(1) & 2$_3$--1$_2$ (OSO), 8$_7$--7$_6$ (JCMT; nd) \\
L1157     &  1.6$\times 10^{-9}$  & 0.25     &  3+(1) & 2$_3$--1$_2$ (OSO), 2$_2$--1$_1$, 4$3$--3$_2$ (NRAO$^a$), 8$_7$--7$_6$ (JCMT; nd) \\
CB244     &  9.0$\times 10^{-10}$ & 3.1      &  3+(1) & 2$_3$--1$_2$ (OSO), 2$_2$--1$_1$, 4$3$--3$_2$ (NRAO$^a$), 8$_7$--7$_6$ (JCMT; nd) \\
L1551     &  1.9$\times 10^{-10}$ & 0.95     &  2     & 2$_3$--1$_2$ (OSO), 4$_3$--3$_2$ (NRAO$^a$) \\
L1489     &  2.0$\times 10^{-9}$  & $\ldots$ &  1+(1) & 2$_3$--1$_2$ (OSO), 8$_7$--7$_6$ (JCMT; nd) \\
TMR1      &  4.1$\times 10^{-9}$  & $\ldots$ &  1+(1) & 2$_3$--1$_2$ (OSO), 8$_7$--7$_6$ (JCMT; nd) \\
TMC1A     &  2.3$\times 10^{-10}$ & $\ldots$ &  1     & 2$_3$--1$_2$ (OSO) \\
TMC1      &  4.1$\times 10^{-9}$  & 0.70     &  2     & 2$_3$--1$_2$ (OSO), 4$_3$--3$_2$ (NRAO$^a$) \\
L1544     &  4.8$\times 10^{-10}$ & $\ldots$ &  1+(1) & 2$_3$--1$_2$ (OSO), 8$_7$--7$_6$ (JCMT; nd) \\ 
L1689B    &  2.6$\times 10^{-9}$  & 3.3      &  2+(1) & 2$_3$--1$_2$, 6$_5$-5$_4$ (SEST), 8$_7$--7$_6$ (JCMT; nd) \\   \hline
\end{tabular}

$^a$2$_2$-1$_1$ and 4$_3$-3$_2$ NRAO observations from
\cite{buckle03}; see further discussion in text. $^{b}$8$_7$--7$_6$
line very wide; larger uncertainty in derived abundance.
\end{table*}
\begin{table}
\caption{Inferred abundances for SO$_2$ based on observations of the
$3_{1,3}-2_{0,2}$ line from the Onsala 20m and SEST telescopes.}
\begin{tabular}{lllll}\hline\hline
Source      & Abundance       \\ \hline
L1448-I2    &$<$6.3$\times 10^{-11}$  \\
L1448-C     &$<$2.2$\times 10^{-10}$  \\
N1333-I2    &$<$2.5$\times 10^{-10}$  \\
N1333-I4A   &\phantom{$<$}1.8$\times 10^{-10}$  \\
N1333-I4B   &\phantom{$<$}6.6$\times 10^{-10}$  \\
L1527       &$<$2.4$\times 10^{-10}$  \\
VLA1623     &\phantom{$<$}1.7$\times 10^{-9}$   \\
L483        &$<$5.3$\times 10^{-11}$  \\
L723        &$<$1.3$\times 10^{-10}$  \\
L1157       &$<$8.4$\times 10^{-11}$  \\
CB244       &$<$1.0$\times 10^{-10}$  \\
L1489       &$<$2.3$\times 10^{-9}$   \\
L1544       &$<$9.0$\times 10^{-11}$  \\ 
L1689B      &\phantom{$<$}6.3$\times 10^{-10}$  \\ \hline
\end{tabular}
\end{table}
\begin{table*}
\caption{Inferred abundances for H$^{13}$CO$^+$ and reduced $\chi^2$
where applicable.}\label{abund_htcop}
\begin{tabular}{lllll}\hline\hline
Source    & Abundance    & $\chi^2_{\rm red}$ & $n_{\rm lines}$ & Lines \\ \hline
L1448-I2  &  9.9$\times 10^{-12}$$^{,a}$& $\ldots$ & 1     & 3--2 (JCMT) \\
L1448-C   &  1.3$\times 10^{-10}$       & 0.21     & 4     & 1--0 (OSO), 3--2 (JCMT, CSO$^b$), 4--3 (CSO$^b$) \\
N1333-I2  &  4.7$\times 10^{-11}$$^{,a}$& 0.59     & 4     & 3--2 (JCMT, CSO$^b$), 4--3 (JCMT, CSO$^b$) \\
N1333-I4A &  6.1$\times 10^{-12}$$^{,a}$& 1.2      & 3     & 3--2 (JCMT, CSO$^b$), 4--3 (CSO$^b$) \\
N1333-I4B &  8.8$\times 10^{-12}$$^{,a}$& 2.7      & 2+(1) & 3--2 (JCMT, CSO$^b$), 4--3 (CSO$^b$; nd) \\
L1527     &  8.5$\times 10^{-12}$       & 0.41     & 5     & 1--0 (OSO), 3--2 (JCMT, CSO$^b$), 4--3 (JCMT, CSO$^b$) \\
VLA1623   &  2.2$\times 10^{-10}$$^{,a}$& 2.0      & 3     & 3--2 (JCMT, CSO$^b$), 4--3 (JCMT) \\
L483      &  2.8$\times 10^{-11}$       & 0.63     & 4     & 1--0 (OSO), 3--2 (JCMT, CSO$^b$), 4--3 (JCMT) \\
L723      &  5.8$\times 10^{-11}$       & 0.61     & 3     & 1--0 (OSO), 3--2, 4--3 (JCMT) \\
L1157     &  8.4$\times 10^{-12}$$^{,a}$& 0.29     & 2     & 3--2 (JCMT), 4--3 (JCMT) \\
CB244     &  7.3$\times 10^{-11}$       & 0.53     & 3     & 1--0 (OSO), 3--2, 4--3 (JCMT) \\
L1551     &  2.3$\times 10^{-11}$       & 1.3      & 2     & 3--2, 4--3 (JCMT) \\
L1489     &  2.5$\times 10^{-10}$$^{,a}$& $\ldots$ & 1     & 3--2 (JCMT) \\
TMR1      &  3.9$\times 10^{-10}$$^{,a}$& $\ldots$ & 1     & 3--2 (JCMT) \\ 
TMC1      & $<$1.4$\times 10^{-10}$$^{,c}$ & $\ldots$ & 2     & 3--2, 4--3 (JCMT$^c$) \\
TMC1A     & $<$1.1$\times 10^{-11}$$^{,c}$ & $\ldots$ & 2     & 3--2, 4--3 (JCMT$^c$) \\
L1544     &  5.6$\times 10^{-12}$$^{,a}$& $\ldots$ & 1+(1) & 3--2 (CSO$^b$, JCMT; nd) \\ 
L1689B    &  1.7$\times 10^{-11}$$^{,a}$& 0.20     & 2     & 3--2 (CSO$^b$, JCMT) \\   \hline
\end{tabular}

$^a$Fits including 1--0 line intensities given in
Table~\ref{htcop_lowj}. See also discussion in text. $^b$CSO
measurements from \cite{gregersen97, gregersen00a, gregersen00b} -
except the 4--3 line towards N1333-I2 (Blake, priv. comm.). $^c$Upper
limits on line intensities from \cite{hogerheijde97}. Using their
HCO$^+$ main isotope 3--2 and 4--3 line intensities, HCO$^+$
abundances of 4.7$\times 10^{-9}$ and 2.2$\times 10^{-10}$ are found for TMC1 and
TMC1A, respectively.
\end{table*}

\begin{table*}
\caption{Inferred abundances for H$^{13}$CO$^+$ including the full set
of lines.}\label{htcop_lowj}
\begin{tabular}{lllll} \hline\hline
Source    & Abundance &  $\chi^2_{\rm red}$$^a$ & $I_{\rm obs}$$^b$ & $I_{\rm mod}$$^c$ \\ \hline
L1448-I2  & 1.1$\times 10^{-11}$ & 15   & 1.58 & 0.34 \\
N1333-I2  & 4.5$\times 10^{-11}$ & 1.6  & 1.77 & 0.96$^d$\\
N1333-I4A & 6.5$\times 10^{-12}$ & 6.6  & 2.32 & 0.38 \\
N1333-I4B & 9.4$\times 10^{-12}$ & 9.4  & 2.10 & 0.40 \\
VLA1623   & 2.2$\times 10^{-10}$ & 4.1  & 3.09 & 1.76 \\
L1157     & 9.2$\times 10^{-12}$ & 8.1  & 0.89 & 0.17 \\
L1489     & 3.1$\times 10^{-10}$ & 6.0  & 0.96 & 0.44 \\
TMR1      & 3.9$\times 10^{-10}$ & 9.4  & 1.13 & 0.40 \\
L1689B    & 1.8$\times 10^{-11}$ & 6.9  & 1.24 & 0.41 \\
L1544     & 6.7$\times 10^{-12}$ & 16   & 0.92 & 0.17 \\ \hline
\end{tabular}

$^a$Reduced $\chi^2$ including lines from Table~\ref{abund_htcop}
together with 1--0 lines from the Onsala 20~m telescope or
SEST. $^b$Observed 1--0 line intensity ($\int T_{\rm MB}\,{\rm
d}v$). $^c$Modeled 1--0 line intensity ($\int T_{\rm MB}\,{\rm d}v$)
with abundance from Table~\ref{abund_htcop}. $^d$See also discussion
in \cite{n1333i2art}.
\end{table*}
\begin{table*}
\caption{Inferred abundances for H$^{13}$CN and reduced $\chi^2$ where
applicable.}
\begin{tabular}{lllll}\hline\hline
Source    & Abundance       & $\chi^2_{\rm red}$ & $n_{\rm lines}$ & Lines \\ \hline
L1448-I2  &   2.1$\times 10^{-11}$  & $\ldots$ &  1     & 1--0 (OSO), 3--2 (JCMT; nd) \\
L1448-C   &   9.0$\times 10^{-11}$  & 2.5      &  2     & 1--0 (OSO), 3--2 (JCMT) \\
          &   7.7$\times 10^{-11}$  & $\ldots$ &  1     & 3--2 (JCMT) \\
N1333-I2  &   2.5$\times 10^{-11}$  & 7.1      &  2     & 1--0 (OSO), 3--2 (JCMT) \\
          &   2.9$\times 10^{-11}$  & $\ldots$ &  1     & 3--2 (JCMT) \\
N1333-I4A &   5.6$\times 10^{-12}$  & 18       &  2     & 1--0 (OSO), 3--2 (JCMT) \\
          &   5.1$\times 10^{-12}$  & $\ldots$ &  1     & 3--2 (JCMT) \\
N1333-I4B &   3.2$\times 10^{-11}$  & 1.9      &  2     & 1--0 (OSO), 3--2 (JCMT) \\
          &   2.8$\times 10^{-11}$  & $\ldots$ &  1     & 3--2 (JCMT) \\
L1527     &   3.2$\times 10^{-11}$  & $\ldots$ &  1     & 1--0 (OSO), 3--2 (JCMT; nd) \\
VLA1623   &   2.5$\times 10^{-10}$  & $\ldots$ &  1+(1) & 1--0 (SEST), 3--2 JCMT; nd) \\
L483      &   2.4$\times 10^{-11}$  & 1.0      &  2     & 1--0 (OSO), 3--2 (JCMT) \\
          &   2.9$\times 10^{-11}$  & $\ldots$ &  1     & 3--2 (JCMT) \\
L723      &   3.6$\times 10^{-11}$  & $\ldots$ &  1     & 1--0 (OSO), 3--2 (JCMT; nd) \\
L1157     &   2.0$\times 10^{-11}$  & $\ldots$ &  1     & 1--0 (OSO), 3--2 (JCMT; nd) \\
CB244     &   3.1$\times 10^{-11}$  & $\ldots$ &  1     & 1--0 (OSO), 3--2 (JCMT; nd) \\
L1489     &$<$1.7$\times 10^{-10}$  & $\ldots$ & (2)    & 1--0 (OSO; nd), 3--2 (JCMT; nd) \\
TMR1      &   3.7$\times 10^{-10}$  & $\ldots$ & 1+(1)  & 1--0 (OSO), 3--2 (JCMT; nd) \\
L1544     &   7.1$\times 10^{-11}$  & $\ldots$ & 1+(1)  & 1--0 (OSO), 3--2 (JCMT; nd) \\ 
L1689B    &   1.3$\times 10^{-11}$  & $\ldots$ & 1+(1)  & 1--0 (SEST), 3--2 (JCMT ; nd) \\ \hline
\end{tabular}

See description in Table~\ref{abund_first}.
\end{table*}
\begin{table}
\caption{Inferred abundances for CN using the 1--0 hyperfine
transitions and reduced $\chi^2$ where applicable.}\label{abund_last}
\begin{tabular}{llll}\hline\hline
Source    & Abundance       & $\chi^2_{\rm red}$ & $n_{\rm lines}$ \\ \hline
L1448-I2  & 8.3$\times 10^{-10}$    & 0.13     &  4  \\
L1448-C   & 2.4$\times 10^{-9}$     & 2.4      &  4  \\
N1333-I2  & 2.4$\times 10^{-9}$     & 2.3      &  4  \\
N1333-I4A & 6.1$\times 10^{-10}$    & 1.4      &  4  \\
N1333-I4B & 8.6$\times 10^{-10}$    & 1.5      &  4  \\
L1527     & 1.6$\times 10^{-9}$     & 0.93     &  4  \\
VLA1623   & 3.5$\times 10^{-9}$     & 2.0      &  4  \\
L483      & 3.3$\times 10^{-10}$    & 2.3      &  4  \\
L723      & 7.2$\times 10^{-10}$    & 1.8      &  4  \\
L1157     & 5.1$\times 10^{-10}$    & 1.1      &  4  \\
CB244     & 1.0$\times 10^{-9}$     & 3.7      &  4  \\
L1551     & 1.1$\times 10^{-9}$     & 3.6      &  4  \\
L1489     & 4.6$\times 10^{-9}$     & $\ldots$ & 1+(3) \\
TMR1      & 5.0$\times 10^{-9}$     & $\ldots$ & 1+(3) \\
TMC1      & 1.3$\times 10^{-8}$     & 5.1      &  4  \\
TMC1A     & 2.1$\times 10^{-9}$     & 3.0      &  4  \\
L1544     & 6.8$\times 10^{-10}$    & 1.7      &  4  \\
L1689B    & 1.6$\times 10^{-10}$    & 3.9      &  4  \\ \hline
\end{tabular}
\end{table}

For a range of the molecules (especially CS and HCO$^+$) the main
isotopes are not well suited for determining chemical abundances since
the lines rapidly become optically thick. Moreover the emission from
these species in the envelope is in some cases hard to disentangle,
since the line profiles show clear signs of wing emission due to
outflows and asymmetries attributed to infalling motions
\citep{gregersen97,gregersen00a,wardthompson01}. The lines from the
weaker isotopes (e.g., C$^{34}$S and H$^{13}$CO$^+$), however, usually
do not suffer from these problems and were therefore used to constrain
the abundances where detected.
\clearpage

\subsection{Shortcomings of the models; drop abundance profiles}\label{shortcom}\label{radial}
\begin{figure}
\resizebox{\hsize}{!}{\includegraphics{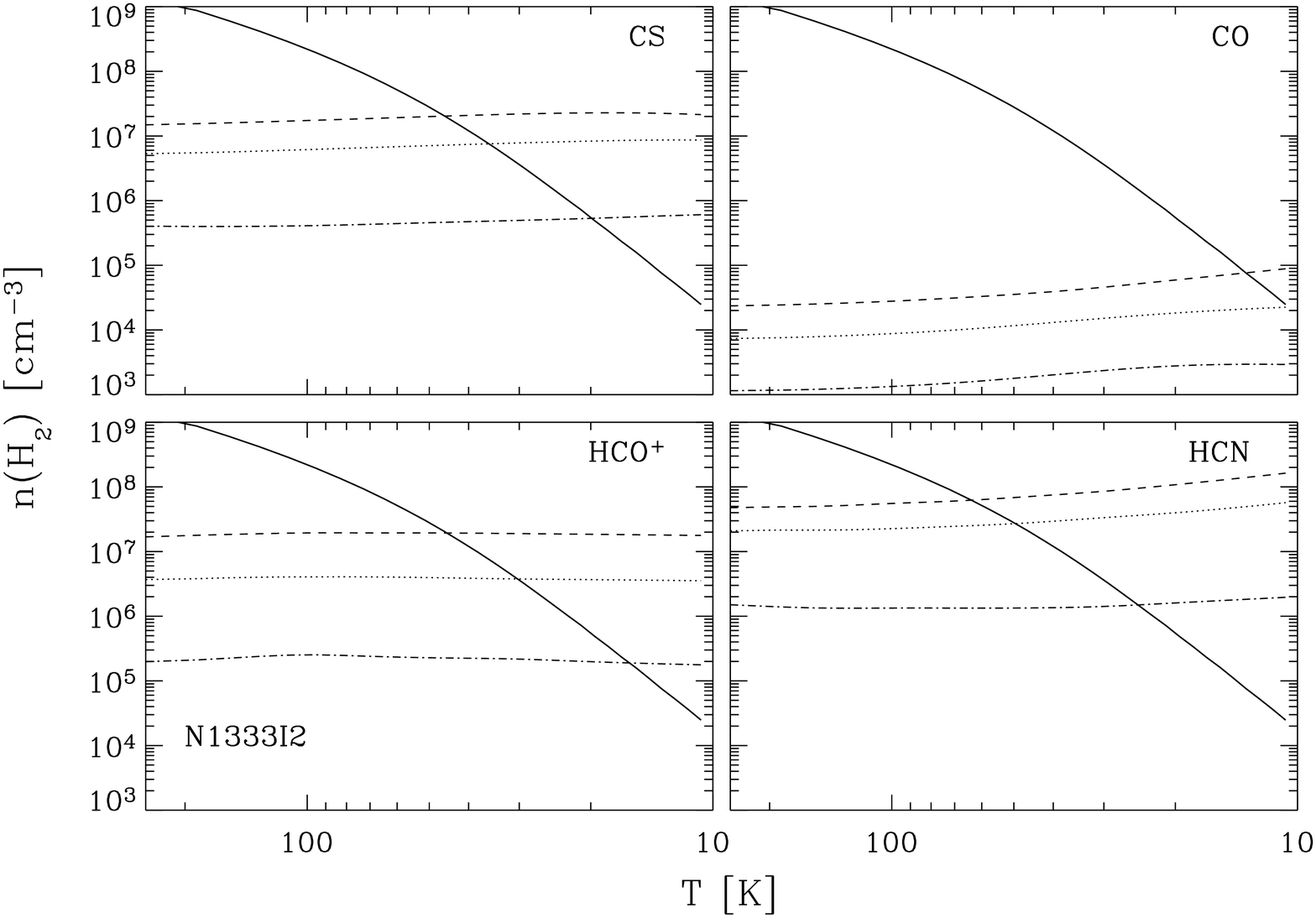}}
\resizebox{\hsize}{!}{\includegraphics{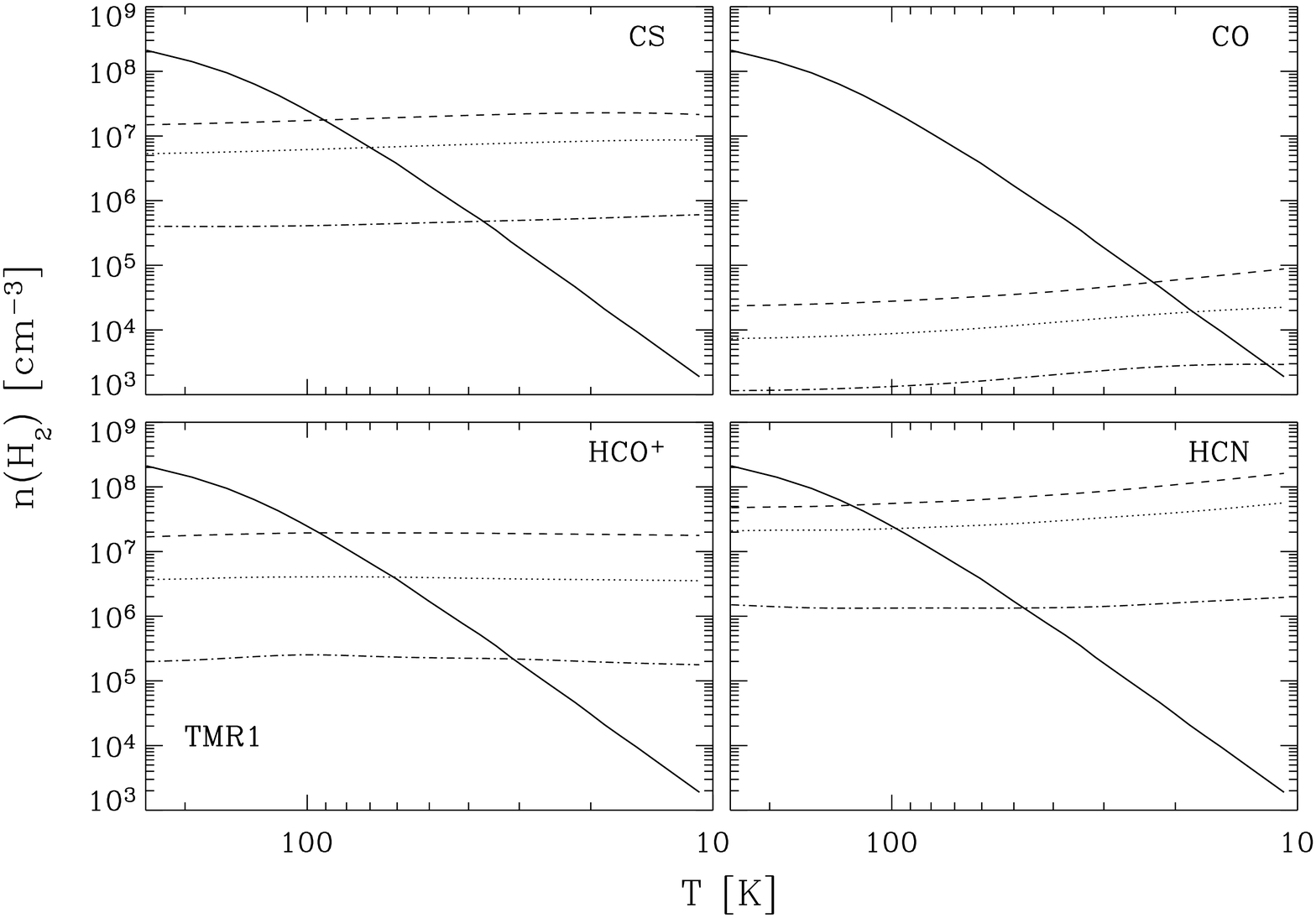}}
\caption{Density as function of temperature for the envelopes around
\object{TMR1} and \object{N1333-I2} (solid line) compared to the
critical densities of the observed transitions of CS, CO, HCO$^+$ and
HCN. The critical densities are indicated in order of increasing
excitation by the dashed-dotted, dotted and dashed lines,
respectively, i.e., showing the 2--1, 5--4 and 7--6 transitions for
CS, the 1--0, 2--1 and 3--2 transitions for CO, and the 1--0, 3--2 and
4--3 transitions for HCO$^+$ and HCN.}\label{critdensplots}
\end{figure}

For a number of species the constant fractional abundance model gives
poor results ($\chi^2 \gtrsim 5$) when fitting both the lowest
rotational lines from the Onsala 20~m and higher excitation lines from
the JCMT. A similar trend was seen in modeling of the CO isotopic
species in Paper~I, where the 1--0 lines were typically underestimated
in models fitted to the 2--1 and 3--2 lines. This trend is
particularly pronounced for H$^{13}$CO$^+$ and the nitrogen-bearing
species (HCN, H$^{13}$CN, CN and HNC), whereas the low $J$ lines of
CS, for example, can be fitted well by a constant abundance. This may
be due to the critical density of the observed transitions which
should be compared to the typical freeze-out and desorption timescales
for the given densities and temperatures. Fig.~\ref{critdensplots}
shows the density for two objects, \object{N1333-I2} and
\object{TMR1}, as function of temperature (i.e., depth) compared to
the critical densities of various transitions of CS, CO, HCO$^+$ and
HCN \citep[e.g.,][]{jansenphd} for the same temperatures. Since the
critical densities of the CS/C$^{34}$S 2--1 lines are higher than
those of the HCO$^+$ and CO 1--0 lines, CS is less sensitive to the
outer region of the envelope where depletion and contribution from the
surrounding cloud may be important. This may explain why these
transitions can be modeled in the constant abundance framework. The
observed 4--3 transitions of, i.p., HCN and HNC have the highest
critical densities and these lines therefore probe the innermost part
of the envelope.

In the outer regions of the envelope the depletion timescale for CO is
comparable to the lifetime of the protostars ($\sim 10^4-10^5$~years)
at the temperatures where the molecule can freeze-out. This could
explain the failure of the constant abundance models in describing the
lowest $J$ lines for CO (and thereby also HCO$^+$; see discussion in
Sect.~\ref{hcop}): in prestellar cores
\citep[e.g][]{caselli99,tafalla02} a trend is seen of decreasing CO
abundances with increasing density toward the center. Since the
temperature in the bulk of the material in these objects is low enough
for CO to be frozen out, the explanation for the radial dependence is
a difference in density and thus the freeze-out timescale. Therefore
the time for CO to freeze-out in the outermost regions may simply be
too long to result in appreciable amounts of depletion. For the
protostellar cores the difference is the heating by the central
source, which induces a temperature gradient toward the center. CO is
therefore expected to be frozen out in a small region, where the
density is high enough that the freeze-out timescale is short, yet the
temperature low enough that CO is not returned to the gas-phase.

\begin{figure}
\resizebox{\hsize}{!}{\rotatebox{90}{\includegraphics{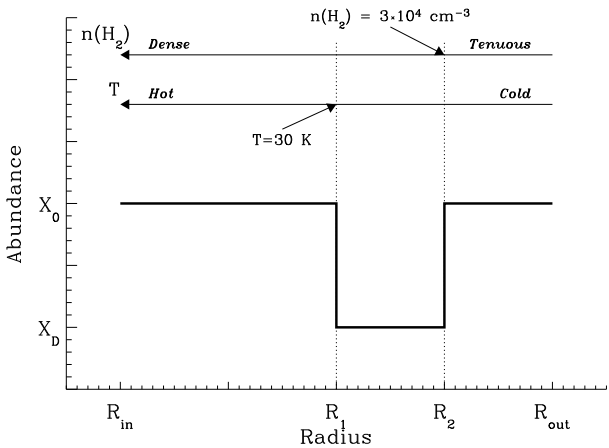}}}
\caption{Simulated abundance profile in ``drop''
models.}\label{drop}
\end{figure}

A simple way of testing this can be performed by introducing a
``drop'' chemical structure as illustrated in Fig.~\ref{drop},
with a constant undepleted CO abundance $X_0$ in the parts of the
envelope with densities lower than $3\times 10^{4}\,{\rm cm}^{-3}$ or
temperatures higher than 30~K. A lower CO evaporation temperature of
$\sim 20$~K is ruled out by the 3--2 line intensities
\citepalias{jorgensen02}. The undepleted abundance, $X_0$, is taken to
be the same in the inner and outer regions of the envelope to avoid
adding another free parameter. The abundance in the region with
temperatures lower than 30~K and densities higher than $3\times
10^{4}\,{\rm cm}^{-3}$, $X_{\rm D}$, can then be adjusted to fit the
observations.

\begin{figure}
\resizebox{\hsize}{!}{\rotatebox{90}{\includegraphics{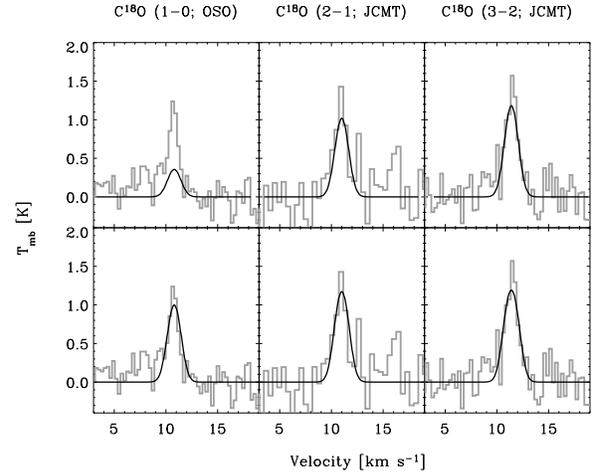}}}
\caption{Fitted C$^{18}$O line-profiles for L723. Upper panels: constant
fractional abundance of 3.9$\times 10^{-9}$ from Paper~I. Lower panels: drop
model with abundances of $X_0$=2.0$\times 10^{-7}$ for the undepleted material
(with either temperatures higher than 30~K or densities lower than
3$\times 10^{-4}$) and an abundance of $X_{\rm D}$=1.5$\times 10^{-8}$ for material
with CO frozen out.}\label{c18o_abundance_jumps}
\end{figure}

Fig.~\ref{c18o_abundance_jumps} shows a comparison for L723 between
the models with a constant fractional abundance from Paper~I and a
model with two abundance jumps described above. The latter model has
two free parameters (besides the Doppler broadening, which does not
alter the results), $X_0$ and $X_{\rm D}$. In the constant abundance
model, the C$^{18}$O abundance is 3.9$\times 10^{-8}$, while in the ``drop''
model, the undepleted abundance $X_{\rm 0}$ is 2$\times 10^{-7}$ and the
depleted abundance $X_{\rm D}$ is 2$\times 10^{-8}$. Similar fits to the C$^{18}$O
abundances of one of the class~I objects, L1489, provide equally good
results - again allowing the 1--0 lines to be fitted together with the
2--1 and 3--2 lines. The fitted abundances in the case of L1489 are
$X_0$ of 5$\times 10^{-7}$ and $X_{\rm D}$ of 5$\times 10^{-8}$.

The fact that the 1--0, 2--1 and 3--2 lines can all be fitted in the
drop models is not unexpected since an extra free parameter is
introduced compared to the results presented in
\citetalias{jorgensen02}, which is used to fit only one extra
line. Still, it should be emphasized that the chemical structure in
the drop models has its foundation in results from the pre-stellar
cores and is thus not completely arbitrary. As expected, the constant
fractional abundances found for both L723 and L1489 in
\citetalias{jorgensen02} are a weighted average of $X_{\rm D}$ and
$X_0$ from the drop models. While the constant abundances were
significantly different for L723 and L1489 (1.9$\times 10^{-5}$ and
1.0$\times 10^{-4}$, respectively), those in the drop models are more similar:
the factor 2.5 difference in derived abundances can be explained
through the uncertainties and approximations in the physical and
chemical description.

Besides these limitations, the constant fractional abundances
characterize the overall envelope chemistry as a good first-order
approximation. The more general trends in abundance variation from
source-to-source can thereby be used for a statistical comparison with
the caveat that the selected transitions may be probing different
temperature and density regimes in which the chemistry may vary. It is
important to note that none of the abundances are correlated with the
distances to the sources or the slopes of their density profiles,
indicating that the uncertainties in these parameters do not introduce
signficant systematic errors.

\object{VLA1623} shows high abundances of most molecular species
compared to the average class 0 objects. As mentioned in Paper~I the
envelope model of this particular object is highly uncertain since it
is located in a dense ridge of material and molecular tracers with low
critical densities, in particular the CO lines and the low $J$ 3~mm
transitions of the other species, may be sensitive to this component
rather than the envelope itself.

\object{TMC1A} stands out among the remainder of the class I objects
with significantly lower abundances in all
molecules. \cite{hogerheijde98} likewise found that the envelope mass
estimated through 1.1~mm continuum observations was a factor 5 higher
than the mass estimated on the basis of $^{13}$CO, C$^{18}$O and HCO$^+$
measurements. One possibility is that \object{TMC1A} does have a more
massive envelope and thus lower abundances due to depletion such as
seen for CO in \citetalias{jorgensen02}. Alternatives could be that
the density in the envelope of this object has been overestimated from
the models of the dust continuum emission or that the molecular line
emission is tracing material not directly associated with the bulk
material in the protostellar envelope.

This could be a general problem for more sources: are there systematic
errors of the envelope dust mass leading to false trends in
abundances?  A systematic overestimate of the mass (i.e., density) for
the class 0 objects would lead to systematically lower abundances,
similar to the depletion effects observed for CO in
\citetalias{jorgensen02}. On the other hand a change in abundance as
seen, e.g., for CO, would require that the density scale for the class
0 objects is off by approximately an order of magnitude, and the
submillimeter dust emission and molecular lines would have to trace
quite unrelated components. This is contradicted by the relative
success of the models in simultaneously explaining observations of
both line and continuum emission from single-dish telescopes
(\citetalias{jorgensen02}, \citealt{schoeier02}, this paper) and
higher resolution interferometers \citep{n1333i2art,hotcorepaper}.

\subsection{Effect of velocity field}\label{velfield}
Most observed lines are simple Gaussians with typical widths of
1~km~s$^{-1}$ (FWHM). Still, for some molecules significant variations are
found between the widths for different rotational transitions and
thereby the broadening due to systematic and/or turbulent motions
required to model the exact line profiles. This indicates either
systematic infall in the envelope as expected from the line profiles
of some of the optically thick species or a variation of the turbulent
velocity field with radius.

The problem with the current models is that the power-law density
profile adopted in \citetalias{jorgensen02} does not give direct
information about the velocity field, as would be obtained by fitting
a specific collapse model like the inside-out collapse model by
\cite{shu77}. A velocity field can, however, still be associated with
the derived density distribution, using the mass continuity
equation. This equation:
\begin{equation}
\frac{\partial \rho}{\partial t}+\rho\nabla\cdot{\mathbf v} = 0
\end{equation}
becomes for a spherical symmetric envelope:
\begin{equation}
\frac{\partial \rho}{\partial t}+\frac{1}{r^2}\frac{\partial}{\partial t}(r^2\rho v_r)=0
\end{equation}
where $v_r$ is the radial velocity. Assuming $\rho \propto r^{-p}$, a
power-law velocity distribution $v\propto r^{-q}$ and a static
envelope density distribution ($\frac{\partial\rho}{\partial t}=0$)
results in:
\begin{equation}
\frac{1}{r^2}\frac{\partial}{\partial t}(r^2 r^{-p} r^{-q})=0
\end{equation}
or
\begin{equation}
r^{2-p-q}=const. \Leftrightarrow q=2-p
\end{equation}
So for a given power-law density distribution, $n({\rm H_2})=n_0
(r/r_0)^{-p}$, it is possible to introduce a corresponding power-law
distribution for the velocity field, $v=v_0 (r/r_0)^{-q}$. Here a
characteristic infall velocity $v_0$ is introduced as an additional
free parameter, which can be fitted by comparison of the line profiles
to the turbulent linewidth.

\begin{figure}
\resizebox{\hsize}{!}{\includegraphics{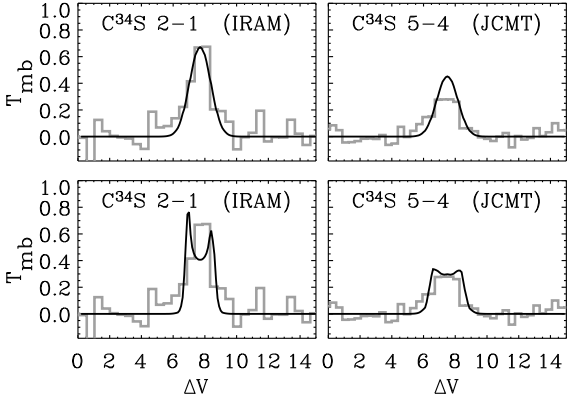}}
\caption{Modeling of the velocity field in the envelope around
\object{N1333-I2}: in the upper panel C$^{34}$S model lines are compared with
observations for a constant broadening of 0.8~km~s$^{-1}$, as in
\citetalias{jorgensen02}. In the lower panel, a model with no
turbulent broadening, but a power-law velocity field with
$v_0=2.5$~km~s$^{-1}$ at the inner radius, $r_0=23.4$~AU, is adopted. In
both plots a constant abundance of 1.4$\times 10^{-10}$ was
assumed.}\label{velfield_lineprofile}
\end{figure}

In Fig.~\ref{velfield_lineprofile} such a comparison is shown for the
C$^{34}$S observations for the ``typical'' class 0 object, \object{N1333-I2}
\citep[see also][]{n1333i2art}. The observed line widths are seen to
constrain the velocity field in terms of the combination of turbulent
broadening and magnitude of the systematic velocity field. For a
parameterization of the velocity field, an estimate of the mass
accretion rate $\dot{M}$ can be derived from:
\begin{equation}
\dot{M}=4\pi\, r_0^2\, \mu\ m_{\rm H}\,n_0\, v_0
\end{equation}
where $\mu$ is the mean molecular weight, 2.33. For \object{N1333-I2}
the upper limit on $v_0$ (at the inner radius, r$_0=23.4$~AU) of
2.5~km~s$^{-1}$, i.e., assuming no turbulent broadening, translates to
a mass accretion rate of 3$\times 10^{-5}$ M$_\odot$ yr$^{-1}$. This
agrees with typical mass accretion rates inferred for the youngest
protostars \citep[e.g.,][]{shu77,bontemps96,difrancesco01}. The
advantage of using the optically thin species to constrain the
velocity field is that they do not suffer from confusion with, e.g.,
outflows, but only pick-up the bulk envelope material as illustrated
by high angular resolution interferometer studies
\citep[e.g.,][]{n1333i2art}. On the other hand, complementary
information about the velocity field is obtained from the detailed
line-profiles of the optically thick, strongly self-absorbed lines,
e.g., the relative strength of red- and blue peaks and the depth of
the self-absorption feature (see, e.g., \cite{evans99} and
\cite{myersppiv} for recent reviews of this topic).

\begin{figure}
\resizebox{\hsize}{!}{\includegraphics{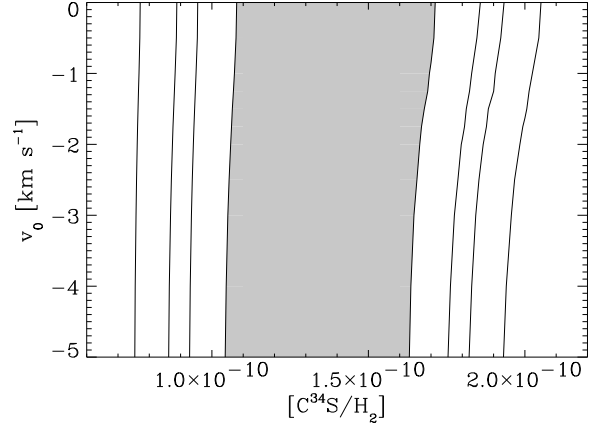}}
\caption{Dependence of the C$^{34}$S abundance on velocity field for
an infalling envelope around \object{N1333-I2}. The grey region indicates
1$\sigma$ confidence and the almost vertical lines indicate the
2$\sigma$, 3$\sigma$ and 4$\sigma$ confidence levels. It is seen that
the derived abundances do not depend on the assumed velocity field for
this optically thin species.}\label{chisqr_velfield}
\end{figure}

Fig.~\ref{chisqr_velfield} illustrates the important point that the
derived abundances do not depend critically on the adopted velocity
field for optically thin species like C$^{34}$S, illustrating that the
static envelope structure provides an adequate description to derive
their overall chemical properties. This is in agreement with the
conclusion reached in \citetalias{jorgensen02}. Note that the
confidence levels on the derived abundance in
Fig.~\ref{chisqr_velfield} only correspond to the calibration
error. Systematic errors due to uncertainties in the adopted model,
collisional data etc. are not taken into account, so the abundances
derived may still be subject to uncertainties not apparent from this
figure.

In summary, although systematic velocities probably exist in all
envelopes besides the turbulent motions, the derived abundances do not
depend critically on the detailed treatment of the velocity field as
long as predominantly optically thin lines are considered, as in this
paper. We will therefore for the remainder of this paper stick with
the assumption of a non-infalling envelope with a constant turbulent
broadening reproducing the approximate width of the lines.

\section{Discussion}\label{discussion}
\subsection{General trends and empirical correlations}
A direct comparison of the derived abundances for the main isotopes
can be found in Table~\ref{abundoverview}. Where possible the
abundances are calculated using the optically thin isotopic species
with the standard isotopic ratios listed in Table~\ref{isotoperatio}.

\begin{table}
\caption{Adopted isotope ratios.}\label{isotoperatio}
\begin{tabular}{lll} \hline\hline
Isotope ratio & Value & Reference \\ \hline
$^{12}$C/$^{13}$C & 70        & \cite{wilson94} \\ 
$^{16}$O/$^{18}$O & 540       & \cite{wilson94} \\
$^{18}$O/$^{17}$O & 3.6       & \cite{penzias81},\citetalias{jorgensen02}$^{a}$ \\
$^{32}$S/$^{34}$S & 22        & \cite{chin96} \\ \hline
\end{tabular}

$^{a}$The $^{18}$O/$^{17}$O ratio is not used for the abundances
derived in this paper, but was used for the CO abundances in
\citetalias{jorgensen02} and is therefore included here for
completeness.
\end{table}

\begin{table*}
\caption{Overview of derived abundances for main isotopic and
deuterated species.}\label{abundoverview}
\begin{tabular}{llllllllllll} \hline\hline
Source & CO & CS & SO & HCO$^+$ & DCO$^+$ & N$_2$H$^+$ & HCN & DCN & HNC & CN & HC$_3$N \\ 
                      & $\times 10^{-5}$ & $\times 10^{-9}$ & $\times 10^{-9}$ & $\times 10^{-9}$ & $\times 10^{-11}$ & $\times 10^{-9}$ & $\times 10^{-9}$ & $\times 10^{-11}$ & $\times 10^{-10}$ & $\times 10^{-10}$ & $\times 10^{-10}$ \\ \hline
\multicolumn{12}{c}{Class 0 (envelope mass $> 0.5 M_\odot$)} \\[1.0ex]
\object{L1448-I2}     & 0.61  & 0.90  & 0.70    & 0.69 & 0.79      & $>$1.0  & 8.0       & 0.33     & 0.35     & 1.8     & 1.9   \\
\object{L1448-C}      & 3.7   & 2.4   & 1.4     & 9.1  & 9.8       & 3.9     & 5.4       & 4.9      & 13       & 20      & 12    \\
\object{N1333-I2}     & 2.4   & 3.1   & 3.4     & 3.3  & 1.6       & 5.0     & 2.0       & 2.1      & 1.8      & 3.0     & 4.3   \\
\object{N1333-I4A}    & 0.79  & 1.0   & 4.6     & 0.43 & 1.2       & $>$1.0  & 0.36      & 0.31     & 0.28     & 0.37    & 0.72  \\
\object{N1333-I4B}    & 1.3   & 1.2   & 3.0     & 0.62 & 2.5       & 3.2     & 2.0       & 1.0      & 1.4      & 1.4     & 1.1   \\
\object{L1527}        & 3.9   & 0.33  & 0.14    & 0.60 & 2.9       & 0.25    & 1.2       & 1.5      & 3.2      & 24      & 8.9   \\[1.5ex]
\object{VLA1623}      & 16    & 4.0   & 12      & 15   & 17        & $>$3.0  & 6.6       & 4.7      & 10       & 8.9     & 3.8   \\
\object{L483}         & 1.4   & 0.68  & 0.29    & 2.0  & 1.1       & 0.75    & 2.0       & 0.94     & 3.9      & 8.3     & 1.8   \\
\object{L723}         & 1.9   & 2.2   & 2.4     & 4.1  & 1.6       & 1.3     & 1.0       & $<$0.91  & 5.1      & 8.6     & 2.7   \\
\object{L1157}        & 0.62  & 0.81  & 1.6     & 0.59 & 0.95      & $>$1.0  & 0.066     & $<$0.28  & 0.61     & 0.65    & 1.3   \\
\object{L1551-I5}     & 3.0   & 0.81  & 0.19    & 1.6  & 1.5       & 3.1     & $\ldots$  & $\ldots$ & 0.72     &$\ldots$ & 2.1   \\
\object{I16293-2422}$^{c}$  & 3.3   & 3.0   & 4.4     & 1.4  & 1.3       & 0.14$^d$& 1.1       & 1.3      & 0.69     & 0.80    & 1.5   \\[1.0ex]
\multicolumn{12}{c}{Class I (envelope mass $< 0.5 M_\odot$)} \\[1.0ex]
\object{L1489}        & 10    & 2.8   & 2.0     & 18   & $<2.3$    & 0.15    & 0.65      & $<$6.4   & 13       & 22      & 8.7   \\
\object{TMR1}         & 20    & 10    & 4.1     & 27   & $<5.0$    & 0.35    & 1.6       & $<$15    & 8.1      & 47      & 35    \\
\object{TMC1A}        & 2.3   & 0.49  & 0.23    & 0.22 & $<0.65$   & 3.9     & $\ldots$  & $\ldots$ & 0.38     &$\ldots$ & 100   \\
\object{TMC1}         & 20    & 6.5   & 4.1     & 4.7  & $<8.5$    & $>$1.0  & $\ldots$  & $\ldots$ &$<$7.7    &$\ldots$ & 4.9   \\
\object{CB244}        & 3.7   & 1.6   & 0.90    & 5.1  & 2.1       & 2.0     & 4.9       & $<$1.9   & 7.8      & 20      & 5.9   \\[1.0ex]
\multicolumn{12}{c}{Pre-stellar} \\[1.0ex]
\object{L1544}        & 0.49  & 0.86  & 0.48    & 0.39 & 2.1       & 5.0     &$<$0.35    & 0.77     & 12       & 4.8     & 16    \\
\object{L1689B}       & 2.4   & 2.6   & 2.5     & 1.2  & 2.4       & 0.43    &$<$0.38    & $<$2.1   &$<$4.6    &$<$2.3   & 0.36  \\ \hline
\emph{Averages:}      &       &       &         &      &           &         &           &          &          &         &       \\
'class 0'$^{a}$       & 2.1   & 1.5   & 2.0     & 2.2  & 2.3       & 2.5     & 1.3       & 1.4      & 2.8      & 6.9     & 3.5   \\
'class I'             & 11    & 4.3   & 2.3     & 11   & 2.1$^{b}$ & 1.6     & 2.1       & $\ldots$ & 7.4      & 30      & 31    \\
Pre-stellar           & 1.4   & 1.8   & 1.5     & 0.80 & 2.3       & 2.7     &$<0.36$    & 0.77     & 12       & 4.8     & 8.2   \\ \hline
\end{tabular}

$^{a}$excluding \object{VLA1623}. $^{b}$only detected for
\object{CB244}. $^{c}$\cite{schoeier02}. $^{d}$This paper.
\end{table*}

In general the derived abundances vary by one to two orders of
magnitude over the entire sample. Following the trend seen in Paper~I
of increasing abundances with decreasing envelope masses, the objects
are accordingly separated into groups with envelope masses ($M_{> \rm
10 K}$) higher or lower than 0.5~$M_\odot$, roughly corresponding to
class 0 and class I objects, respectively. This definition only moves
the two borderline class 0/I objects L1551 and CB244 from class I to
class 0 and vice versa compared to the source list given in Table~1 of
Paper~I.
\begin{figure*}
\resizebox{\hsize}{!}{\includegraphics{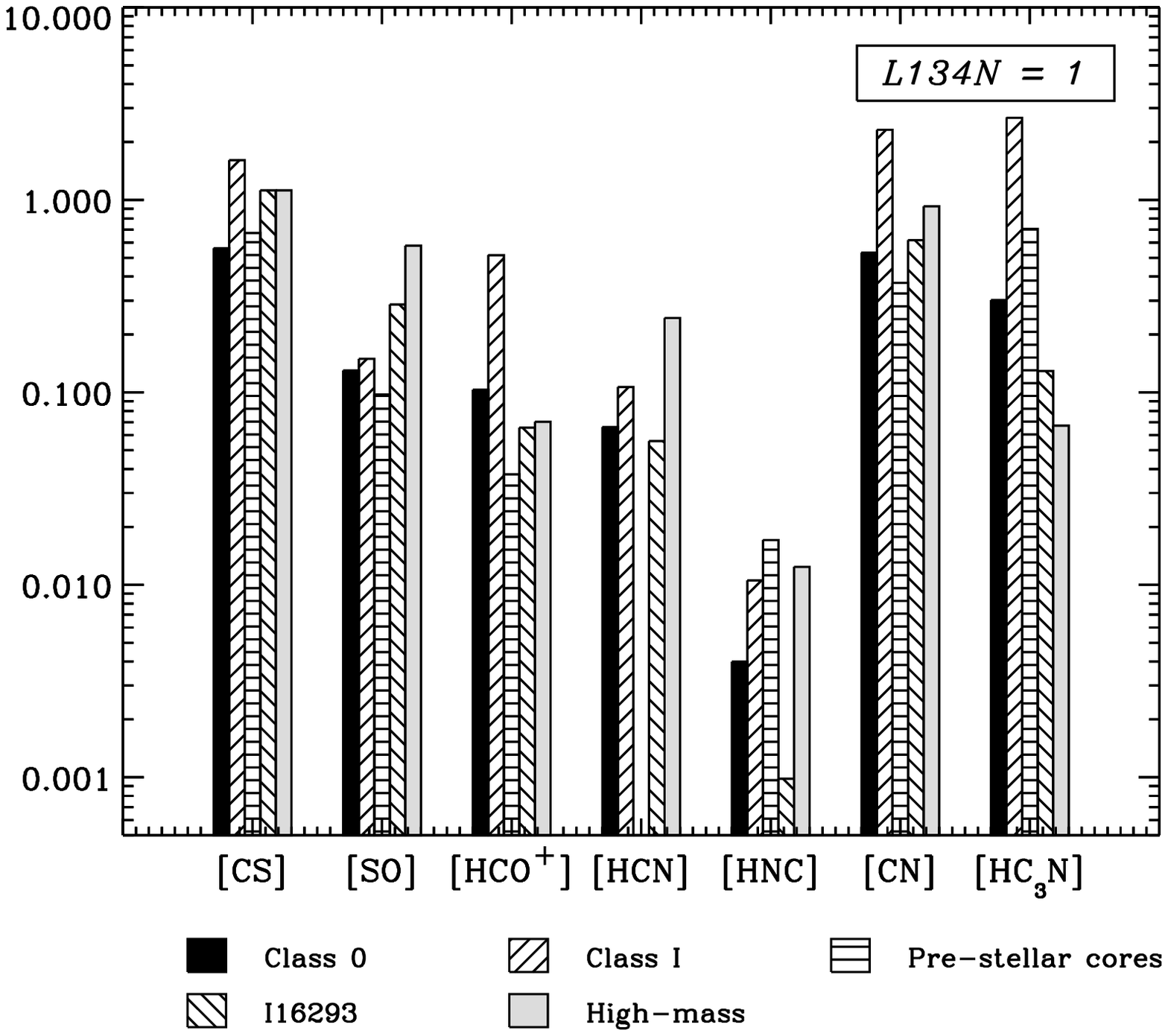}}
\caption{Comparison between average abundances for class 0 and I
objects and pre-stellar cores (this paper), \object{IRAS~16293-2422}
outer envelope \citep{schoeier02}, average abundances for W3(IRS4),
W3(IRS5) and W3(H2O) \citep[all high-mass YSOs;][]{helmich97} and
abundances in the dark cloud L134N \citep{dickens00}. Note that the
L134N abundances have been rescaled assuming a CO abundance of
2.7$\times 10^{-4}$ \citep{lacy94}, as was also assumed by \cite{helmich97}
for the high-mass YSOs. The L134N abundances thereby become:
[CS]~=~2.7$\times 10^{-9}$, [SO]~=~1.5$\times 10^{-8}$, [HCO$^+$]~=~2.1$\times 10^{-8}$,
[HCN]~=~2.0$\times 10^{-8}$, [HNC]~=~7.0$\times 10^{-8}$, [CN]~=~1.3$\times 10^{-9}$, and
[HC$_3$N]~=~1.2$\times 10^{-9}$.}\label{abundhist}
\end{figure*}

On average the class 0 objects have lower abundances than the class I
objects for most species (see Fig.~\ref{abundhist}). The most
pronounced effect is seen for CO, HCO$^+$ and CN where the average
abundances differ by up to an order of magnitude, whereas especially
SO and HCN have close to constant abundances with envelope mass,
albeit with large scatter around the mean. As discussed in the
following sections the variations of abundances with mass are not
identical, however, which indicates chemical effects regulating the
relative abundances for the different molecular species. In order to
quantify this more rigorously and in an unbiased way, the Pearson
correlation coefficients were calculated for each set of abundances
and are listed in Table~\ref{pearsoncoeff}. The Pearson correlation
coefficient is a measure of how well a $(x,y)$ data set is fitted by a
linear correlation compared to the spread of $(x,y)$ points. Values of
$\pm 1$ indicate good correlations (with positive or negative slopes)
whereas a value of 0 indicates no correlation.
\begin{table*}
\caption{Pearson correlation coefficients for the abundances for all
objects.}\label{pearsoncoeff}
\begin{tabular}{llllllllll} \hline\hline
        & Mass   & CO     &HCO$^+$ & CS     & SO     & HCN    & HNC    & CN     & HC$_3$N \\ \hline
Mass    &$\ldots$& -0.74  & -0.51  & -0.46  & -0.18  & -0.11  & -0.40  & -0.71  & -0.71   \\
CO      & -0.74  &$\ldots$& 0.79   & 0.69   & 0.35   & 0.46   & 0.52   & 0.69   & 0.59    \\
HCO$^+$ & -0.51  & 0.79   &$\ldots$& 0.80   & 0.48   & 0.44   & 0.70   & 0.69   & 0.48    \\
CS      & -0.46  & 0.69   & 0.80   &$\ldots$& 0.79   & 0.29   & 0.48   & 0.31   & 0.39    \\
SO      & -0.18  & 0.35   & 0.48   & 0.79   &$\ldots$& -0.05  & 0.14   & -0.27  & -0.03   \\
HCN     & -0.11  & 0.46   & 0.44   & 0.29   & -0.05  &$\ldots$& 0.63   & 0.55   & 0.45    \\
HNC     & -0.40  & 0.52   & 0.70   & 0.48   & 0.14   & 0.63   &$\ldots$& 0.86   & 0.72    \\
CN      & -0.71  & 0.69   & 0.69   & 0.31   & -0.27  & 0.55   & 0.86   &$\ldots$& 0.83    \\
HC$_3$N & -0.71  & 0.59   & 0.48   & 0.39   & -0.03  & 0.45   & 0.72   & 0.83   &$\ldots$ \\ \hline
\end{tabular}
\end{table*}

\begin{figure}
\resizebox{\hsize}{!}{\includegraphics{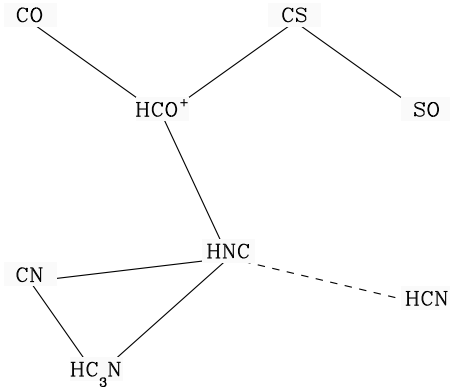}}
\caption{Relations between different molecules as judged from the
Pearson correlation coefficients. The dashed line between HCN and HNC
indicates the strongest correlation for HCN with any of the other
molecules studied. The correlation coefficient for this relation is,
however, lower than the cut of 0.7 adopted for good
correlations.}\label{abundancerelations}
\end{figure}

As can be seen from Table~\ref{pearsoncoeff}, significant differences
exist between the various sets of abundances. Setting an (arbitrary)
cut of $>|0.7|$ to indicate good correlation, the results suggest that
the molecular species are related as indicated in
Fig.~\ref{abundancerelations}. Individual results are shown in
Fig.~\ref{first_abundfig}-\ref{last_abundfig}. The abundances of
groups of species, e.g., the nitrogen- or sulfur-bearing species, are
closely related as expected from naive chemical considerations. HCN is
the only molecule whose abundance does not directly correlate with
that of any other molecule at this level. The closest correlation is
found with its isomer HNC (correlation coefficient of 0.63). The best
correlation between abundance and mass is found for CO followed by CN
and HC$_3$N. Naturally the correlations seen in this comparison may
indicate the structure of the general chemical network rather than
direct relations between the individual molecules: for example the
ranking of correlations for SO is as follows: CS (0.79), HCO$^+$
(0.48) and CO (0.35). As indicated in Fig.~\ref{abundancerelations}
this is exactly the decline in correlation coefficients one would
expect with the relations between these species on the pair-by-pair
comparison basis adopted when constructing
Fig.~\ref{abundancerelations}. Such ``connectivity'' could also be the
cause for the relation between HNC, CN and HC$_3$N - the correlation
between HNC and HC$_3$N may in fact just reflect that both these
molecules are related to CN. The rather low number statistics imply
that care should be taken not to overinterpret the absolute values of
the correlation coefficients, but as a first step they give valuable
hints. To fully understand the underlying chemistry, a more in-depth
consideration on a species by species basis is required, as discussed
in the following sections.

\subsection{CS and SO}
\begin{figure}
\resizebox{\hsize}{!}{\includegraphics{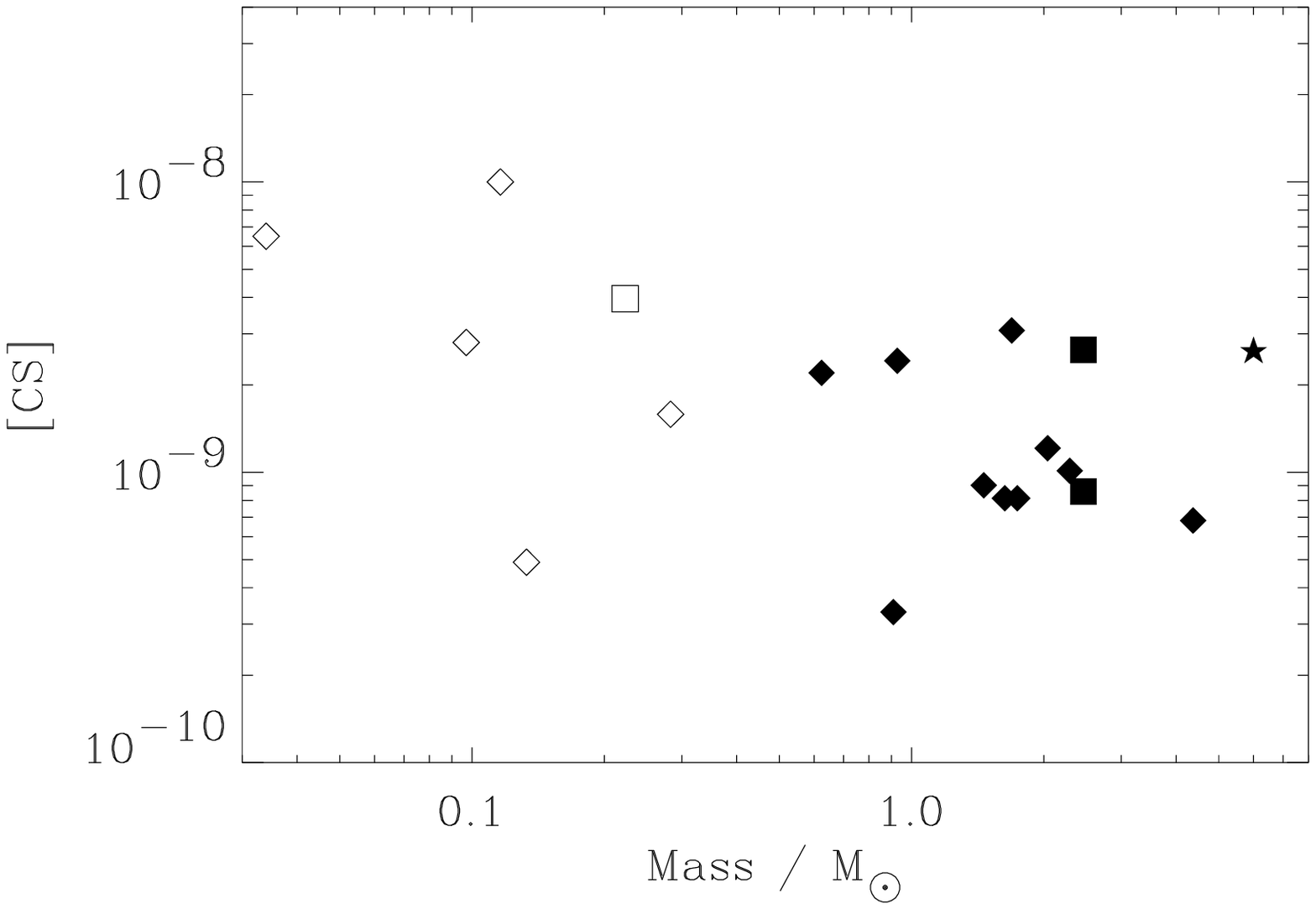}}
\resizebox{\hsize}{!}{\includegraphics{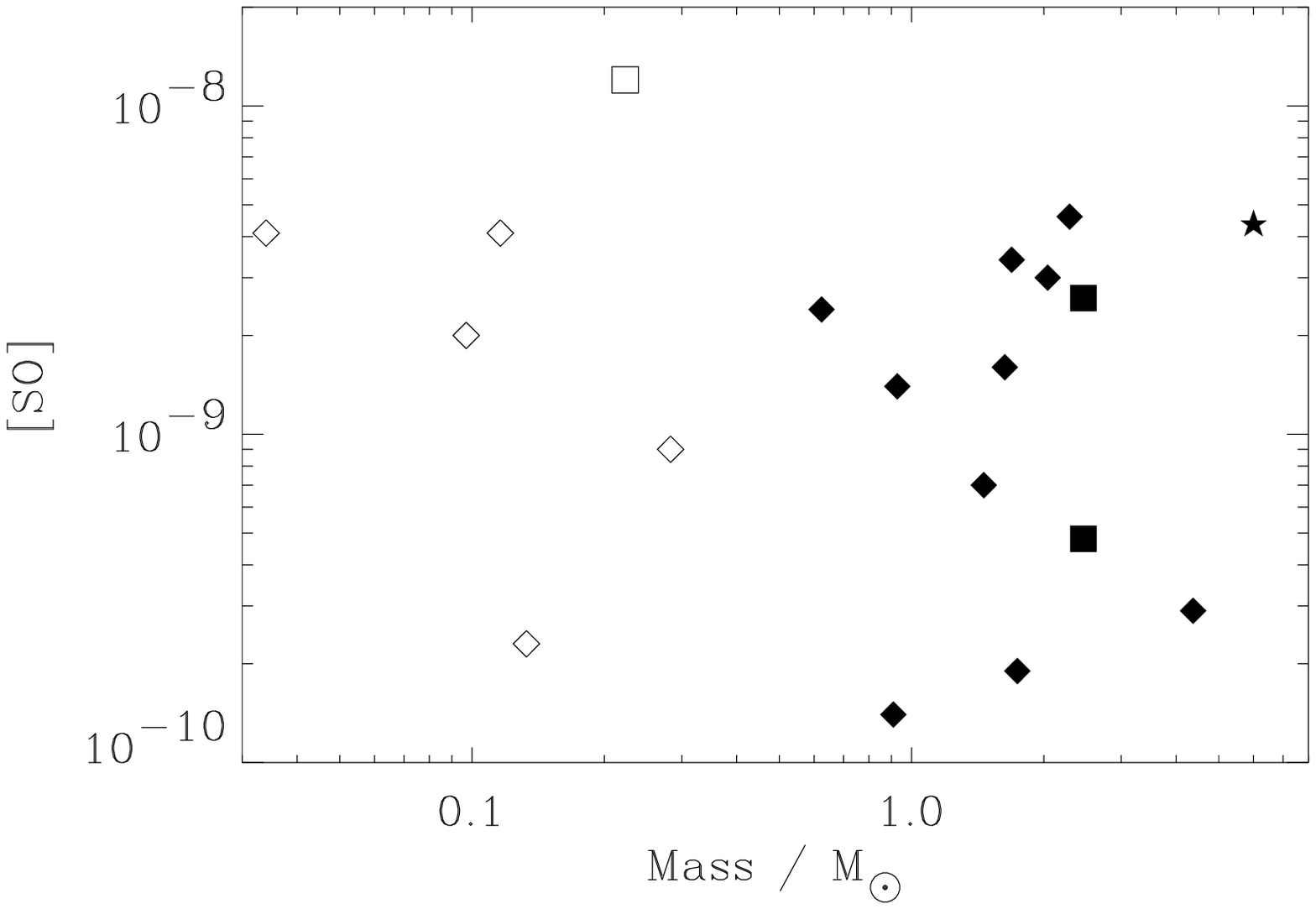}}
\caption{Abundances of CS from optically thin C$^{34}$S isotopic lines
(where detected) and CS lines (upper panel) and of SO (lower panel)
vs. mass. In this figure and in following figures in this paper, the
class 0 objects are indicated by ``$\blacklozenge$'', the class I
objects by ``$\lozenge$'' and the pre-stellar cores by
``$\blacksquare$''. The class 0 objects \object{VLA1623} and
\object{IRAS~16293-2422} have been singled out by ``$\square$'' and
``$\bigstar$'',
respectively.}\label{cs_mass}\label{so_mass}\label{first_abundfig}
\end{figure}

As can be seen from Fig.~\ref{cs_mass} abundances of the
sulfur-bearing species, CS and SO, are close to constant with envelope
mass, contrasting the picture for CO \citepalias{jorgensen02}. CS has
often been used to constrain the density scales in protostellar
envelopes \citep[e.g.,][]{vandertak00} assuming the chemistry to be
homogeneous throughout the envelope. SO also does not show any
significant trends with envelope mass, but has a larger scatter.

CS and SO are indeed found to show relatively small abundance
variations in pure gas-phase models of pre-stellar cores
\citep{bergin97} and in cold exterior regions of high-mass
protostellar envelopes \citep{doty02}. As argued by \cite{bergin97}
and \cite{bergin01}, however, sulfur-bearing species such as CS and SO
should suffer from depletion at densities and temperatures
characteristic for these regions. Maps of CS toward pre-stellar cores
\citep{tafalla02,difrancesco02} and molecular clouds \citep[e.g.,
IC~5146;][]{bergin01} combined with models of the abundances suggest
that this molecule does indeed freeze out toward the inner colder and
denser parts. Typical abundances in such environments range from
$\approx 1\times 10^{-10}$ to a few $\times 10^{-9}$ between the inner (low
abundance) and outer (high abundance) regions. This agrees well with
the average abundances found for the protostellar envelopes analyzed
in this paper, which have a central source of heating. Our CS
abundances are also similar to those inferred for a sample of
high-mass protostars by \cite{vandertak00} using a similar analysis.

An important conclusion regarding the derived CS abundances concerns
the impact of outflow processing of the gas in the envelopes: CS and
SO are seen to be greatly enhanced in shocked gas in protostellar
outflows \citep{bachiller97,i2art}. The small source-to-source
variation in the derived CS abundances, however, illustrates that
although increased CS abundances may be present in small parts of the
envelopes, the bulk of the emission originates in parts of the
envelope unaffected by such processes. The same conclusion was reached
by \cite{n1333i2art} from millimeter interferometer observations of
the C$^{34}$S 2--1 line emission toward NGC~1333-IRAS2.

It has been suggested that comparison between sulfur-bearing species
like SO and CS can be used as chemical probes of the evolutionary
stages in star-forming regions \citep[e.g.,][]{ruffle99} - both when
considering high- \citep{charnley97, hatchell98} and low-mass stars
\citep{buckle03}. The time-dependence of the sulfur-chemistry network
is initiated when significant amounts of H$_2$S are released in the
gas-phase by evaporation of grain-mantles. This is followed by
formation of SO and SO$_2$ (through reactions with H and H$_3$O$^+$
forming S and H$_3$S$^+$ and subsequently through reactions with OH
and O$_2$). At later times most of the sulfur is incorporated into CS,
H$_2$CS and OCS.

In particular \cite{buckle03} estimated abundances of sulfur-bearing
species from SO, SO$_2$ and H$_2$S line observations toward a sample
of class 0 and I objects assuming LTE and a constant CO/H$_2$
abundance ratio. They found that their class I low-mass YSOs had lower
abundances of SO and H$_2$S than class 0 objects, suggesting that this
was a result of their later chemical evolutionary stage. For the
sources in this paper it is seen, however, that there is no
significant difference in SO abundances between class 0 and I
objects. Van der Tak et al. \citeyearpar{vandertak03} surveyed a range
of different sulfur-bearing species toward a sample of high-mass YSOs
and likewise found no systematic trends between known indicators of
the evolutionary stage and the abundances of the sulfur molecules.

\begin{figure}
\resizebox{\hsize}{!}{\includegraphics{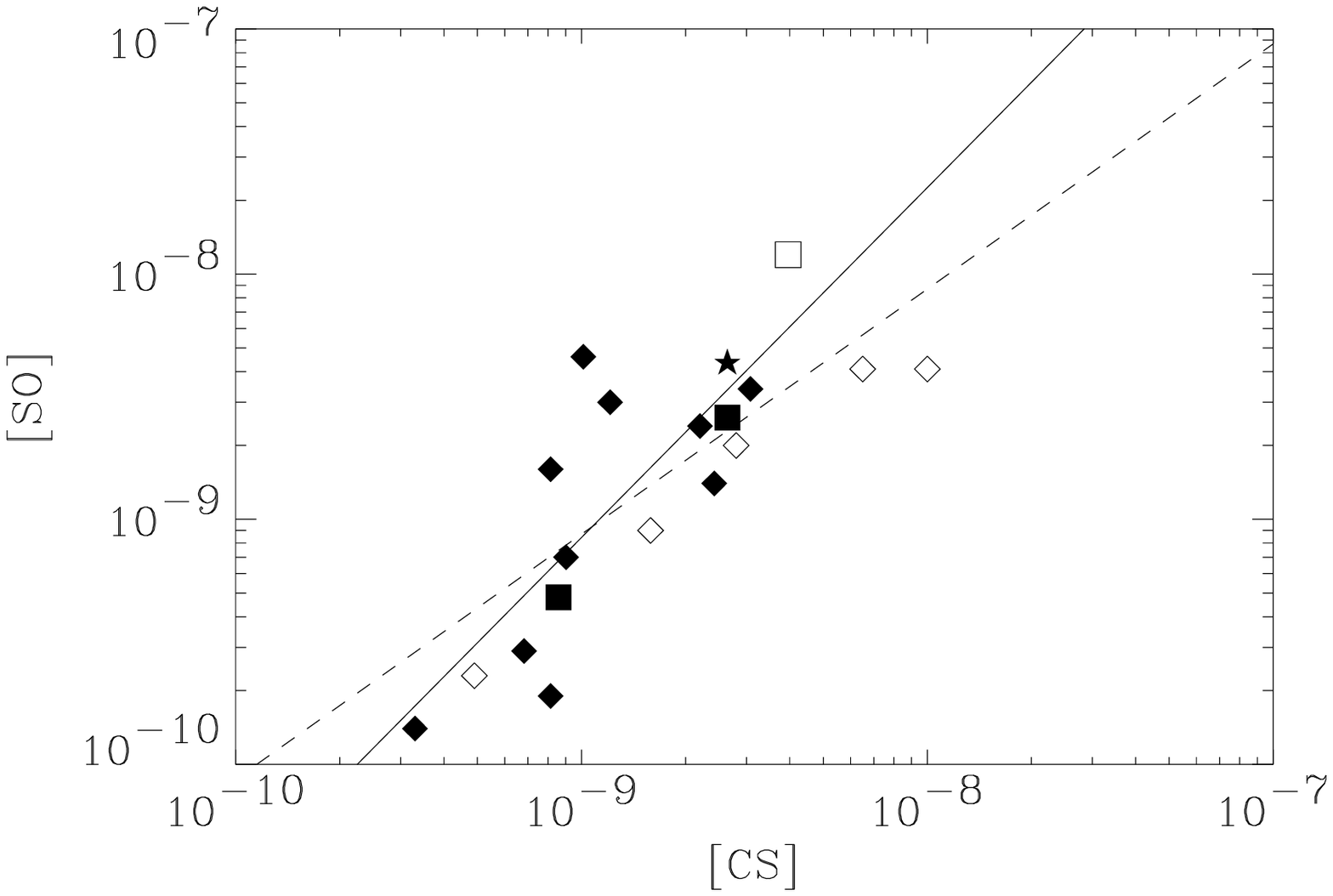}}
\resizebox{\hsize}{!}{\includegraphics{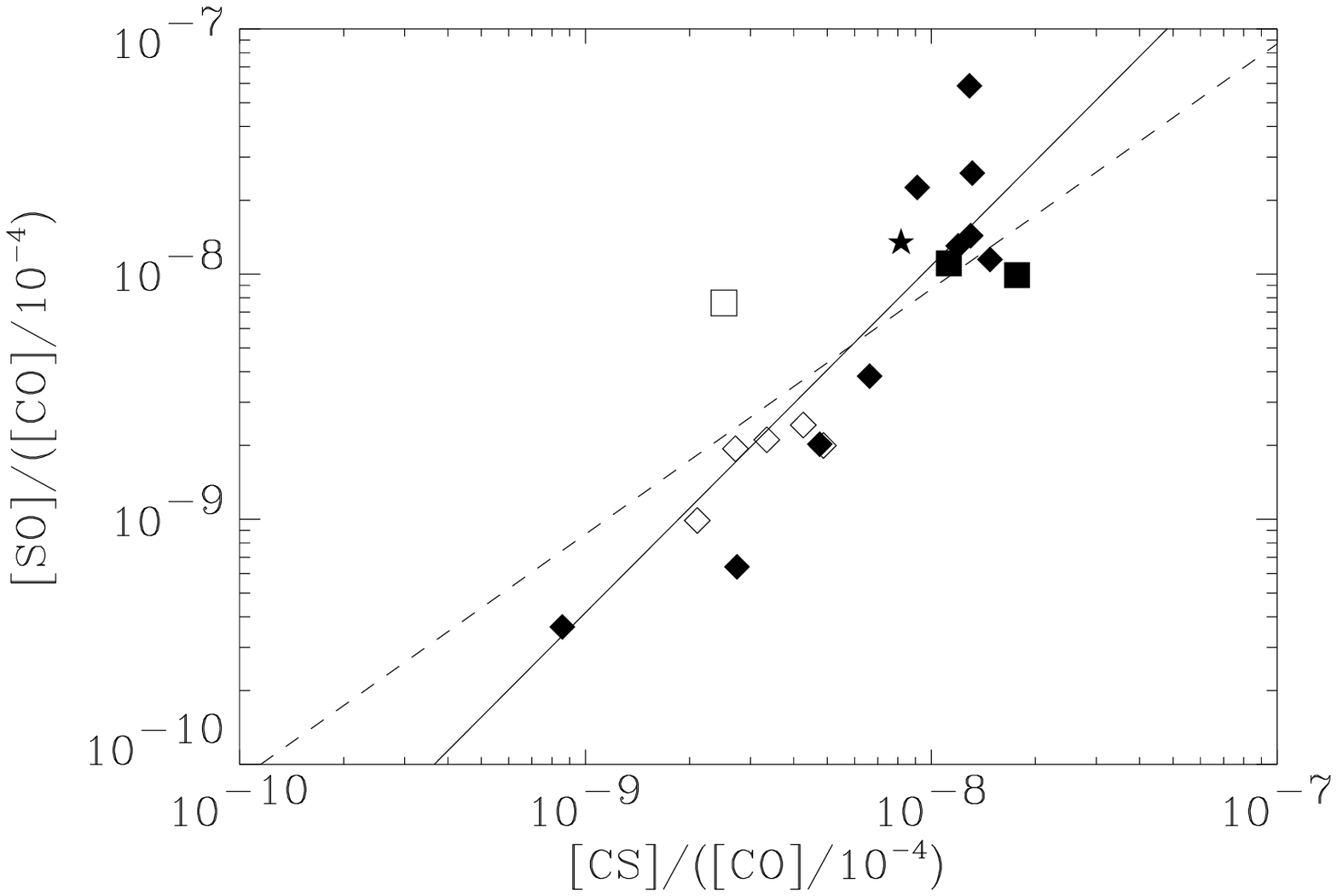}}
\caption{CS vs. SO abundance. The dashed line indicates a linear
relation between the CS and SO abundances, the solid line is the
best-fit correlation. In the lower panel the abundances have been
normalized to a CO abundance of 10$^{-4}$, mimicking the assumption in
\cite{buckle03}. Symbols are defined in
Fig.~\ref{first_abundfig}.}\label{cs_so}
\end{figure}

There is a slight overlap between the objects studied by
\cite{buckle03} and those treated in this paper. For these overlapping
objects the \citeauthor{buckle03} data have been included in our
analysis as indicated in Table~\ref{abundso}, and it is found that our
models can explain their observations. A likely explanation for the
different findings is therefore the CO depletion found for sources
with the more massive envelopes (Paper~I and Sect.~\ref{radial} in
this paper). In fact abundances calculated assuming a [CO/H$_2$]
abundance of 1$\times 10^{-4}$ leads to overestimated abundances for objects
in which CO is depleted, i.e., those with the most massive envelopes
\citepalias{jorgensen02}. The abundances for the class 0 objects in
\cite{buckle03} could therefore be overestimated and their
evolutionary trend an artifact of this assumption. Fig.~\ref{cs_so}
compares the relation between CS and SO abundances relative to the
density scale set in Paper~I and to a CO abundance of
10$^{-4}$. Fixing the CO abundance increases the average SO and CS
abundances for the class 0 objects - to almost an order of magnitude
higher than those for the class I objects. This in fact resembles what
\cite{buckle03} find.

An interesting feature of Fig.~\ref{cs_so} is the correlation between
the CS and SO abundances. Here the normalization to the CO abundance
also serves as a valuable test: if for some reason the absolute
density scale had been systematically overestimated for the most
massive envelopes and underestimated for the least massive envelopes,
a false trend of abundances with mass could result and trends between
abundances such as those seen in Fig.~\ref{cs_so} should arise. In
this case, however, normalization by a ``standard'' abundance should
take out such an effect, but as illustrated in Fig.~\ref{cs_so} this
is not the case. The relation between CS and SO therefore seems to be
real.

Interestingly, the CS/SO abundance ratio has previously also been
suggested to trace evolutionary effects related to cloud conditions
and evolution, e.g., variation of the initial C/O ratio, density
effects, the temporal evolution of a given core or importance of
X-rays \citep{bergin97b,nilsson00}. It is found through time dependent
modeling of the chemistry that the CS/SO ratio increases throughout
the evolution of a molecular cloud starting from an atomic carbon-rich
phase, but stabilizes at late times at a level dependent on the
initial C/O ratio. As illustrated in Fig.~\ref{cs_so}, the
relationship between the CS and SO abundances is clearly non-linear,
implying that one or more of these effects may play a role in
determining the relative abundances of these two molecules. The CS/SO
ratio varies from $\approx$~0.2 to 4, in good agreement with the
results of \cite{nilsson00} who analyzed CS and SO abundances from a
sample of 19 molecular clouds.

SO$_2$ is detected toward only a few sources in the sample. Typically,
the upper limit to the SO$_2$ abundance is found to be a few$\times 10^{-10}$
in this study. The same was seen by \cite{buckle03} who only detected
SO$_2$ emission toward 30\% of their sources, i.p., sources in the
Serpens region. In fact, \citeauthor{buckle03} did not detect SO$_2$
for any of the four sources also in our sample. For these sources
upper limits based on the observations of \citeauthor{buckle03} are a
few$\times 10^{-11}$.

The non-detections can also be compared to the results of
\cite{schoeier02} for \object{IRAS~16293-2422}, for which abundance
jumps, either due to thermal evaporation or outflow-induced shocks,
were found. \citeauthor{schoeier02} argued for an SO$_2$ abundance
jump from 2$\times 10^{-10}$ in the outer envelope to 1$\times
10^{-7}$ in the inner envelope. The SO$_2$ lines in this study are in
fact expected to probe the outer region of the envelope and the
derived upper limit to the abundances do seem to indicate that the
abundances found for \object{IRAS~16293-2422} are higher than those
found here. It is interesting to note that SO$_2$ is only detected
toward regions with high outflow activity (i.p., the objects in
NGC~1333 and \object{VLA1623}) and that the SO 8$_7$-7$_6$ was found
to be very broad (and only detected) toward these objects with widths
of $\sim 5-10$~km~s$^{-1}$ (FWHM) contrasting the other observed
lines. These objects also show the highest SO abundances. Together
with the strong SO and SO$_2$ emission toward the Serpens sources
which are also related to strong outflows, this suggests an
enhancement of sulfur-bearing species in the inner envelopes due to
outflows. Large enhancements of the sulfur-species (together with
CH$_3$OH and SiO) are observed in outflows where these can be studied
well separated from their driving protostar
\citep{bachiller97,i2art}. A deep systematic study of the line
emission from these and other sulfur species (e.g., H$_2$S, HCS$^+$,
H$_2$CS) toward a large sample of objects will shed more light on this
question and thus provide better insight into the sulfur-chemistry in
low-mass protostars.

\subsection{HCO$^+$ and N$_2$H$^+$}\label{hcop}\label{nthp}
HCO$^+$ is of great importance in chemical models of protostellar
environments as it is the primary molecular ion and thus regulator of
the electron density/ionization structure \citep[e.g.,][]{caselli02b}
and the most important destroyer of other molecules
\citep[e.g.,][]{bergin97}.
\begin{figure}
\resizebox{\hsize}{!}{\includegraphics{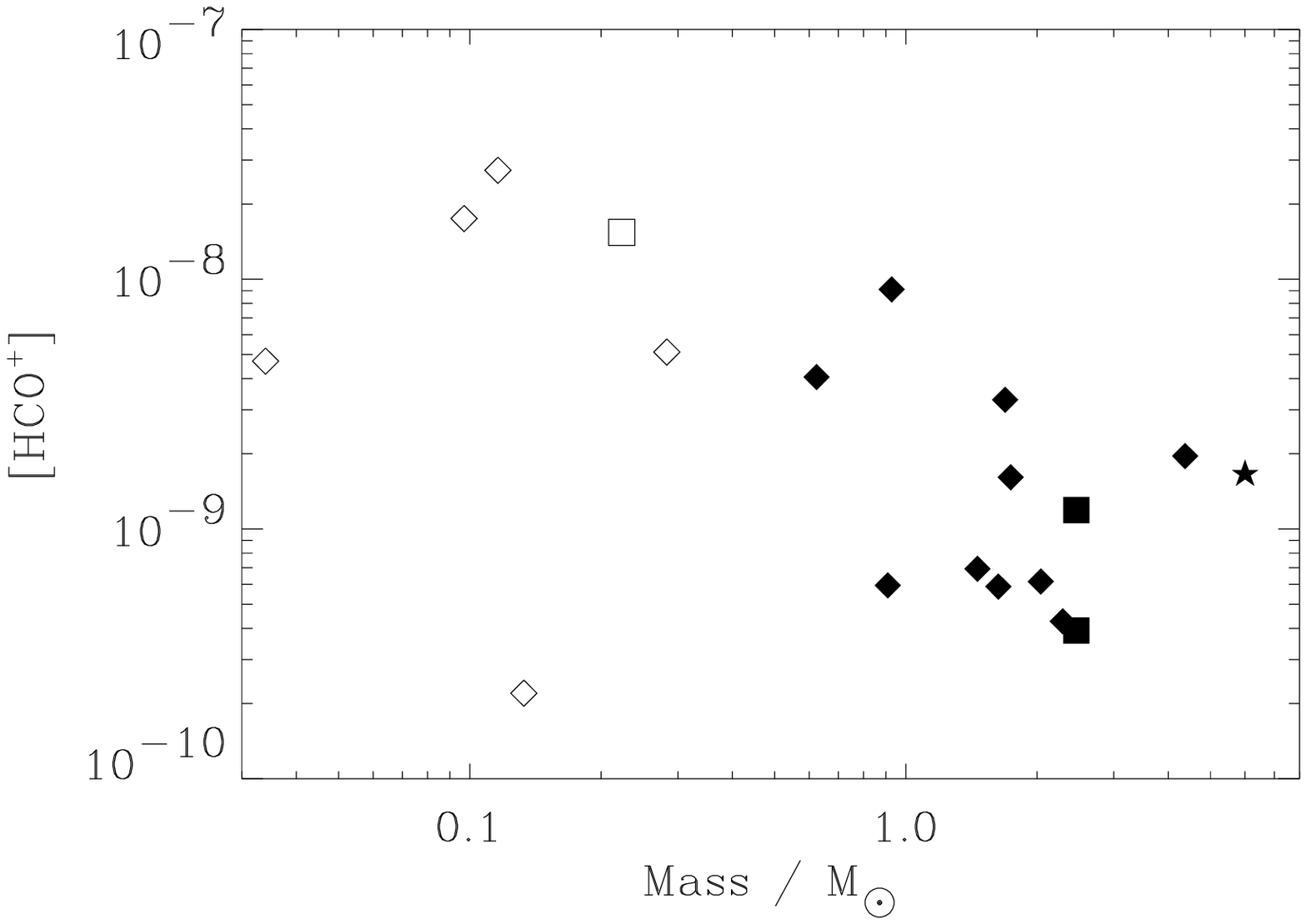}}
\resizebox{\hsize}{!}{\includegraphics{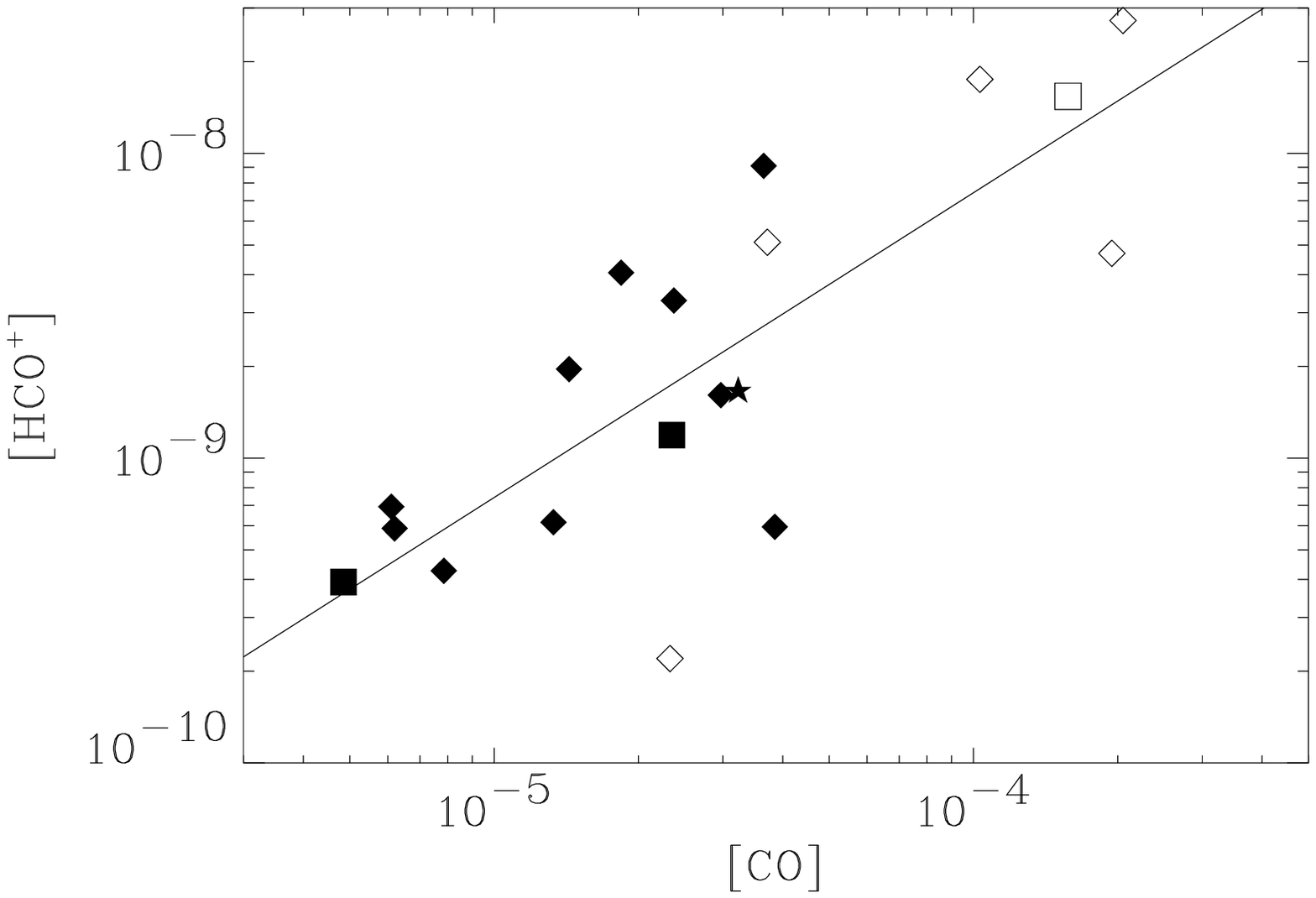}}
\caption{HCO$^+$ abundance vs. mass (upper panel) and vs. CO abundance
(lower panel). In the lower panel has the linear correlation between
the HCO$^+$ and CO abundances been
overplotted. Symbols defined as in Fig.~\ref{first_abundfig}.}\label{hcoP_mass}\label{hcoP_co}
\end{figure}

As shown in the upper panel of Fig.~\ref{hcoP_mass}, the derived
HCO$^+$ abundances show an evolution with mass similar to that found
for CO \citepalias{jorgensen02}. This is even more clearly illustrated
in the lower panel of Fig.~\ref{hcoP_co}, where a tight correlation
between CO and HCO$^+$ abundances is seen. In fact, the CO and HCO$^+$
abundances are linearly dependent with
\[{\rm [HCO^+]}=7.4\times 10^{-5}\times {\rm [CO]}\]
or put differently: a ``standard'' undepleted CO abundance of
$10^{-4}$ corresponds to an HCO$^+$ abundance of $7.4\times 10^{-9}$.

\begin{figure}
\resizebox{\hsize}{!}{\includegraphics{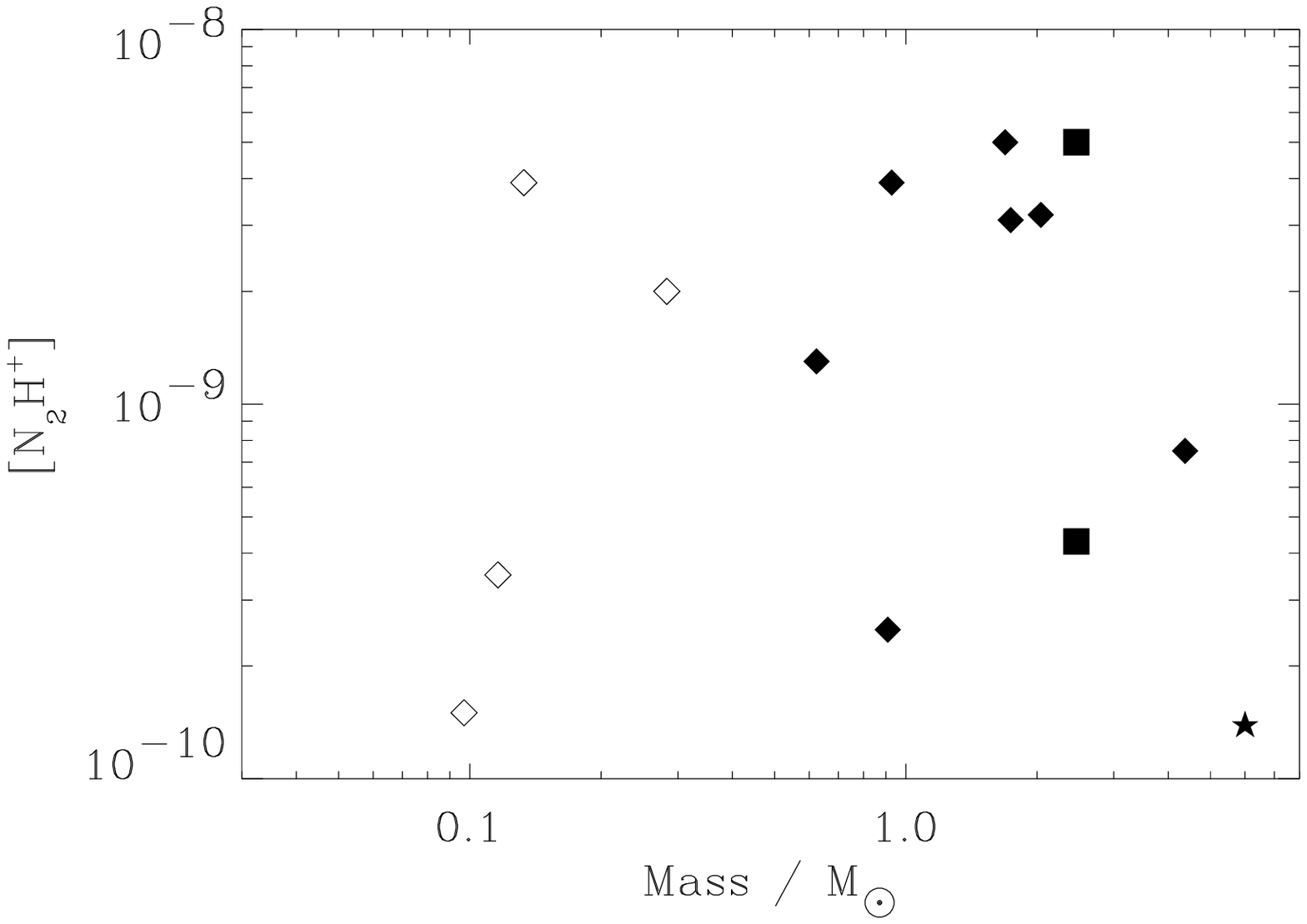}}
\resizebox{\hsize}{!}{\includegraphics{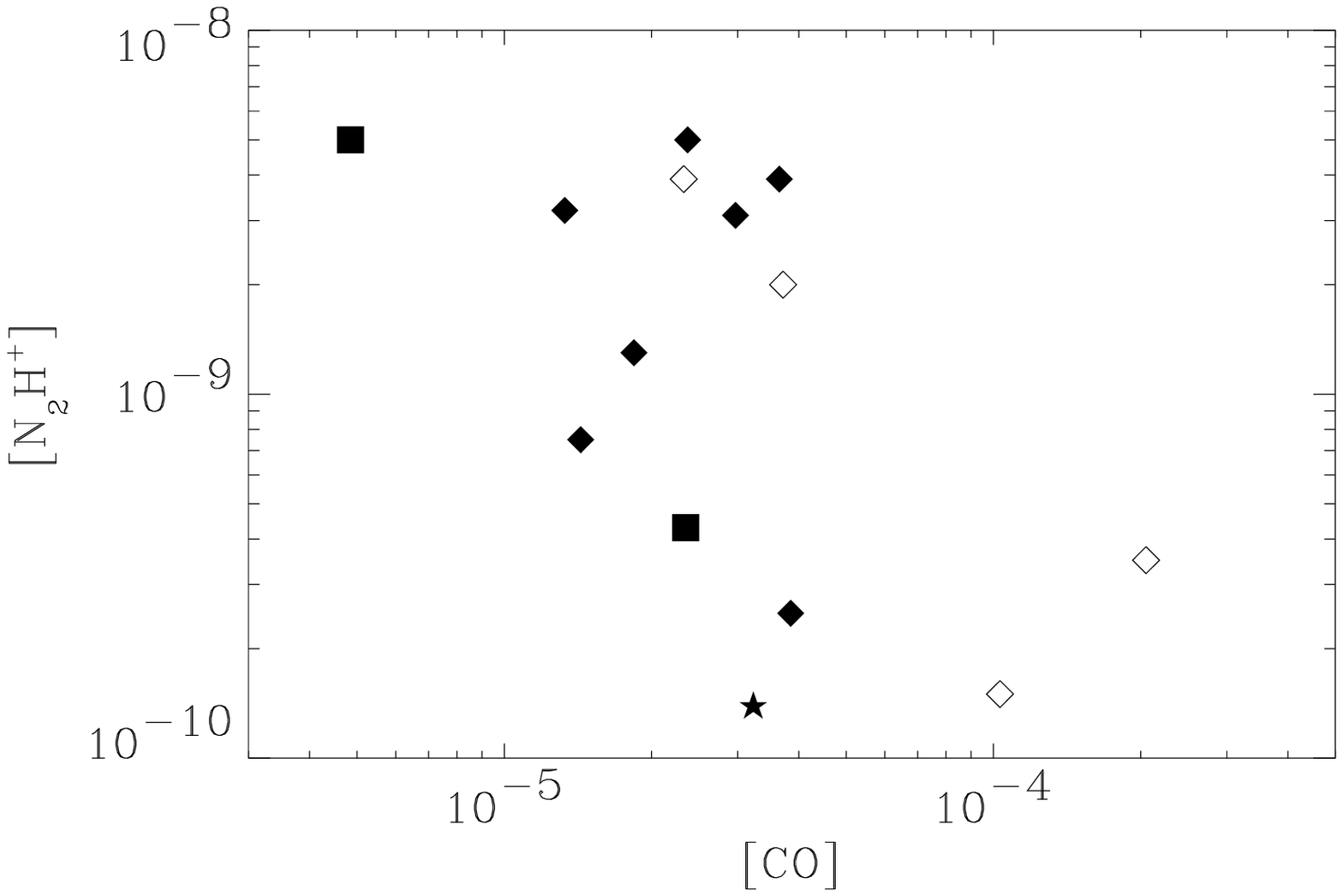}}
\caption{N$_2$H$^+$ abundance vs. mass (upper panel) and vs. CO abundance
(lower panel). Symbols as in
Fig.~\ref{first_abundfig}.}\label{n2hP_mass}\label{n2hP_co}
\end{figure}

It is found that N$_2$H$^+$ marks a clear contrast to HCO$^+$: as
shown in Fig.~\ref{n2hP_mass} the N$_2$H$^+$ abundance decreases with
increasing CO abundance. High angular resolution interferometer maps
of protostellar regions \citep[e.g.,][]{bergin01,n1333i2art} find that
cores with low CO abundances show up stronger when mapped in N$_2$H$^+$.

Both trends can be understood when considering the chemical network in
more detail taking the depletion of CO into account. For both HCO$^+$
and N$_2$H$^+$ the primary formation routes are through reactions with
H$_3^+$, i.e.:
\begin{equation}
{\rm H}_3^+ + {\rm CO}  \rightarrow {\rm HCO}^+ + {\rm H}_2 \label{hcoP_formation}
\end{equation}
\begin{equation}
{\rm H}_3^+ + {\rm N}_2 \rightarrow {\rm N}_2{\rm H}^+ + {\rm H}_2 \label{n2hp_formation}
\end{equation}
For standard CO abundances ([CO/H$_2$]~$\sim 10^{-4}$)
eq.~(\ref{hcoP_formation}) is the dominant removal mechanism for
H$_3^+$, but as CO freezes out this reaction drops in importance and
eq.~(\ref{n2hp_formation}) becomes more important for the removal of
H$_3^+$. The main destruction mechanism for HCO$^+$ is dissociative
recombination, which is also the case for N$_2$H$^+$ when CO is
depleted. However, as CO returns to the gas-phase, destruction of
N$_2$H$^+$ through reactions with CO:
\begin{equation}
{\rm N}_2{\rm H}^+ + {\rm CO} \rightarrow {\rm HCO}^+ + {\rm N}_2
\label{n2hp_removal}
\end{equation}
becomes the dominant removal mechanism for N$_2$H$^+$. In
Appendix~\ref{chemical_network_details} we consider the chemical
network for H$_3^+$, HCO$^+$, and N$_2$H$^+$ in detail. The main
conclusions are that a linear increase of the HCO$^+$ abundance with
CO abundance is expected when CO is depleted. For higher CO
abundances, however, the HCO$^+$ abundance does not depend on [CO]
since a balance between formation through eq.~(\ref{hcoP_formation})
and destruction through dissociative recombination exists. In contrast
the N$_2$H$^+$ abundance is high when CO is depleted but declines rapidly
as $([{\rm CO}])^{-2}$ with the increasing CO abundance as H$_3^+$ is
removed (forming HCO$^+$) and N$_2$H$^+$ is destroyed through
eq.~(\ref{n2hp_removal}).

\begin{figure}
\resizebox{\hsize}{!}{\includegraphics{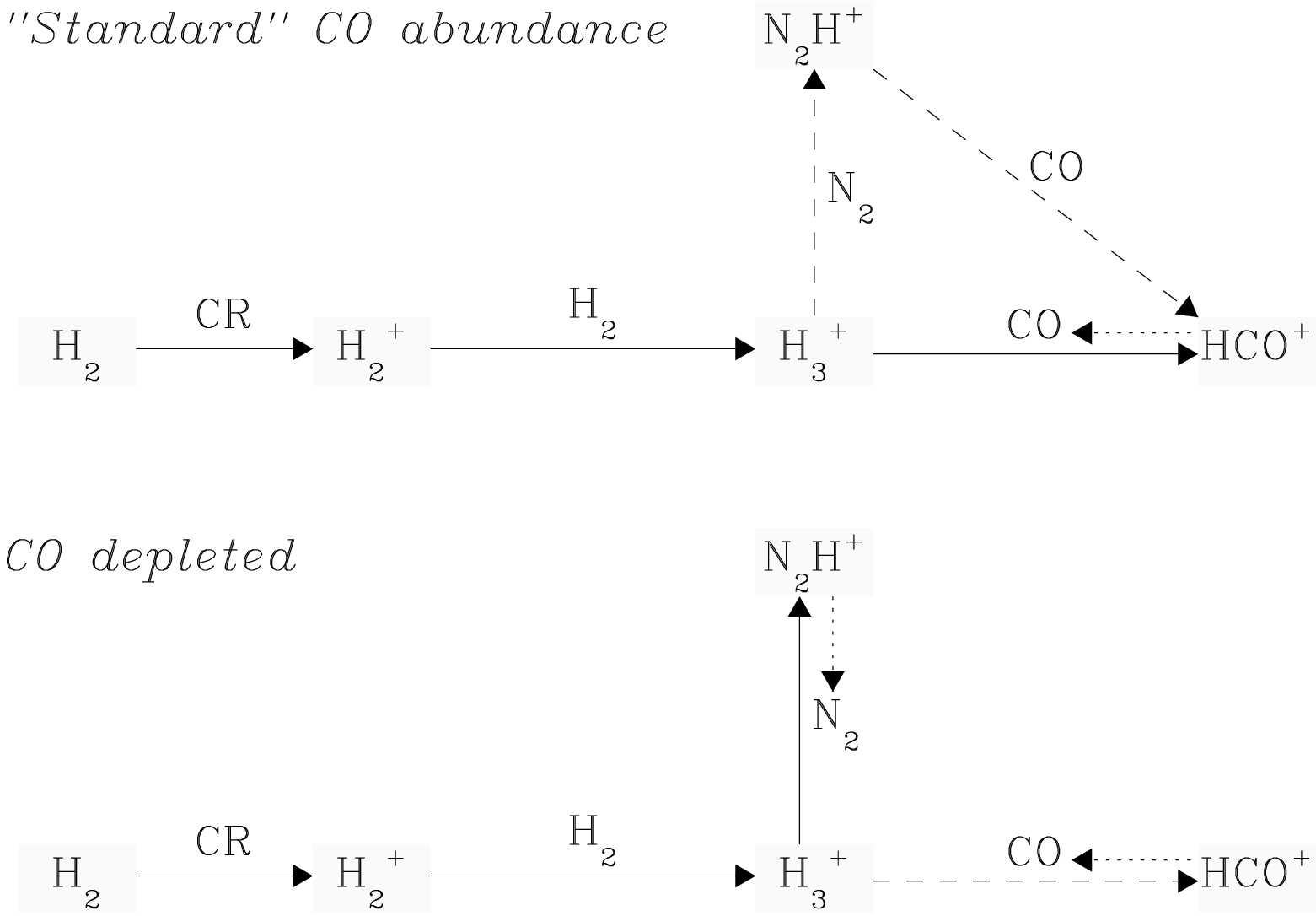}}
\resizebox{\hsize}{!}{\includegraphics{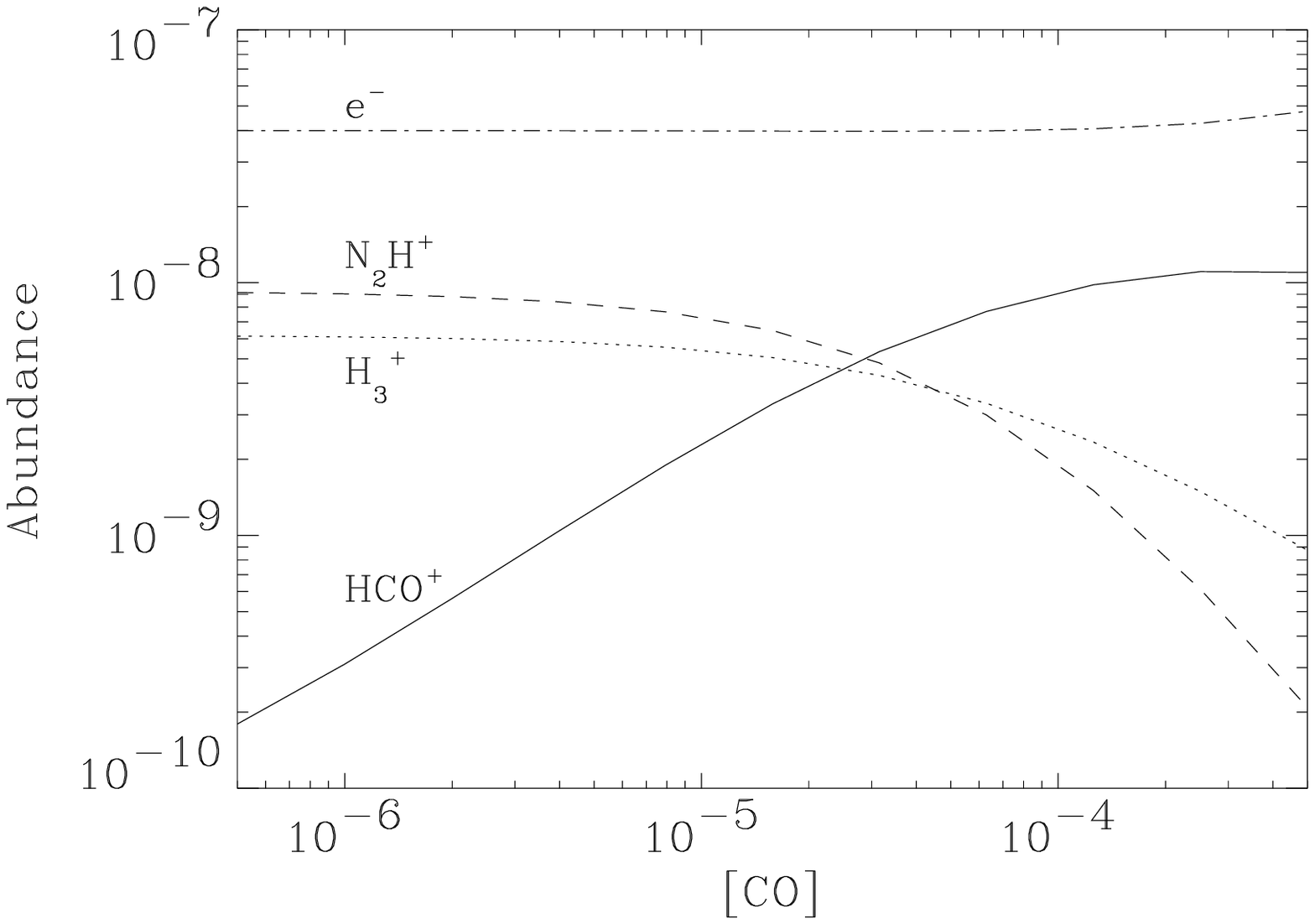}}
\caption{Upper panel: the chemical networks for low CO abundances
(i.e., depletion) and standard CO abundance ([CO]$\sim 10^{-4}$). The
dominant reactions are indicated by solid arrows, secondary reactions
by dashed arrows. Where dissociative recombination is the main
destruction for a molecule (i.e., N$_2$H$^+$ or HCO$^+$) this has been
indicated by a dotted arrow. Lower panel: the electron, N$_2$H$^+$,
H$_3^+$, and HCO$^+$ abundances as functions of CO abundance in a cell
with density $n({\rm H_2})=1\times 10^6~{\rm cm}^{-3}$ and temperature
$T=20$~K.}\label{co_n2hp_hcop}
\end{figure}

To further illustrate these points the upper panel of
Fig.~\ref{co_n2hp_hcop} shows the chemical network for low (depleted)
and standard CO abundances. The lower panel shows the N$_2$H$^+$,
HCO$^+$ and H$_3^+$ abundances as functions of CO abundance calculated
in a cell with density $n({\rm H_2})=1\times 10^6~{\rm cm}^{-3}$ and
temperature $T=20$~K at 10$^4$~years using the chemical code of
S.D.~Doty and adopting the chemistry used in the detailed chemical
modeling of the envelope around \object{IRAS~16293-2422}
\citep{doty04}. The figure clearly shows the linear relationship
between the CO and HCO$^+$ abundances for CO values lower than
$\approx$~2$\times 10^{-5}$ and likewise the rapid decline of
N$_2$H$^+$ for higher CO abundances. The absolute values of the
abundances and the exact CO abundance dividing between the ``low'' and
``standard'' [CO] regions is regulated by the exact details of the
chemistry (e.g., the initial N$_2$ abundance) and the cosmic ray
ionization rate, but the overall trends remain the same. Thus trends
of a linear increase of HCO$^+$ abundance with increasing CO abundance
can be understood in a limit where CO is depleted and
Eq.~(\ref{hcoP_formation}) is no longer the dominant removal mechanism
for H$_3^+$.

\subsection{HCN, HNC and CN}
\begin{figure*}
\resizebox{\hsize}{!}{\includegraphics{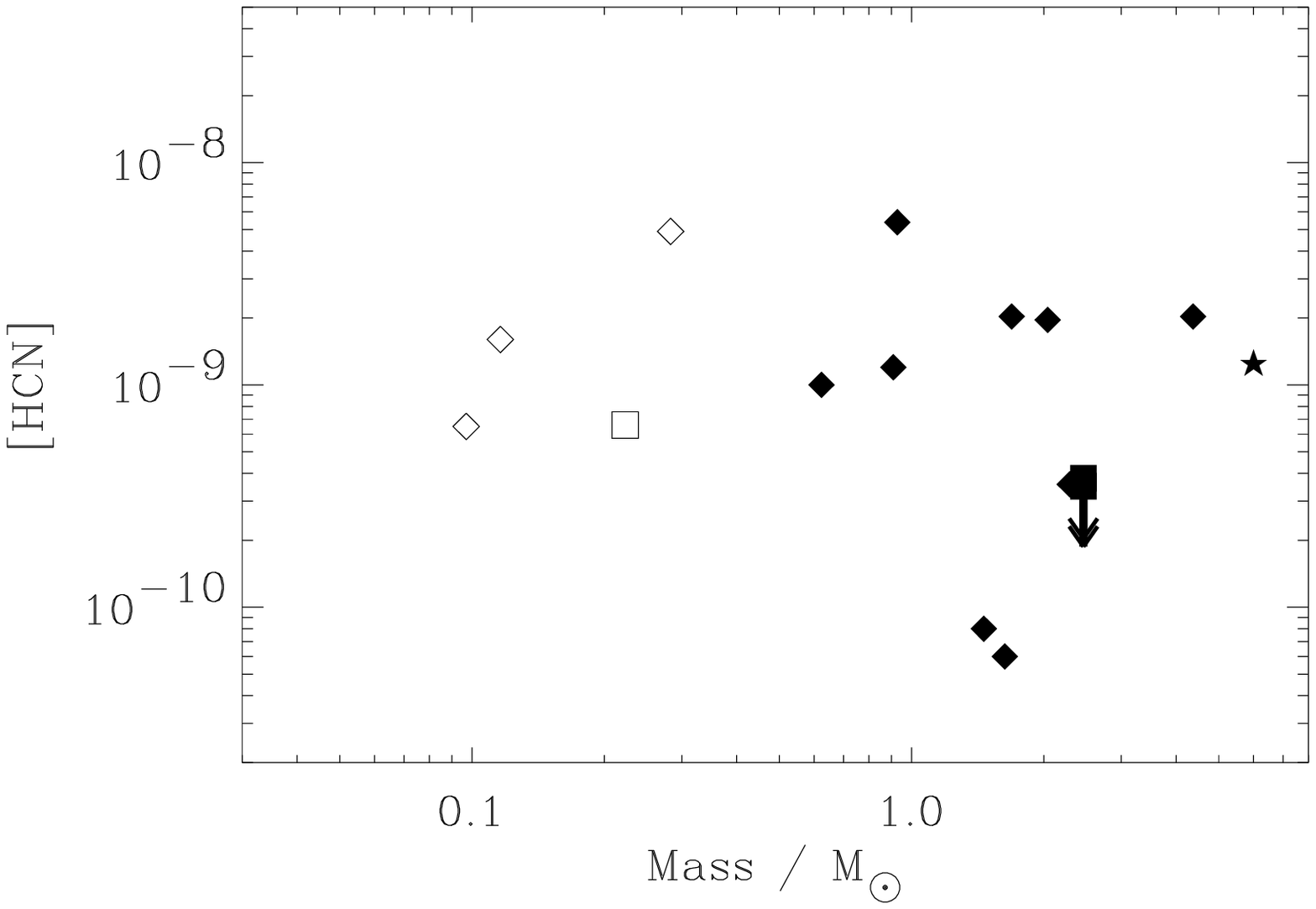}\includegraphics{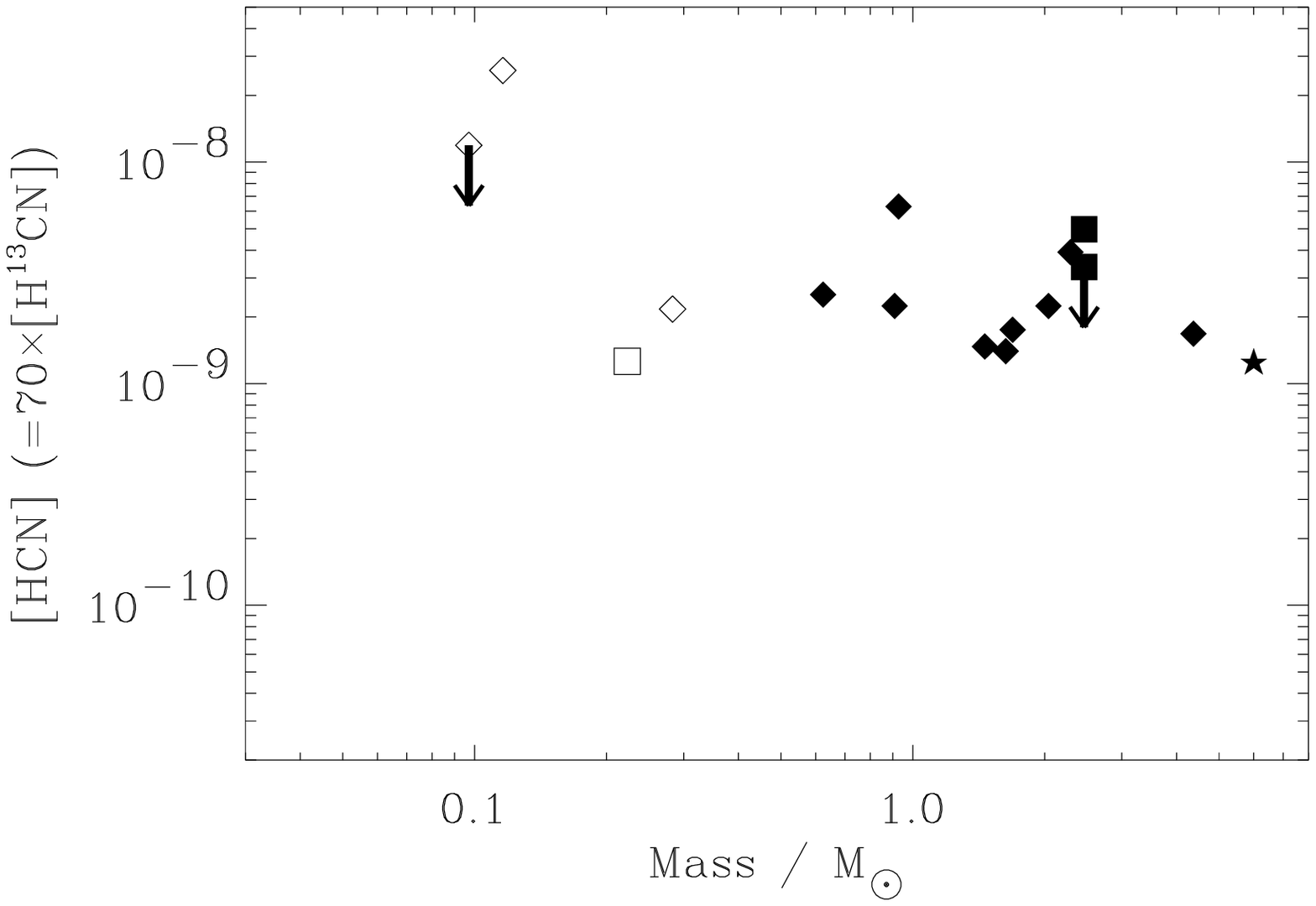}}
\resizebox{\hsize}{!}{\includegraphics{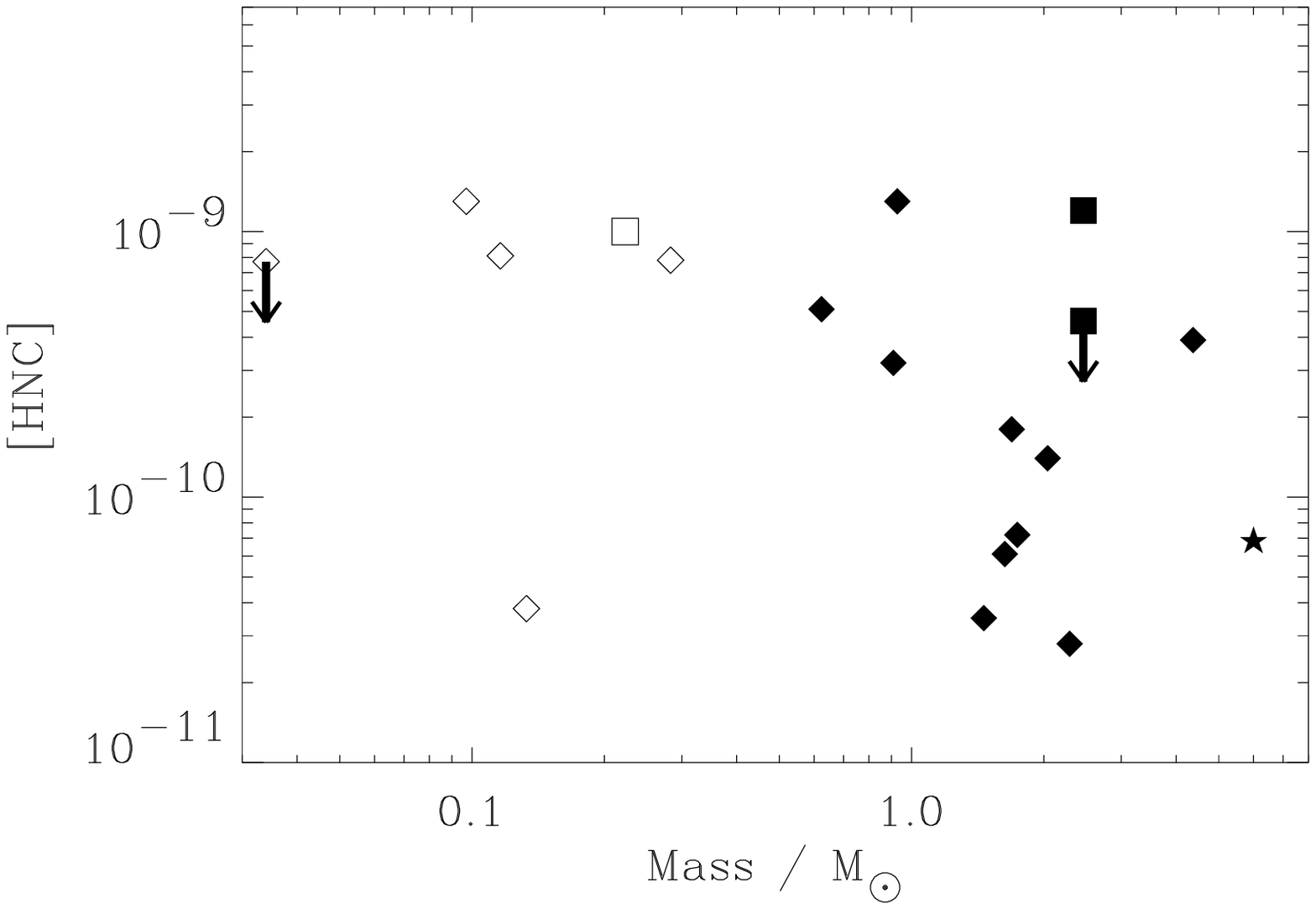}\includegraphics{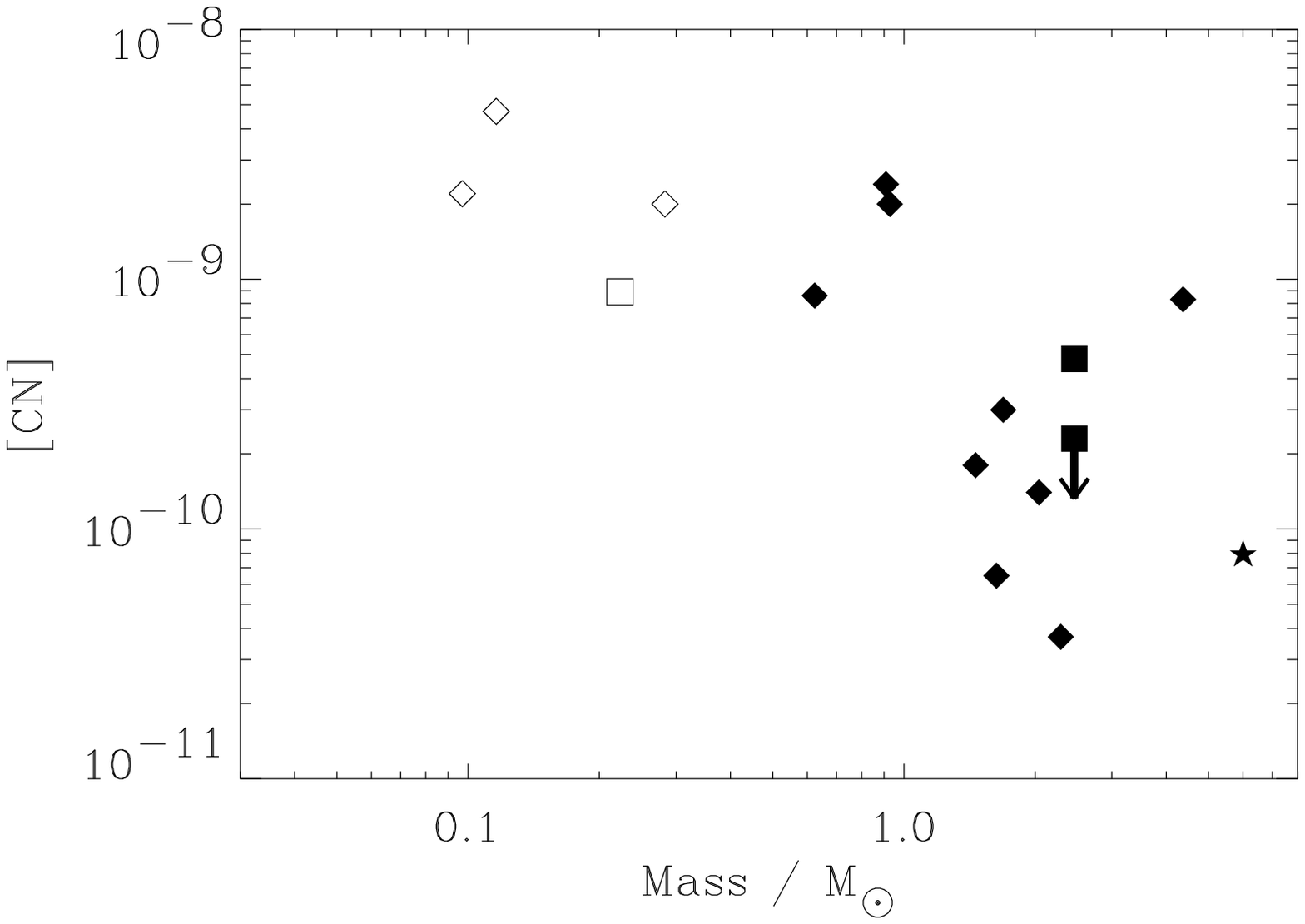}}
\caption{HCN abundances derived on the basis of main isotopic species
and H$^{13}$CN (upper panels, left and right) and CN and HNC
abundances (lower panels) vs. mass. As in previous figures, the class
0 objects are indicated by ``$\blacklozenge$'', the class I objects by
``$\lozenge$'' and the pre-stellar cores by ``$\blacksquare$'' with
the class 0 objects \object{VLA1623} and \object{IRAS~16293-2422}
singled out by ``$\square$'' and ``$\bigstar$'',
respectively.}\label{hcn_mass}
\end{figure*}

HCN is the molecule with the most striking lack of correlation with
mass or CO abundances, as can be seen in
Fig.~\ref{hcn_mass}. Chemically HCN and its geometrical isomer, HNC,
are naturally thought to be closely related and [HNC]/[HCN] ratios of
unity or slightly higher are typically observed toward molecular
clouds \citep{hirota98,dickens00}.

\begin{figure}
\resizebox{\hsize}{!}{\includegraphics{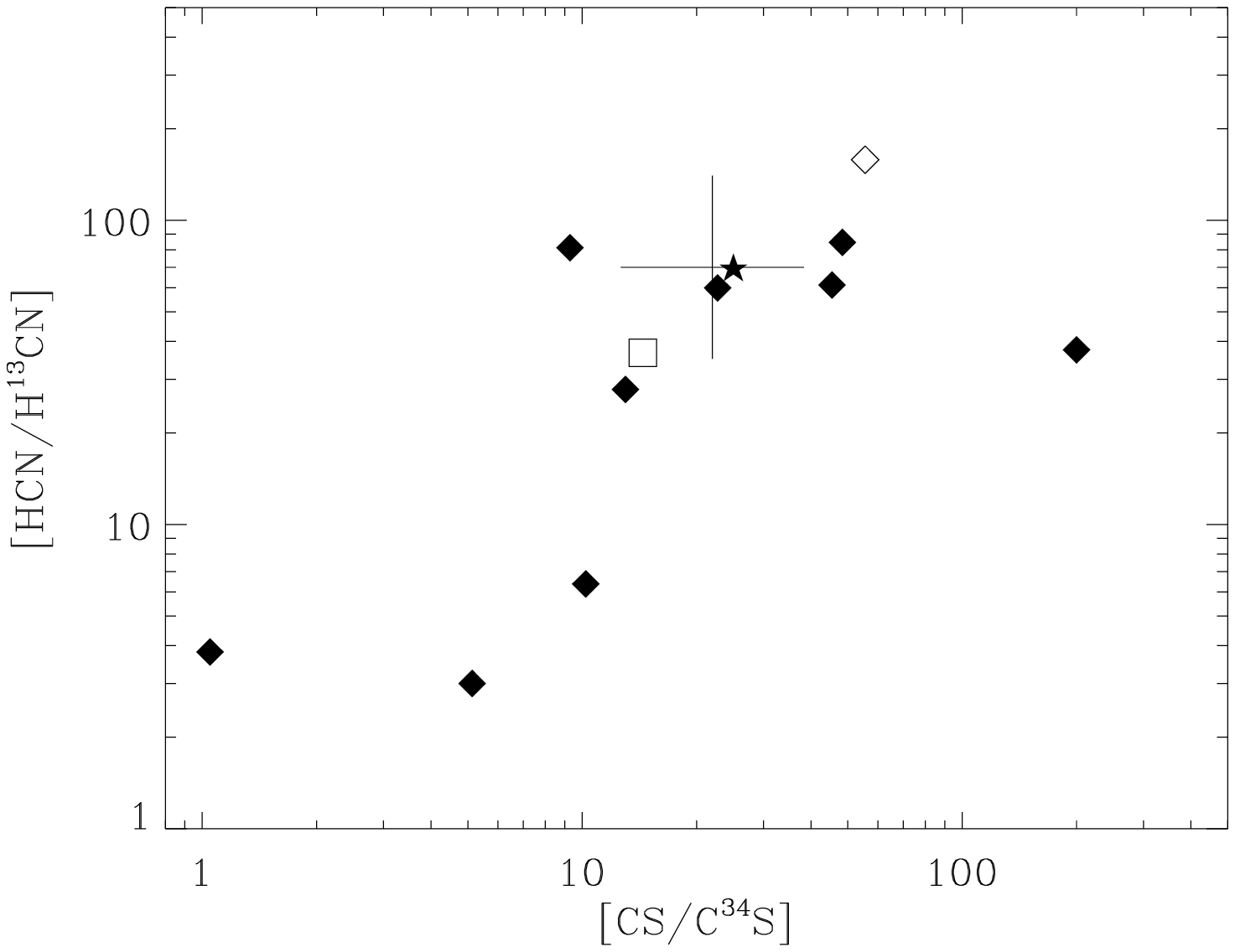}}
\resizebox{\hsize}{!}{\includegraphics{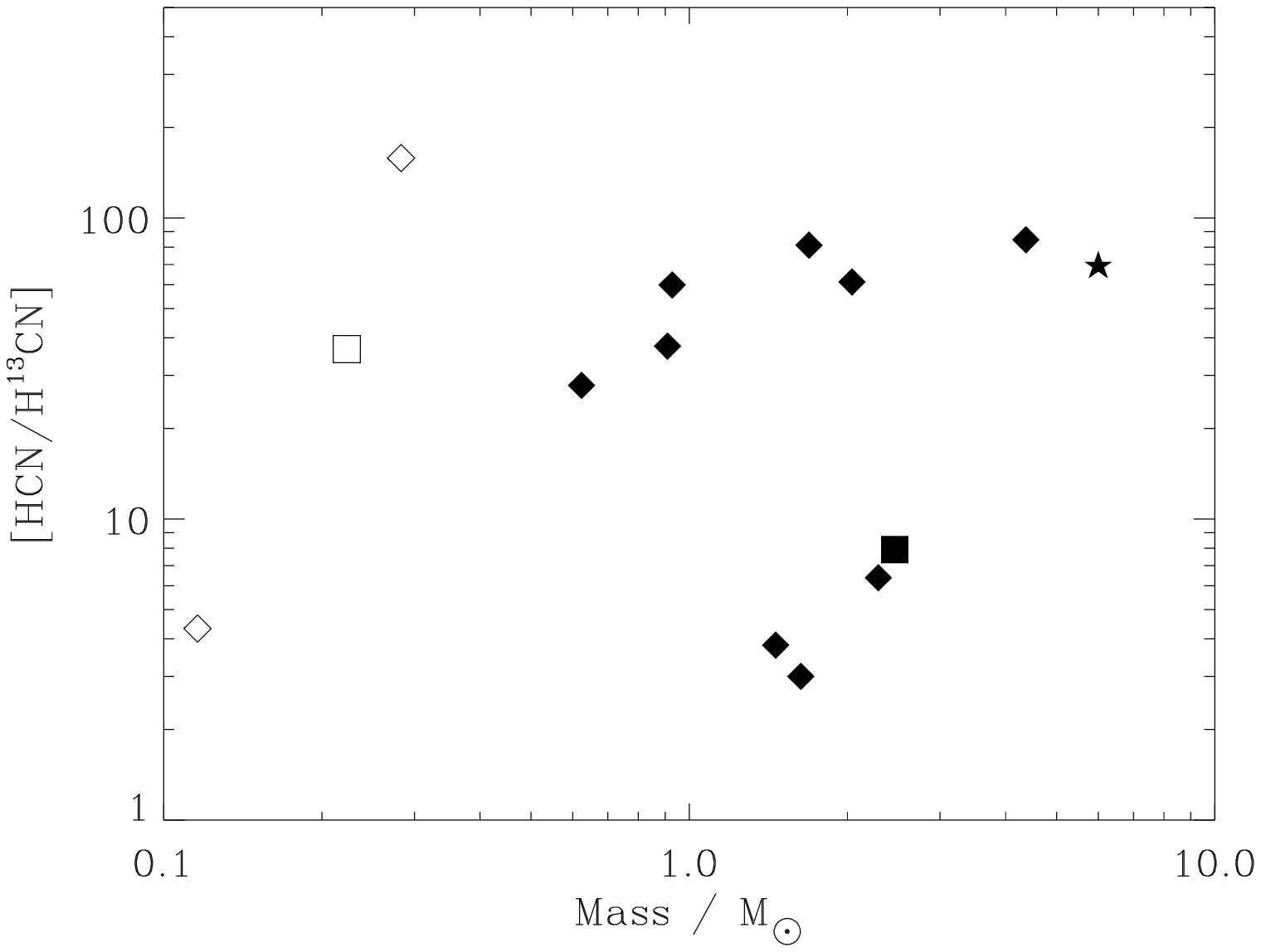}}
\resizebox{\hsize}{!}{\includegraphics{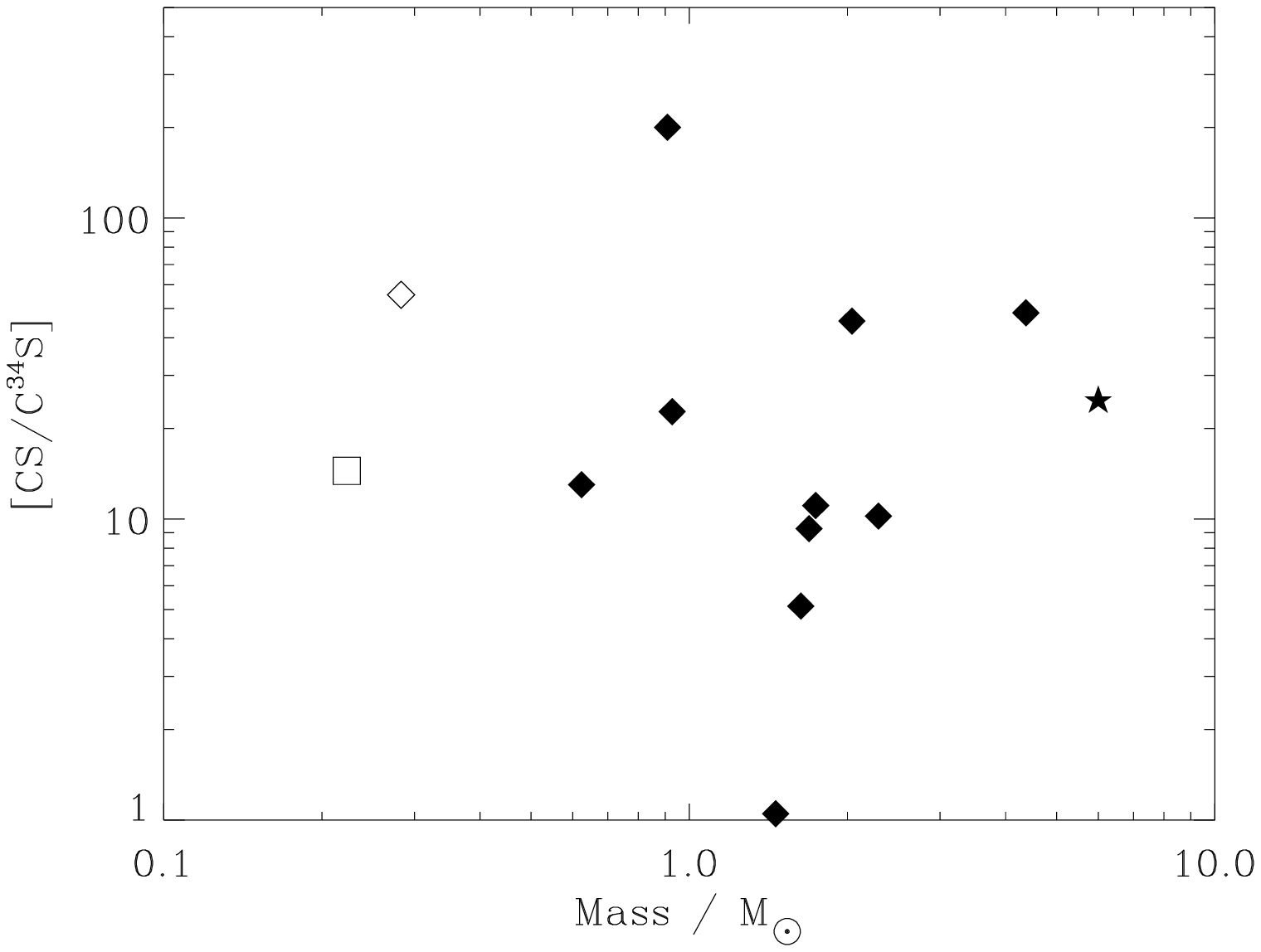}}
\caption{The ratio of the CS and C$^{34}$S abundances plotted vs.  HCN
and H$^{13}$CN ratio. The big cross mark the predictions from the standard
isotopic ratio of $^{12}$C:$^{13}$C of 70 and $^{32}$S:$^{34}$S of
22. Symbols as in Fig.~\ref{first_abundfig}.}\label{ratio_compare}
\end{figure}

For both HCN and HNC it is found that the 1--0 lines trace material
with higher abundances - or additional material outside what can be
described by the single power-law density models. As seen in
Fig.~\ref{ratio_compare} the [HCN/H$^{13}$CN] ratio is significantly
lower than 70 quoted by \cite{wilson94}. One exception is the case of
N1333-I4B which has a high estimated HCN abundance, possibly related
to confusion with the outflow. This ratio is, however, not correlated
with mass, as would be expected in case of an error in the opacity
treatment of the lines. The explanation is more likely that the
H$^{13}$CN abundances are heavily biased toward determinations based
on the low $J$ lines observed with the Onsala telescope since the
higher $J$ lines are only detected toward a small fraction of the
sources. Since the abundances derived on the basis of the isotopic
H$^{13}$CN thereby probe the outermost, less depleted regions this
should lower the estimated [HCN/H$^{13}$CN] ratios.

A higher degree of CO depletion could be expected to lead to a removal
of gas-phase carbon and oxygen and thereby a decline of the
[HNC]/[HCN] and [CN]/[HCN] ratios. On the other hand it is found that
neither the [CN]/[HCN] nor the [HNC]/[HCN] ratio correlate with the
degree of CO depletion. Another option is destruction of HNC at higher
temperatures through neutral-neutral reactions. This would be in
agreement with the result that the Orion molecular clouds have
significantly lower HNC abundances relative to HCN \citep{schilke92}
than the dark clouds surveyed by \cite{hirota98}.

\begin{figure}
\resizebox{\hsize}{!}{\includegraphics{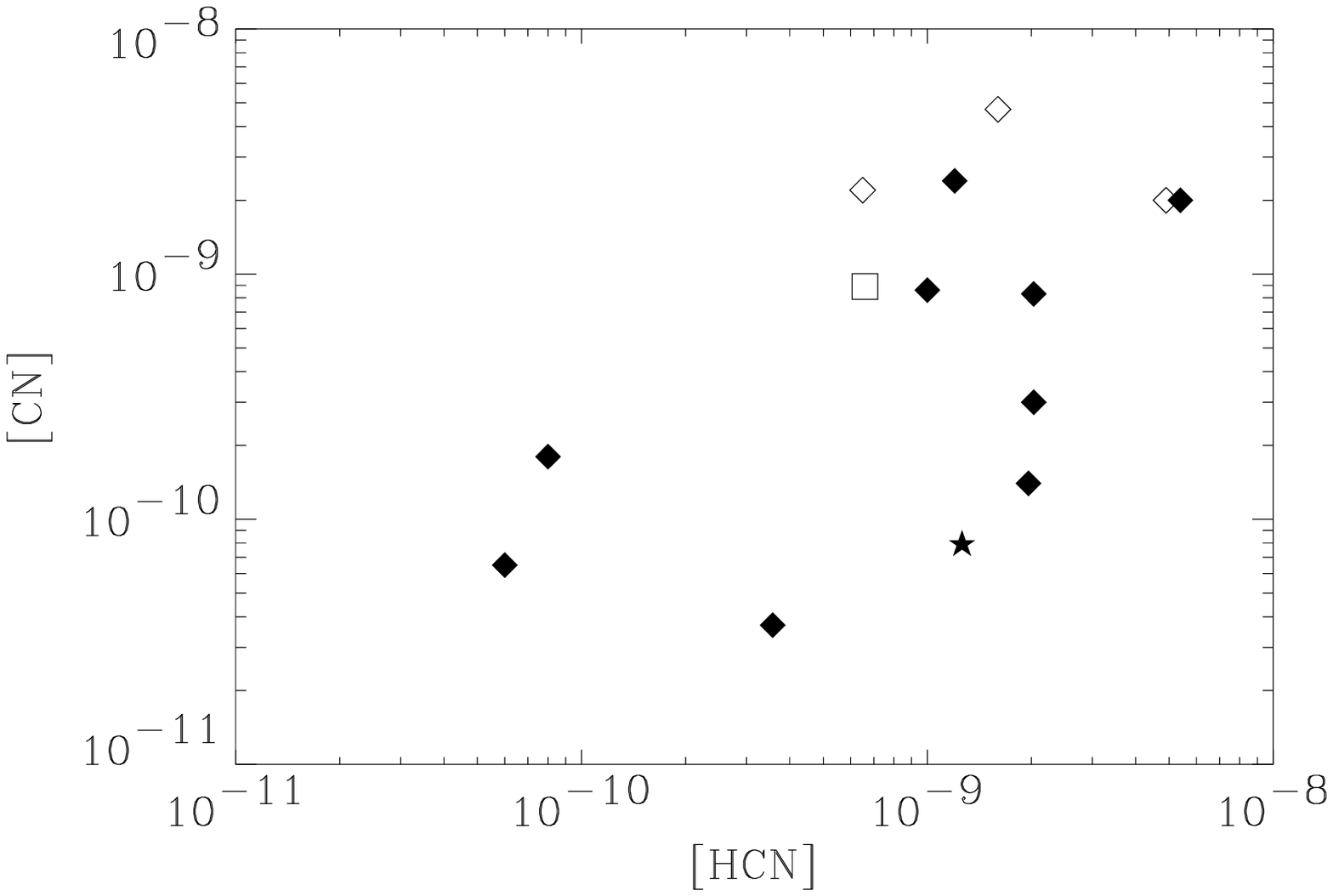}}
\resizebox{\hsize}{!}{\includegraphics{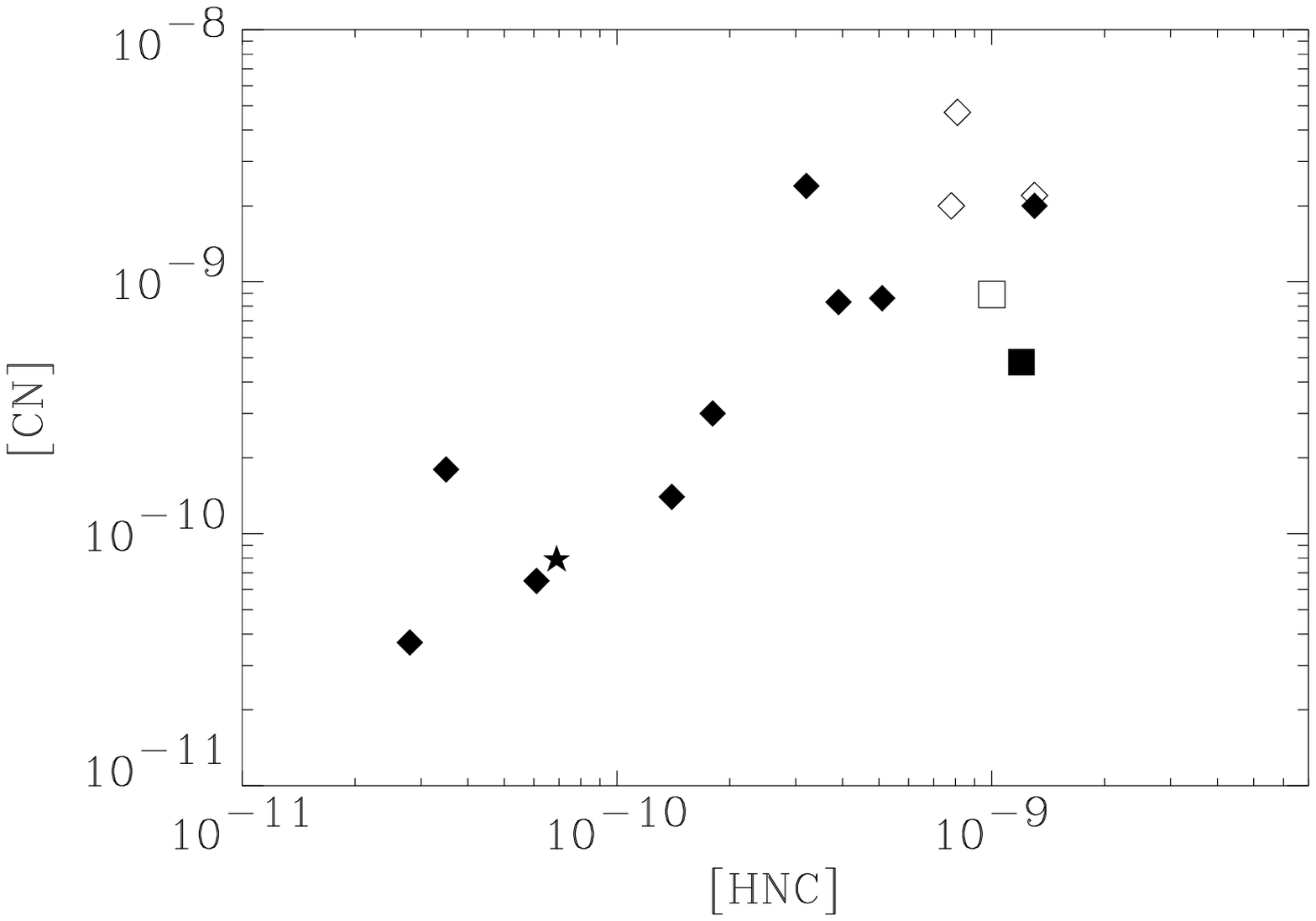}}
\resizebox{\hsize}{!}{\includegraphics{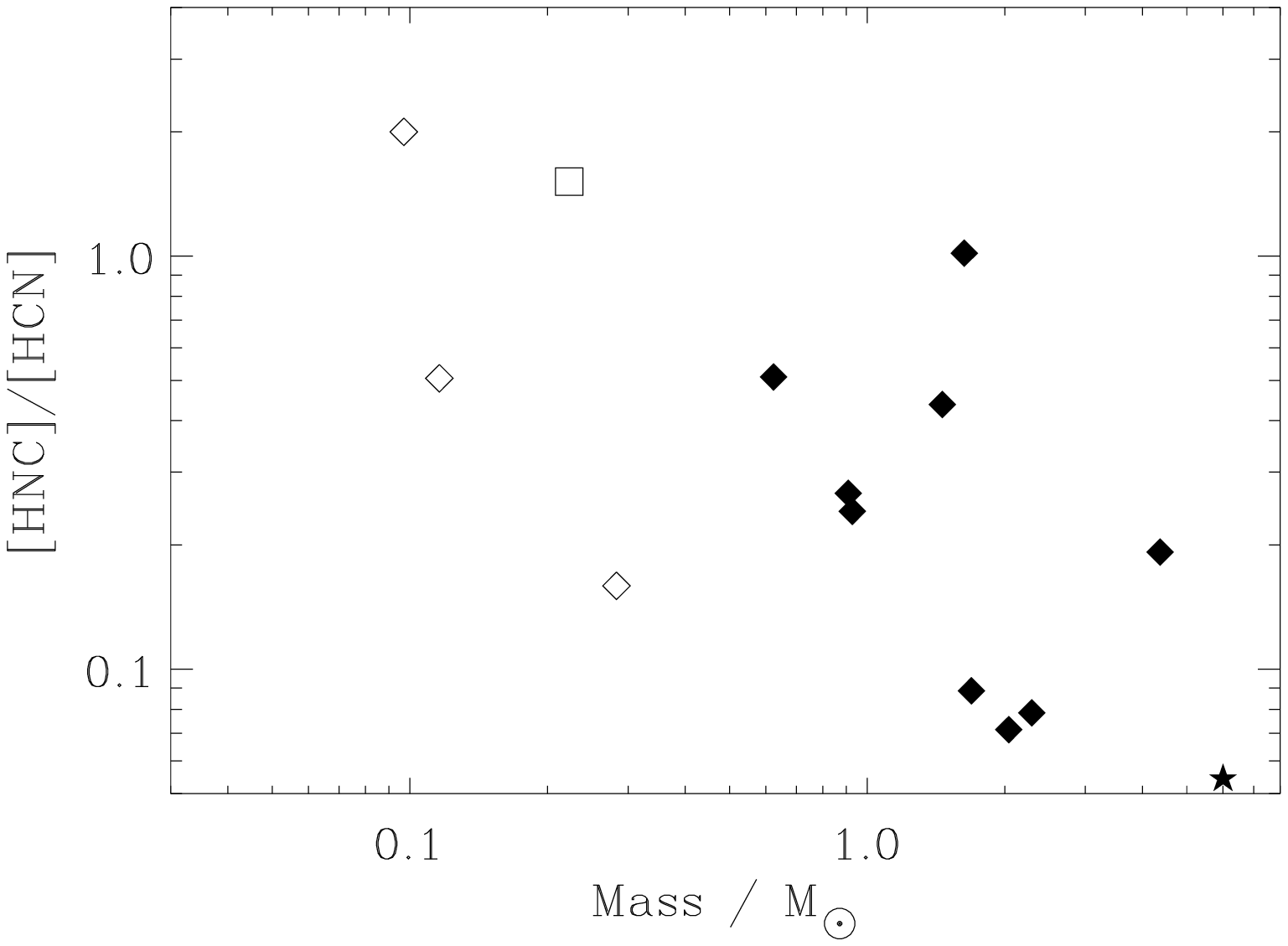}}
\caption{[CN] vs. [HCN] (upper panel) and vs. [HNC] (lower
panel). Symbols as in Fig.~\ref{first_abundfig}.}\label{hcn_hnc_cn}
\end{figure}

Fig.~\ref{hcn_hnc_cn} illustrates the close correlation between the
HNC and CN abundances also indicated by the correlation coefficients
(Table~\ref{pearsoncoeff} and Fig.~\ref{abundancerelations}). HNC and
CN are expected to be related, with HCNH$^+$ as an intermediate
product, through the reactions:
\begin{equation}
{\rm HNC} + {\rm H_3^+} \rightarrow {\rm HCNH^+}+{\rm H_2}
\end{equation}
\begin{equation}
{\rm HCNH^+} + {\rm e^-} \rightarrow {\rm CN}+{\rm H_2}
\end{equation}
These reactions are according to the UMIST database \citep{umist99}
the dominant formation and removal mechanisms for the three species at
20~K and 1$\times 10^{6}$~cm$^{-3}$. The main formation mehcanism for
HCN at this temperature and density is also through dissociative
recombination for HCNH$^+$ but this is secondary compared to the
formation of CN, which could explain the weaker correlation between
HCN and the other nitrogen-bearing species.

\subsection{HC$_3$N}\label{cssorat}
\begin{figure}
\resizebox{\hsize}{!}{\includegraphics{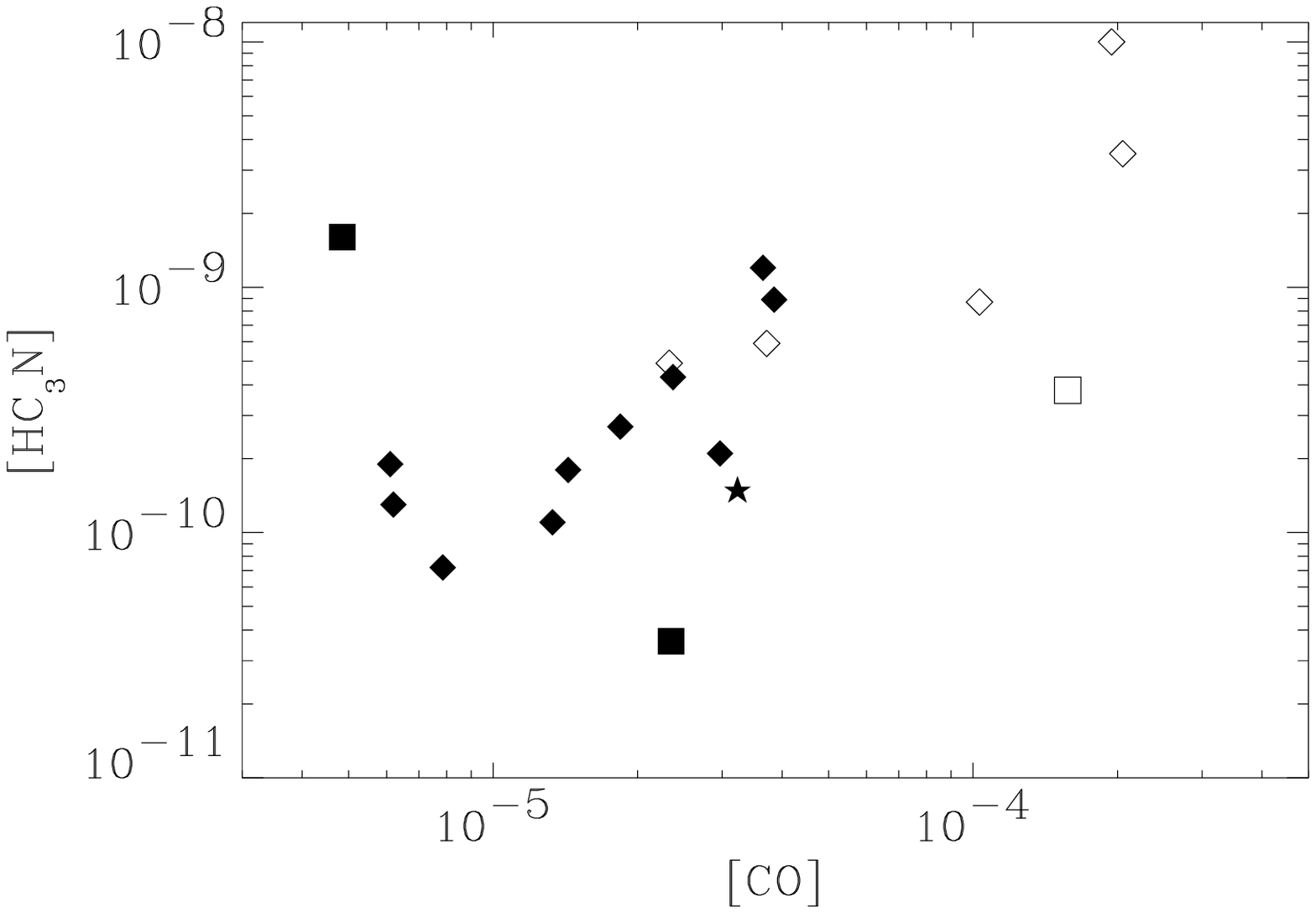}}
\resizebox{\hsize}{!}{\includegraphics{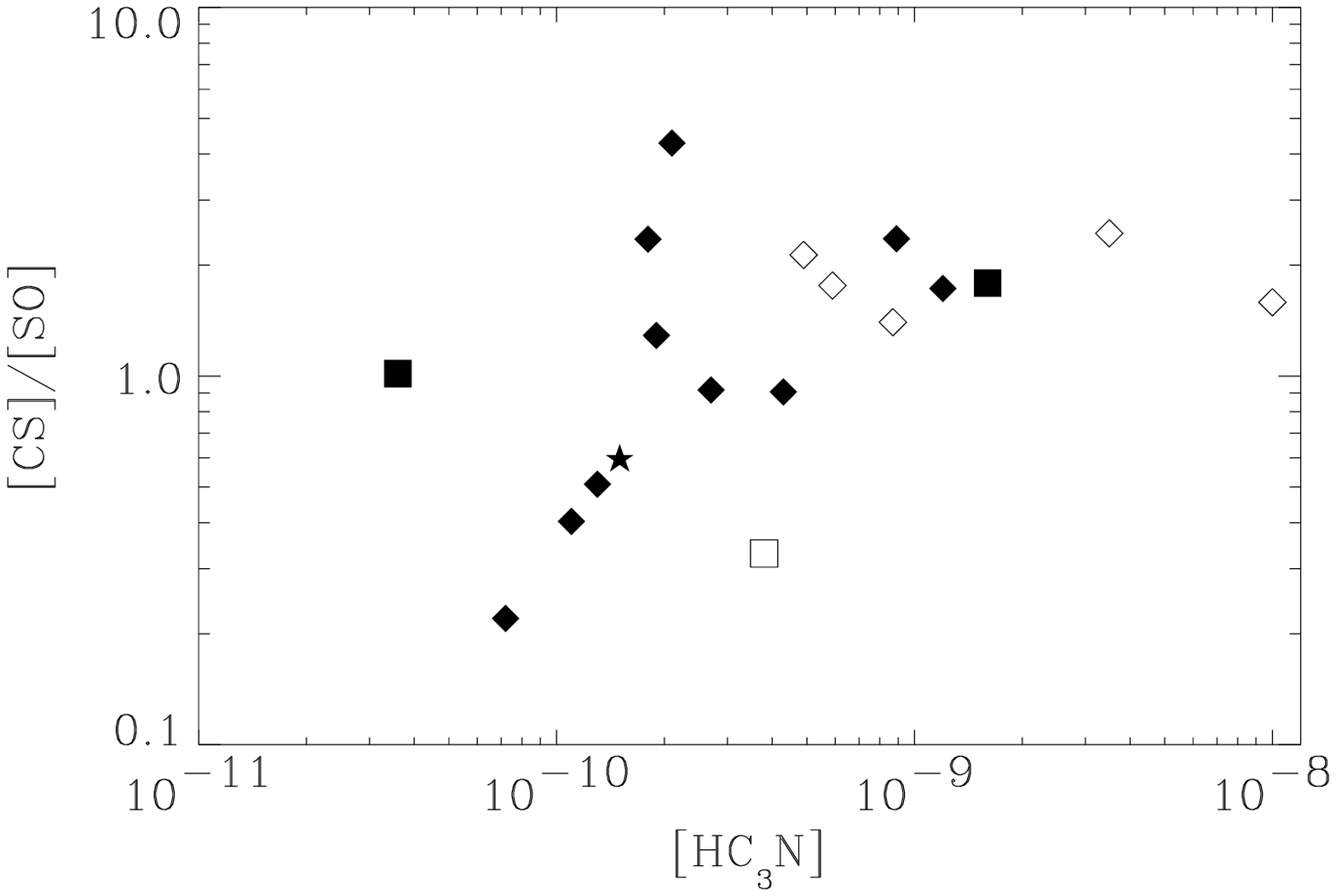}}
\caption{[HC$_3$N] vs. mass and [CO] (upper panel) and vs. [CS]/[SO]
ratio (lower panel). Symbols as in
Fig.~\ref{first_abundfig}.}\label{hctn}
\end{figure}

The HC$_3$N abundance has been suggested to be an indicator of the
temporal evolution or the degree of depletion
\citep[e.g.,][]{hirahara92,ruffle97,caselli98} in dark clouds and
pre-stellar cores. The HC$_3$N abundance peaks early in the evolution
of dark clouds when a substantial amount of carbon is in atomic form
in the gas-phase, but also increases with increasing depletion (i.e.,
potentially at ``later'' stages). Depletion tends to remove atomic
oxygen from the gas-phase, which otherwise has a tendency to destroy
ions necessary for the formation of species such as
HC$_3$N. Fig.~\ref{hctn} compares the HC$_3$N abundance with the CO
abundance and the CS/SO abundance ratio. As can be seen, [HC$_3$N] is
not particularly higher in objects with a larger degree of CO
depletion - except for the pre-stellar cores when these are considered
separately (see Sect.~\ref{prestellar}).

For the protostars in our sample, however, Fig.~\ref{hctn} shows that
the HC$_3$N abundance is related to the [CS]/[SO] ratio - with lower
ratios of the two sulfur-bearing molecules corresponding to lower
HC$_3$N abundances. This can be understood in a scenario where the
HC$_3$N abundance is indeed a tracer of atomic carbon, since the CS/SO
ratio would likewise be increased by higher amounts of atomic carbon,
as suggested by the models of \cite{bergin97b}. The question is then
whether this should be taken as an indicator of chemical ``youth''. As
can be seen in Fig.~\ref{hctn}, the dynamically ``older'' class I
objects have higher HC$_3$N abundances and [CS]/[SO] ratios, which
apparently would contradict this suggestion.

\begin{figure}
\resizebox{\hsize}{!}{\includegraphics{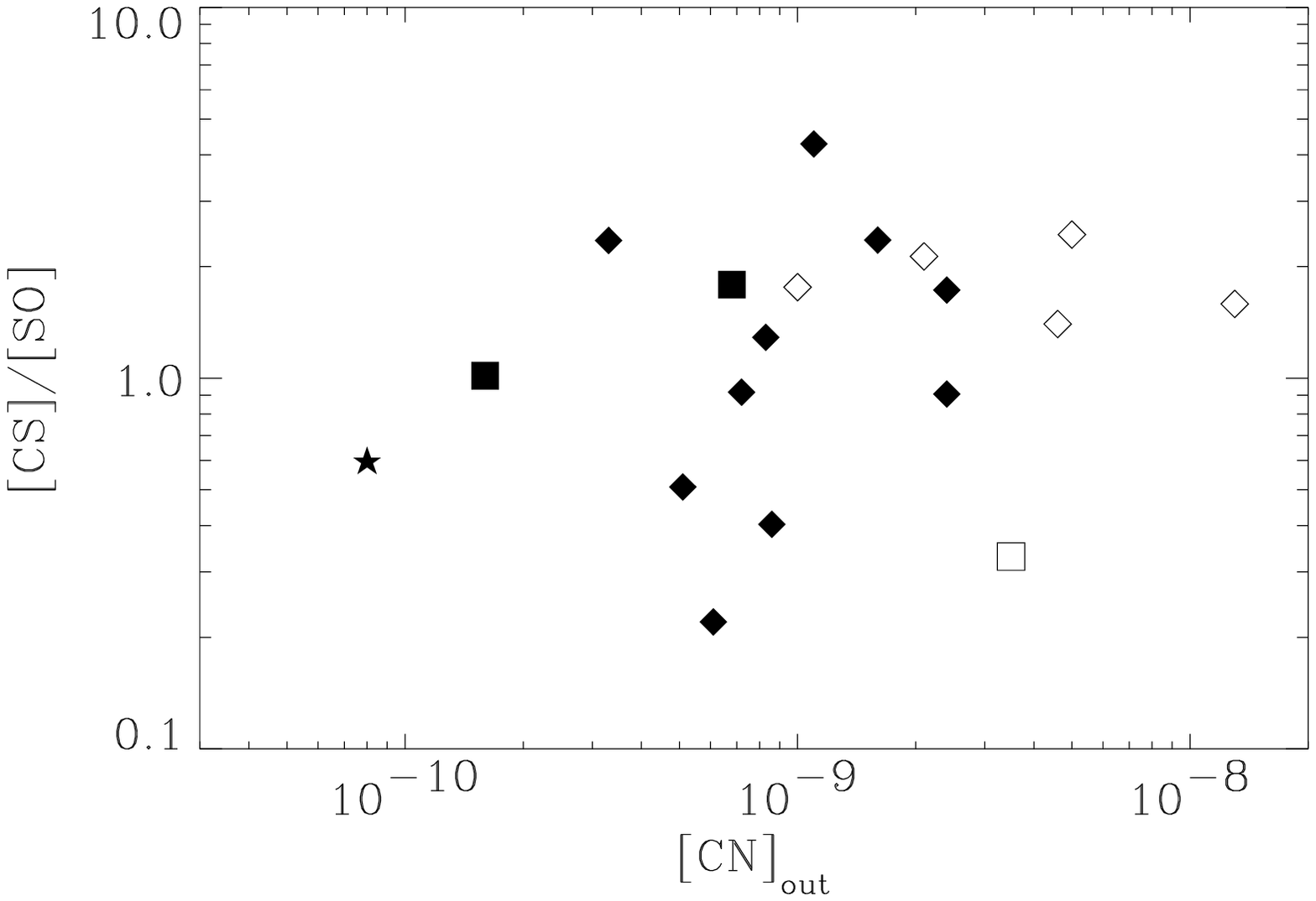}}
\resizebox{\hsize}{!}{\includegraphics{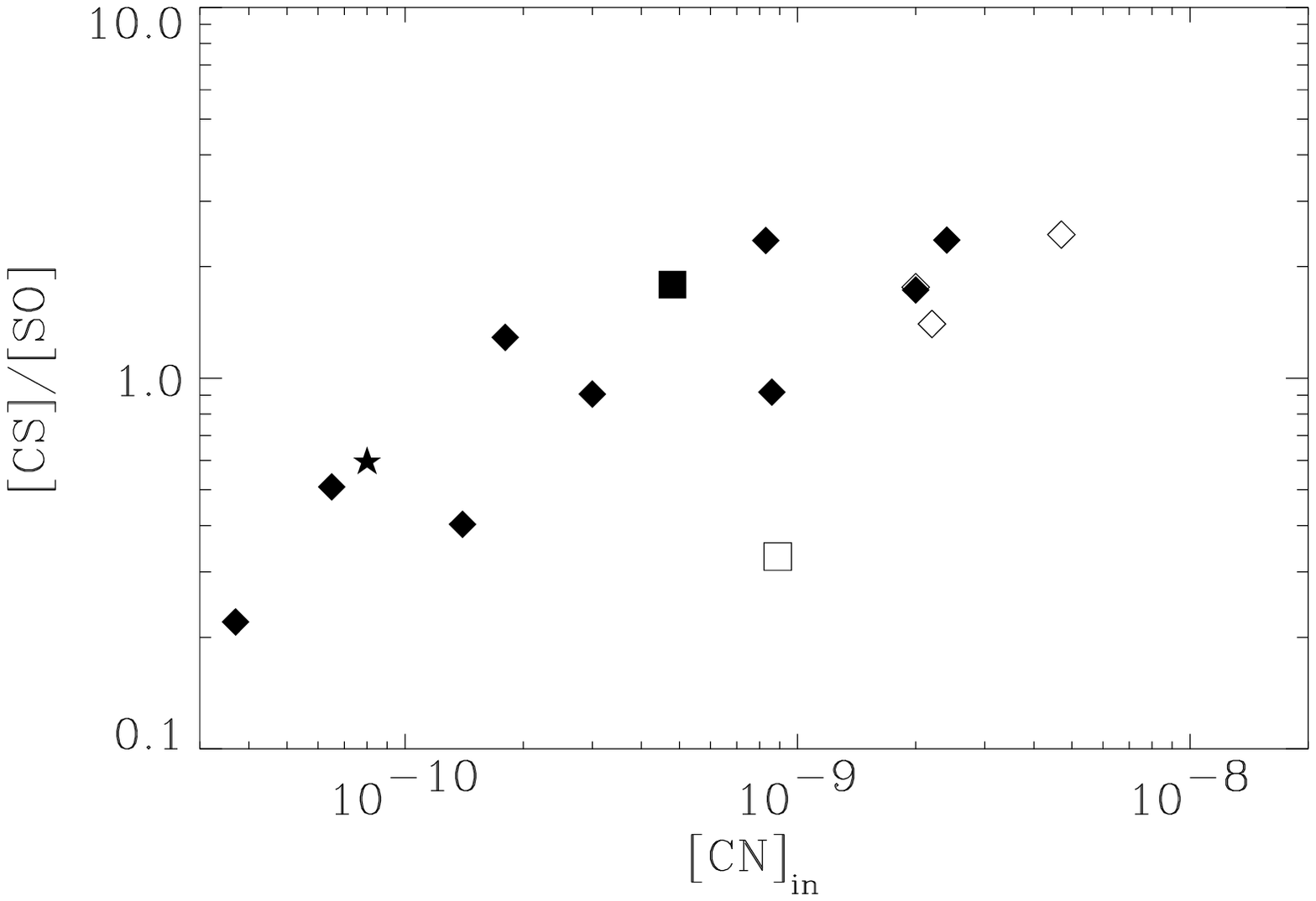}}
\caption{CS/SO abundance ratio vs. abundance of CN constrained by the
1--0 lines (upper) and 3--2 lines (lower) probing the outer and inner
regions of the envelope, respectively. Symbols as in
Fig.~\ref{first_abundfig}.}\label{cn_csso}
\end{figure}

An alternative explanation could be that the amount of atomic carbon
is enhanced by the impact of UV radiation from the outside due to the
interstellar radiation field. We can, however, argue that this is not
the case from the CN line observations. As noted above the HC$_3$N and
CN abundances are found to be interlinked, which is not difficult to
understand if one considers the dominant formation and destruction
mechanisms for HC$_3$N in gas where the degree of CO depletion is low:
\[{\rm CN} + {\rm C_2 H_2} \rightarrow {\rm HC_3N} + {\rm H}\]
\[{\rm C^+} + {\rm HC_3N}  \rightarrow {\rm C_3H^+} + {\rm CN}\]
If the trend seen between the HC$_3$N and the CS/SO ratio is related
to the radiation field, the correlation should be more clear for
abundances constrained by the CN 1--0 lines which probe the outermost
region of the envelope. As illustrated in Fig.~\ref{cn_csso}, however,
the trend is much stronger for the CN abundances constrained by the CN
3--2 lines that have a significantly higher critical density and thus
probe the inner, more shielded region of the envelope. UV radiation
from the central star could increase the amount of atomic carbon in
the inner envelope and would lead to trends similar to that observed,
but whether UV radiation can penetrate far enough out into the
envelope to be important is still not clear. A more detailed study of
the molecular species in the chemical network for the nitrogen-bearing
species together with more detailed modeling including the radial
variation of the molecular abundances is needed to fully address these
questions.

\subsection{Deuterium fractionation}\label{deuterium}
\begin{figure}
\resizebox{\hsize}{!}{\includegraphics{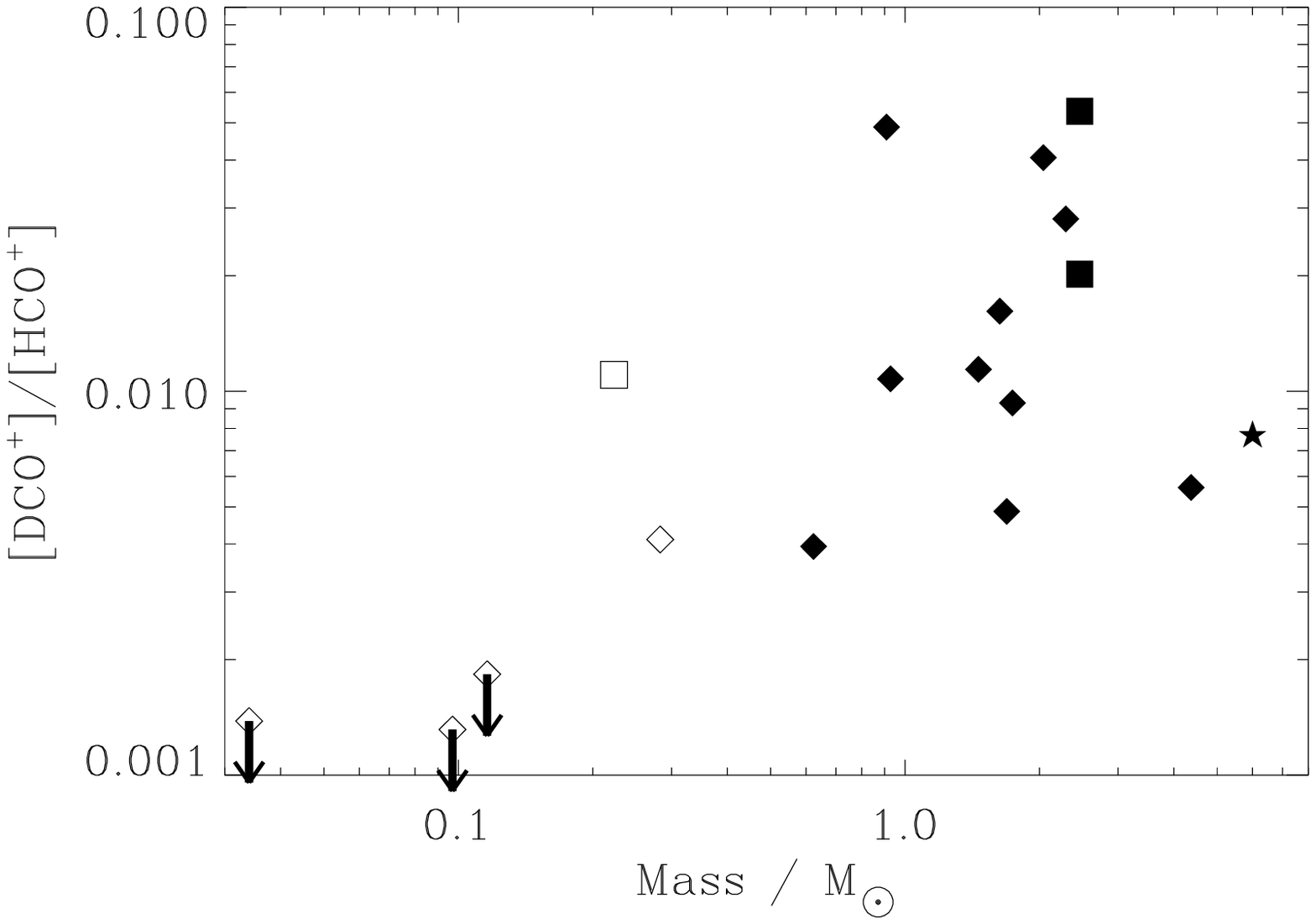}}
\resizebox{\hsize}{!}{\includegraphics{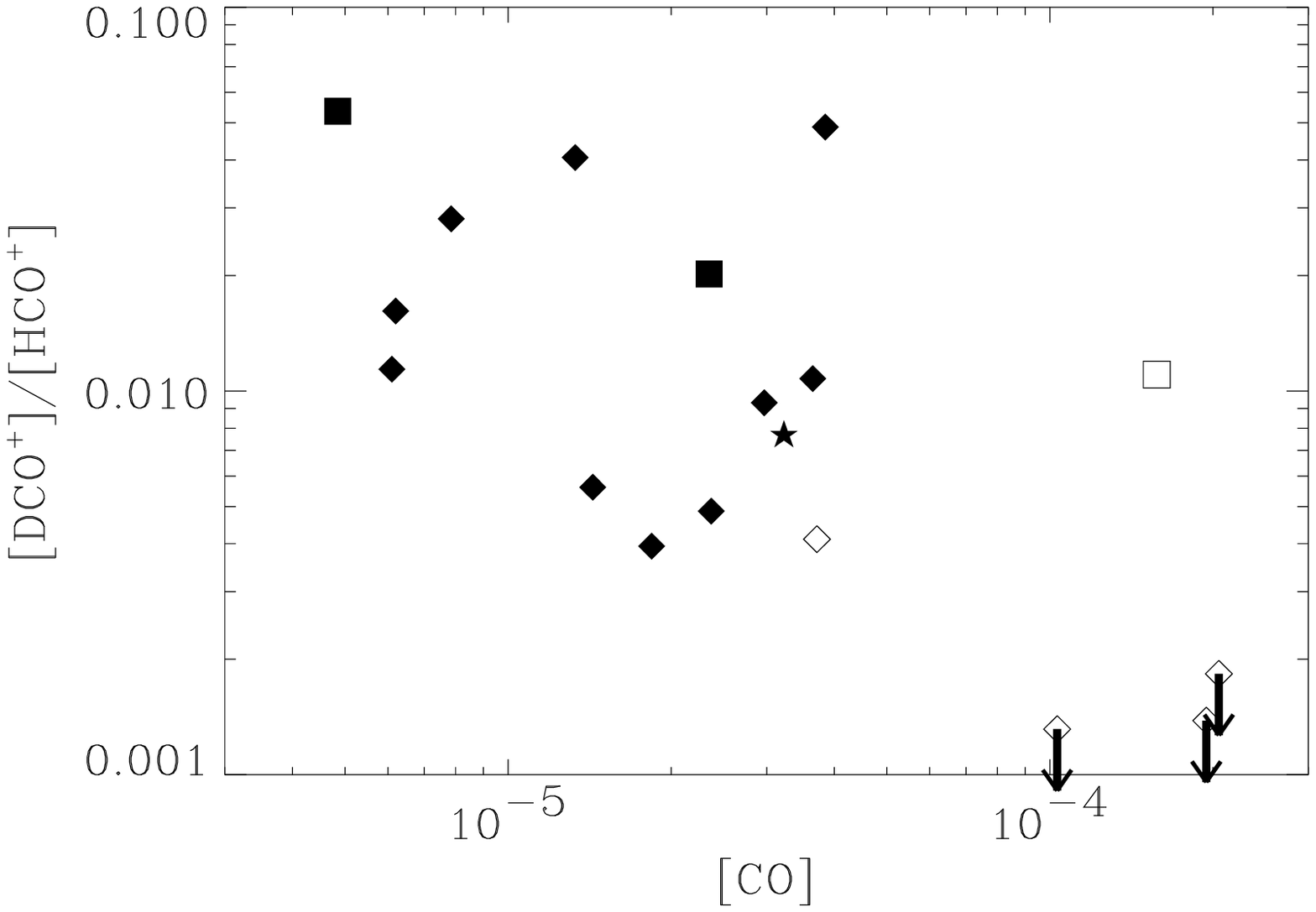}}
\caption{[DCO$^+$]/[HCO$^+$] ratio vs. mass (upper panel) vs. and [CO]
(lower panel). Symbols as in Fig.~\ref{first_abundfig}.}\label{dcoP_mass}
\end{figure}

The deuterium fractionation of HCO$^+$ is seen from plots of the
[DCO$^+$]/[HCO$^+$] ratio in Fig.~\ref{dcoP_mass}. The fact that the
DCO$^+$ emission predominantly originates in the cold outer gas is
evidenced by the narrow line widths for all sources: the turbulent
broadening required to model the DCO$^+$ line widths is only
0.3-0.5~km~s$^{-1}$. The prestellar cores clearly show the highest
[DCO$^+$]/[HCO$^+$] ratio of $\sim$5\% in agreement with findings by,
e.g., \cite{caselli02b}. The class 0 sources show [DCO$^+$]/[HCO$^+$]
ratios ranging from 0.004 to 0.05. DCO$^+$ is not detected for the
class I sources corresponding to the upper limits on the
[DCO$^+$]/[HCO$^+$] ratio of $\sim 0.001$. As can be seen from the
lower panel of Fig.~\ref{dcoP_mass}, the [DCO$^+$]/[HCO$^+$] ratio
does seem to be correlated with the degree of CO depletion.

The typical deuterium abundance ratios are in general significantly
higher than the ``cosmic'' D/H ratio of $10^{-5}$. Both gas-phase
reactions and grain-surface reactions have been invoked to describe
the deuterium fractionation at low temperatures in pre- and
protostellar environments. In gas-phase models by \cite{roberts00a}
such a trend is indeed expected. In pure gas-phase models the primary
mechanism for driving the fractionation of HCO$^+$ is the small
zero-point energy in the reaction:
\begin{equation}
{\rm H}_3^+ + {\rm HD} \leftrightarrows {\rm H}_2{\rm D}^++{\rm H}_2 \label{dcoP_formation}
\end{equation}
that predominantly drives the D into ${\rm H}_2{\rm D}^+$ relative to
H$_3^+$, and which subsequently reacts with CO to form
DCO$^+$. Depletion of CO causes (\ref{dcoP_formation}) to be the
dominant mechanism for removal of H$_3^+$ and since DCO$^+$ is
produced subsequently, the formation of HCO$^+$
(eq.~(\ref{hcoP_formation}) in Sect.~\ref{hcop}) will be less
productive compared to the deuterated versions. Models by
\cite{roberts00b} and \cite{roberts03} show that a high degree of
depletion may be required in order to produce formation of the doubly
and triply deuterated species observed in protostellar environments
\citep{ceccarelli98,lis02,parise02,vandertak02}. Other studies
indicate an increase of the deuteration with CO depletion, e.g., the
survey of the deuteration of D$_2$CO for a sample of pre-stellar cores
by \cite{bacmann03}.

\begin{figure}
\resizebox{\hsize}{!}{\includegraphics{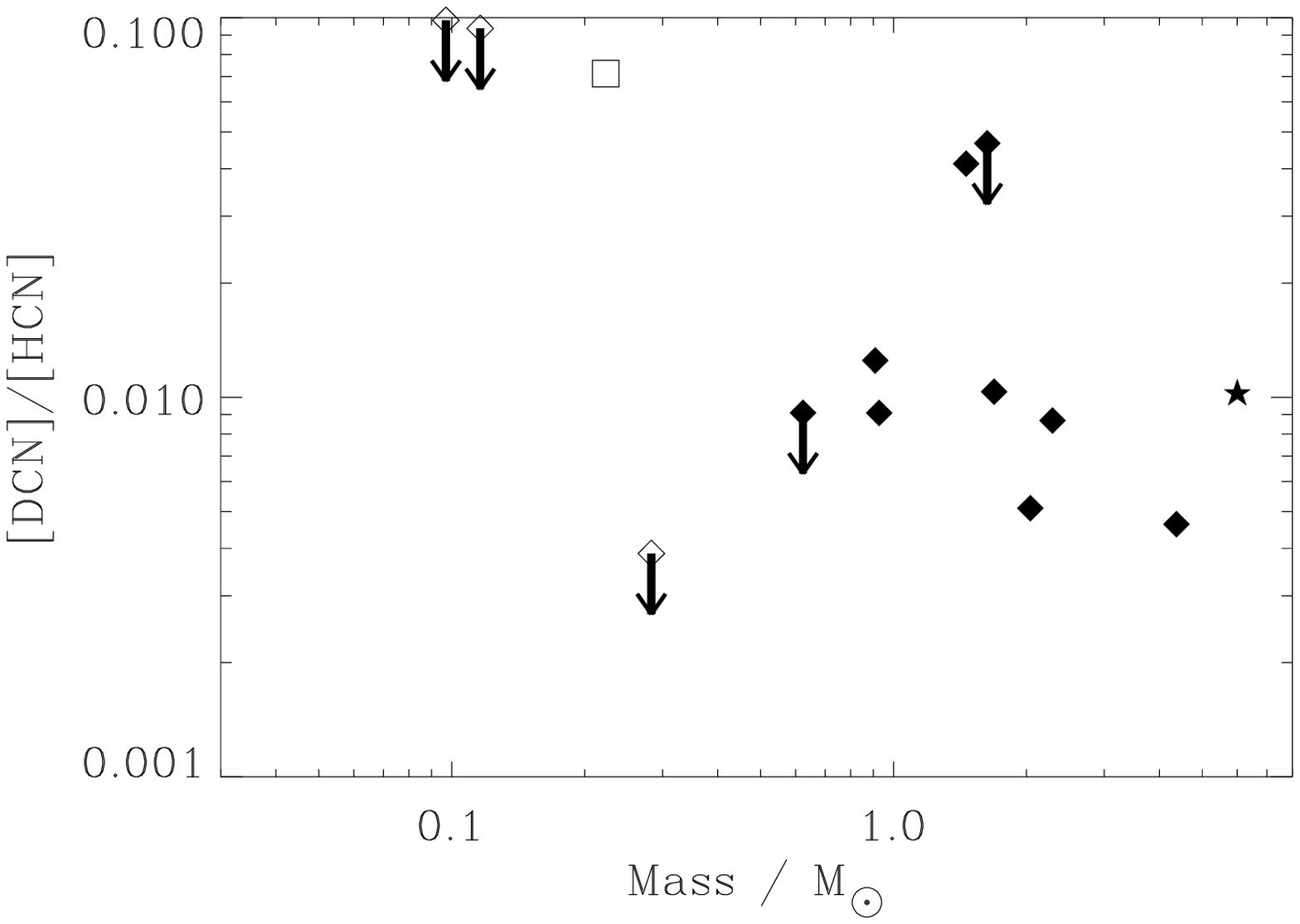}}
\resizebox{\hsize}{!}{\includegraphics{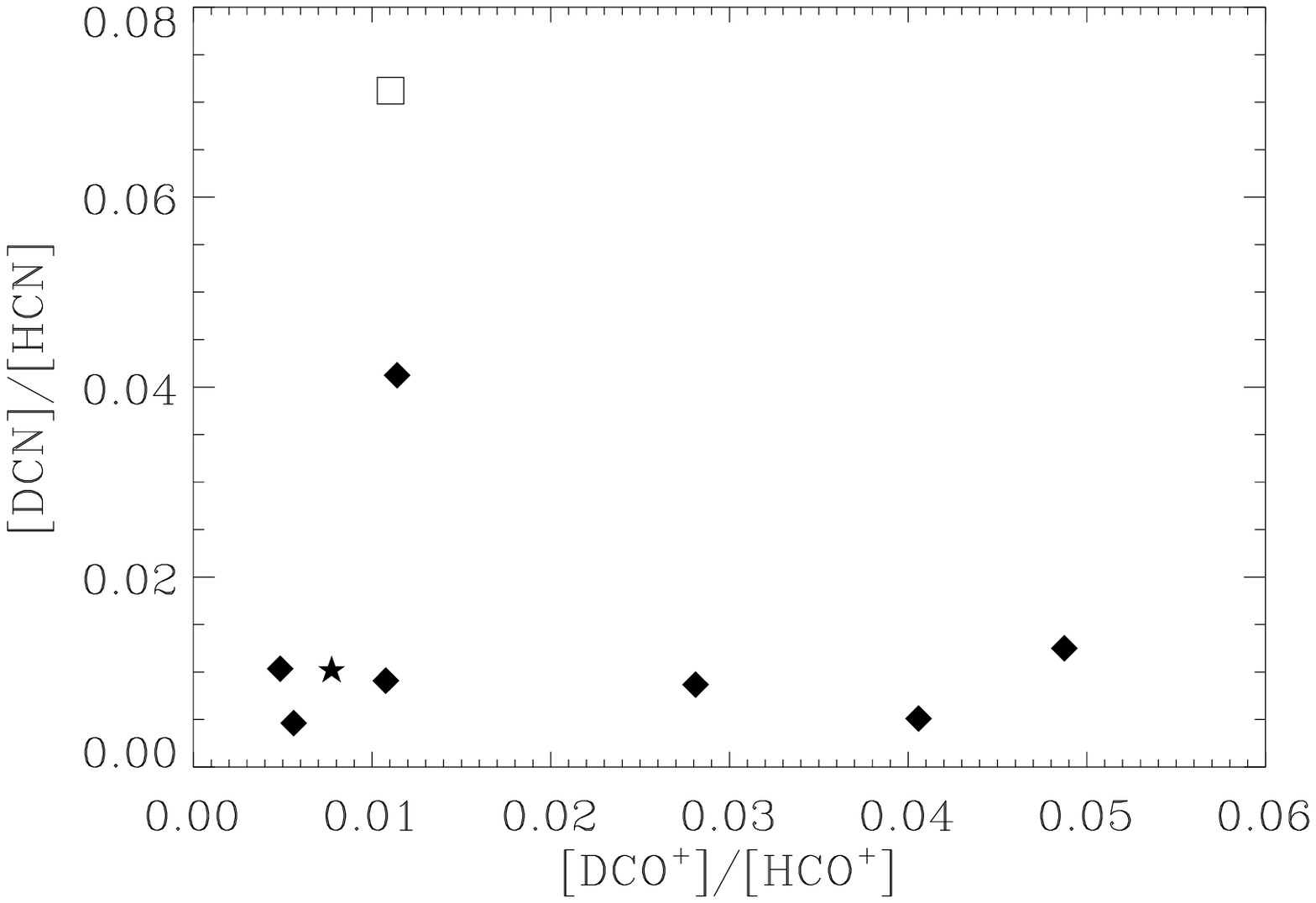}}
\caption{[DCN]/[HCN] abundance vs. mass (upper panel) and
[DCO$^+$]/[HCO$^+$] ratio (lower panel). Symbols as in
Fig.~\ref{first_abundfig}.}\label{deutfrac}\label{last_abundfig}
\end{figure}

Figure~\ref{deutfrac} compares the [DCN]/[HCN] ratio with envelope
mass and [DCO$^+$]/[HCO$^+$] ratio. It appears that the [DCN]/[HCN]
and [DCO$^+$]/[HCO$^+$] ratios are not correlated. In fact the absence
of DCN in the pre-stellar cores is striking considering their strong
DCO$^+$ emission. This could imply that the [DCN]/[HCN] ratio is
higher for the warmer envelopes. This is in agreement with gas-phase
deuteration of HCN, which may occur at slightly higher temperatures
than that of HCO$^+$: in particular deuteration through ${\rm
CH_3^+}+{\rm HD} \rightarrow {\rm CH_2D^+}$ may be more important for
temperatures higher than $\approx 30$~K
\citep[e.g.,][]{turner01}. Alternatively, the [DCN]/[HCN] ratio may
have been established earlier in the protostellar evolution, frozen
out onto the dust grains and released back at higher temperatures than
is the case for the [DCO$^+$]/[HCO$^+$] ratio.

\subsection{The pre-stellar cores}\label{prestellar}
The two pre-stellar cores in our sample, \object{L1689B} and \object{L1544}, were also
studied by \cite{lee03} together with an additional core, L1512.
\citeauthor{lee03} found a high degree of depletion of CO and HCO$^+$
for \object{L1544} but close to ``standard'' abundances for these molecules in
\object{L1689B}. This is consistent with the results in this
paper. \citeauthor{lee03} also observed N$_2$H$^+$ 1--0 and found it
to be weak in \object{L1689B} compared to \object{L1544}. This is in good agreement with
our results which show N$_2$H$^+$ to be an order of magnitude more
abundant in \object{L1544} than in \object{L1689B}, strengthening the N$_2$H$^+$ and CO
anti-correlation discussed in Sect.~\ref{nthp}. Of the other molecules
in this paper, the sulfur-bearing species (CS and SO) are also found
to have lower abundances in \object{L1544} indicating a higher degree of
overall depletion. The nitrogen-bearing species in contrast show an
opposite trend with high abundances in \object{L1544}. In particular, HC$_3$N
is close to a factor of 50 higher in \object{L1544} than in \object{L1689B}, supporting
the suggestion that HC$_3$N traces the degree of depletion in the
pre-stellar stages \cite[e.g.][]{ruffle97,caselli98}. Also the higher
degree of HCO$^+$ deuteration in \object{L1544} than in \object{L1689B} is consistent
with the higher degree of depletion \object{L1544} than in \object{L1689B} (ref. the
discussion in Sect.~\ref{deuterium}).

\subsection{Comparison to other star-forming regions}
In Fig.~\ref{abundhist} the average abundances for the class 0 and I
objects and pre-stellar cores are compared to the abundances of other
star-forming regions, the outer envelope around the class 0 object
\object{IRAS~16293-2422} \citep{schoeier02}, the average abundance for
the 3 high-mass YSOs W3(IRS4), W3(IRS5) and W3(H2O) \citep{helmich97}
and the ``C'' position of the dark cloud L134N \citep{dickens00}.

As mentioned in the introduction the class 0 object
\object{IRAS~16293-2422} is the most studied low-mass protostar in
terms of the chemistry of its protostellar envelope \cite[see,
e.g.,][]{blake94,vandishoeck95,ceccarelli98,ceccarelli00a,ceccarelli00b,schoeier02,
parise02,cazaux03}, because of its rich spectrum and its warm inner
region where ices have evaporated. This naturally raises the question
whether \object{IRAS~16293-2422} is indeed a typical class 0 object:
is the richness of its spectrum simply caused by it being the closest
object with the most massive envelope, or is it caused by other
effects, such as the interaction of its outflow with the nearby
envelope? As it can be seen from Table~\ref{abundoverview} and
Fig.~\ref{first_abundfig}-\ref{last_abundfig} \object{IRAS~16293-2422}
has a fairly standard set of outer envelope abundances for CO, CS and HCN. On the
other hand it shows lower abundances (factors 4-20) of
especially HNC, CN and N$_2$H$^+$ and high abundances of SO and SO$_2$
compared to the typically upper limit found for the objects in this
study. It does not, however, stand markedly out considering the
scatter in abundances within the larger group of class 0 objects. It
therefore seems that \object{IRAS~16293-2422}, despite possibly being
affected by outflows on smaller scales \citep[e.g.,][]{hotcorepaper}
and having a ``hot inner region''
\citep[e.g.,][]{ceccarelli98,schoeier02} has a cold outer envelope that is
similar to that of the other class 0 objects in terms of the overall
abundances. Otherwise the most striking feature of
Fig.~\ref{abundhist} is the significantly higher HNC abundance (two
orders of magnitude) in L134N compared to the other sources and
molecules. The high- and the low-mass YSOs differ slightly with SO and
HCN abundances higher by up to a factor 5 and HC$_3$N lower by a
factor 5-10 in the high-mass YSOs.

\section{Conclusion}
The molecular inventories for the envelopes around a sample of
low-mass protostars have been established. Using models for the one
dimensional physical structure of the envelopes from
\cite{jorgensen02} (Paper~I), the abundances of a range of molecular
species are constrained through Monte Carlo line radiative transfer
modeling of single-dish submillimeter and millimeter observations. The
main conclusions are:

\begin{enumerate}
\item For most sources and molecules, the high excitation 0.6-1.4~mm
  lines are well represented by a constant fractional abundance. The
  low excitation 3~mm lines of HCO$^+$ and the nitrogen-bearing
  species, however, are signifcantly underestimated by such models,
  similar to the trend seen for CO in Paper~I. Varying freeze-out
  timescales in the regions of the envelopes corresponding to
  different densities can explain this. All lines for these species
  can be accounted for with a ``drop'' profile where the envelopes
  have ``standard'' molecular abundances in the outermost low density
  region and in the innermost high temperature region, but drops in an
  intermediate zone where the molecule can freeze-out.
\item Effects of the envelope velocity field are tested and confirmed
to be insignificant when determining the abundances from optically
thin species. An upper limit to the magnitude of the velocity field
and thereby the mass accretion rate can be inferred from comparison to
the widths of those lines.
\item An empirical chemical network is constructed through
calculations of correlation coefficients for pairs of
abundances. Strong correlations are found between, e.g., HCO$^+$ and
CO, CS and SO and between the nitrogen-bearing species, HNC, CN and
HC$_3$N.
\item A linear relationship between the abundances of CO and HCO$^+$
is seen, whereas the N$_2$H$^+$ abundances appear lower in the objects
with high CO abundances. This can be understood from analytic
considerations of the chemical network of HCO$^+$ and N$_2$H$^+$,
where both are formed through reactions involving H$_3^+$. A linear
relationship between [CO] and [HCO$^+$] exist in the limit where N$_2$
rather than CO removes H$_3^+$ due to CO freeze-out. Likewise the
N$_2$H$^+$ abundances rapidly decline as soon as the CO abundances
increase, because H$_3^+$ and N$_2$H$^+$ are destroyed through
reactions with CO.
\item The CS/SO abundance ratio is found to be correlated with the
abundances of CN and HC$_3$N, which may reflect the amount of atomic
carbon in the gas-phase.
\item High levels of deuterium fractionation of HCO$^+$
([DCO$^+$]/[HCO$^+$]~=~1-5~\%) are found in the pre-stellar cores and
in the most massive/coldest envelopes around the class 0 objects. The
[DCN]/[HCN] ratio is not correlated with the [DCO$^+$]/[HCO$^+$]
ratio, possibly indicating differences in the gas-phase deuteration of
HCO$^+$ and HCN.
\item Among the two pre-stellar cores in our sample, \object{L1544} shows a
higher degree of depletion of CO and the sulfur-bearing species,
resulting in a chemistry with higher abundances of N$_2$H$^+$, HC$_3$N
and a higher degree of HCO$^+$ deuteration than what is seen in
\object{L1689B}.
\item The outer abundances of the previously studied class 0 object
\object{IRAS~16293-2422} \citep[e.g.,][]{schoeier02} are found to be
in agreement with the average abundances in the studied sample of
class 0 objects, indicating that the chemistry of the outer cold
envelope of this particular object is similar to that of other class 0
objects.
\end{enumerate}

The results of this paper show the potential of multi-transition line
observations to understand radial variations and find empirical
correlations between various species elucidating the chemistry in
star-forming environments. Future observational studies should focus
on deeper systematic searches, e.g., in order to fully understand the
sulfur chemistry and to constrain abundances of species connecting the
nitrogen-bearing species to the overall chemical network. Also
high-resolution observations with, e.g., the SMA, CARMA and ALMA can
be used to test the predictions of radial variations; as discussed
elsewhere \citep{n1333i2art,hotcorepaper} the models presented here
serve as a valuable starting point for interpreting such data. This
study has not addressed the chemistry of the innermost hot core region
where complex organic molecules may be formed through evaporation of
ices and high-temperature gas-phase chemistry. Since these regions are
small (typically $<1"$ for the objects considered here), high angular
resolution observations of high-excitation lines are needed to probe
them. Finally, the results presented here will form the basis for more
detailed modeling of the envelope chemistry.

\begin{acknowledgement}
This work is the result of an extensive observing program at the James
Clerk Maxwell Telescope (JCMT) on Hawaii. The authors are grateful to
the JCMT staff, in particular Remo Tilanus, for excellent technical
assistance and support before, during and after numerous observing
sessions. Thanks also goes to various observers who have carried out
parts of the observations in ``service'' mode. We are especially
grateful to Sebastien Maret and Cecilia Ceccarelli for useful
discussions and for the IRAM 30~m observations. Michiel Hogerheijde
and Floris van der Tak are acknowledged for making their Monte Carlo
code publically available and for helpful discussions. We are also
grateful to Steve Doty for the use of his chemical code. Participation
in conferences, where parts of these results were presented and
discussed, were financially supported by Leids Kerkhoven-Bosscha
Fond. The work of JKJ is funded by the Netherlands Research School for
Astronomy (NOVA) through a Ph.D. stipend and research in
astrochemistry in Leiden is supported by a NWO Spinoza grant. FLS
further acknowledges support from the Swedish Research Council.
\end{acknowledgement}

\bibliographystyle{aa}

\appendix
\section{The chemical network for HCO$^+$, N$_2$H$^+$, and H$_3^+$}\label{chemical_network_details}
This appendix explores the chemical network for HCO$^+$, N$_2$H$^+$,
and H$_3^+$ (Fig.~\ref{co_n2hp_hcop}). As discussed in
Sect.~\ref{hcop} the dominant formation and destruction mechanisms for
HCO$^+$ are:
\begin{equation}
{\rm H^+_3} + {\rm CO}  \rightarrow {\rm HCO^+} + {\rm H_2} \label{a_hcop_form}
\end{equation}
\begin{equation}
{\rm HCO^+} + {\rm e^-} \rightarrow {\rm CO} + {\rm H} \label{a_hcop_dest}
\end{equation}
The main formation mechanism for N$_2$H$^+$ is:
\begin{equation}
{\rm H^+_3} + {\rm N_2} \rightarrow {\rm N_2H^+} + {\rm H_2} \label{a_n2hp_form}
\end{equation}
When the CO abundance is low the main removal mechanism for N$_2$H$^+$ is
through dissociative recombination:
\begin{equation}
{\rm N_2H^+} + {\rm e^-} \rightarrow {\rm N_2} + {\rm H} \label{a_n2hp_dest_depl}
\end{equation}
whereas for ``standard'' CO abundances of 1$\times 10^{-4}$ the main removal
mechanism becomes:
\begin{equation}
{\rm N_2H^+} + {\rm CO} \rightarrow {\rm N_2} + {\rm HCO^+} \label{a_n2hp_dest_std}
\end{equation}
For H$_3^+$ the main formation route is through cosmic ray ionization
of H$_2$:
\[
{\rm H_2} + {\rm CR}  \rightarrow {\rm H_2^+} + {\rm e^-}              
\]
\begin{equation}
{\rm H_2^+} + {\rm H_2} \rightarrow {\rm H_3^+} + {\rm H} \label{a_h3p_form}
\end{equation}
The main removal mechanism for H$_3^+$ is through
eq.~(\ref{a_n2hp_form}) when the CO abundance is low and through
eq.~(\ref{a_hcop_form}) when the CO abundance is close to the standard
value $\sim 10^4$.

We introduce the rate coefficients for each reaction with $k_{\rm
HCO^+}$, $k_{\rm CO}$, and $k_{\rm N_2H^+}$ being the rate
coefficients for eq.~(\ref{a_hcop_form}), (\ref{a_hcop_dest}), and
(\ref{a_n2hp_form}), respectively, and $k_{\rm N_2}$ and $k^\prime_{\rm
N_2}$ being the rate coefficients for eq.~(\ref{a_n2hp_dest_depl}) and
(\ref{a_n2hp_dest_std}), respectively.

We can equate the formation and destruction rates of HCO$^+$ as:
\begin{equation}
k_{\rm HCO^+} \, n_{\rm H_3^+} \, n_{\rm CO} \, = \,  k_{\rm CO} \, n_{\rm HCO^+} \, n_{\rm e} \label{eqna}
\end{equation}
where $n_{\rm HCO^+}$, $n_{\rm H_3^+}$, $n_{\rm CO}$ and $n_{\rm e}$ are
the densities of HCO$^+$, H$_3^+$, CO and electrons,
respectively. 

When the CO abundance is $\sim 10^{-4}$ the main removal mechanism for
H$_3^+$ is eq.~(\ref{a_hcop_form}), so equating the formation and
destruction rate for H$_3^+$ gives:
\begin{equation}
\zeta\, n_{\rm H_2} \, = \, k_{\rm HCO^+} \, n_{\rm H_3^+} \, n_{\rm CO} \label{eqnb}
\end{equation}
where $\zeta$ is the cosmic ray ionization rate, $n_{\rm H_2}$ the
H$_2$ density and it is assumed that each H$_2^+$ molecular ion
produced in eq.~(\ref{a_h3p_form}) immediately reacts with H$_2$ to
form H$_3^+$. From eq.~(\ref{eqna}) and (\ref{eqnb}) we can then
write:
\begin{equation}
\zeta\, n_{\rm H_2} \, = \,  k_{\rm CO} \, n_{\rm HCO^+} \, n_{\rm e} \label{eqnc}
\end{equation}
which dividing by $n_{\rm H_2}^2$, introducing the abundances $[{\rm
X}]=n_{\rm X}/n_{\rm H_2}$ and isolating [HCO$^+$] gives:
\begin{equation}
[{\rm HCO^+}]=\frac{\zeta/n_{\rm H_2}}{k_{\rm CO}\,[{\rm e}]} \label{eqn_hcop_std}
\end{equation}
We can similarly equate the formation and destruction rates for N$_2$H$^+$
(eq.~(\ref{a_n2hp_form}) and (\ref{a_n2hp_dest_std})):
\begin{equation}
k_{\rm N_2H^+} \, n_{\rm H_3^+} \, n_{\rm N_2} \, = \,  k^\prime_{\rm N_2} \, n_{\rm N_2H^+} \, n_{\rm CO} \label{eqnd}
\end{equation}
which together with eq.~(\ref{eqnb}) gives (eliminating $n_{\rm
H_3^+}$):
\begin{equation}
k_{\rm N_2H^+} \, n_{\rm N_2}\,\frac{\zeta\,n_{\rm H_2}}{k_{\rm HCO^+}\,n_{\rm CO}} = k^\prime_{\rm N_2} \, n_{\rm N_2H^+} \, n_{\rm CO} \label{eqne}
\end{equation}
which can be rewritten in terms of the N$_2$H$^+$ abundance:
\begin{equation}
[{\rm N_2H^+}]=\frac{[{\rm N_2}]}{[{\rm CO}]^2}\,\frac{k_{\rm N_2H^+}}{k^\prime_{\rm N_2}\,k_{\rm HCO^+}}\,\zeta/n_{\rm H_2} \label{eqn_n2hp_std}
\end{equation}

In the low [CO] limit the main removal mechanism for H$_3^+$ is by
eq.~(\ref{a_n2hp_form}) rather than by eq.~(\ref{a_hcop_form}), so in this
limit we write for the formation/destruction balance for H$_3^+$
\begin{equation}
\zeta n_{\rm H_2} \, = \, k_{\rm N_2H^+} \, n_{\rm H_3^+} \, n_{\rm N_2} \label{eqnf}
\end{equation}
replacing eq.~(\ref{eqnb}). From eq.~(\ref{eqna}) and (\ref{eqnf}) we can
then eliminate $n_{\rm HCO^+}$:
\begin{equation}
n_{\rm HCO^+}=n_{\rm CO}\frac{k_{\rm HCO^+}}{k_{\rm CO}}\frac{\zeta\,n_{\rm H_2}}{n_{\rm e}\,k_{\rm N_2H^+}\,n_{\rm N_2}}
\end{equation}
or in terms of the abundances:
\begin{equation}
[{\rm HCO^+}]=\frac{k_{\rm HCO^+}}{k_{\rm CO}\,k_{\rm N_2H^+}}\frac{\zeta/n_{\rm H_2}}{[{\rm e}]\,[{\rm N_2}]}\,[{\rm CO}]\label{eqn_hcop_depl}
\end{equation}
Furthermore, in this limit the destruction of N$_2$H$^+$ is by
dissociative recombination (eq.~(\ref{a_n2hp_dest_depl}), so for N$_2$H$^+$
we can write:
\begin{equation}
k_{\rm N_2H^+} \, n_{\rm H_3^+} \, n_{\rm N_2} \, = \, k_{\rm N_2}\,n_{\rm e}\,n_{\rm N_2H^+} \label{eqng}
\end{equation}
and again eliminating $n_{\rm H_3^+}$, rewriting the expression in
terms of abundances, we find:
\begin{equation}
[{\rm N_2H^+}]=\frac{\zeta/n_{\rm H_2}}{k_{\rm N_2}\,[{\rm e}]}\label{eqn_n2hp_depl}
\end{equation}

The main points to be drawn from these equations (i.p.,
eq.~(\ref{eqn_hcop_std}), (\ref{eqn_n2hp_std}), (\ref{eqn_hcop_depl})
and (\ref{eqn_n2hp_depl})) are that the HCO$^+$ abundance does not
depend on the CO abundance in the ``high'' [CO] limit, whereas the
N$_2$H$^+$ abundance declines quickly as $([{\rm CO}])^{-2}$. In the low
[CO] limit, however, the HCO$^+$ abundance increases linearly with
increasing CO abundance whereas the N$_2$H$^+$ abundance is constant.
\end{document}